\documentclass[twocolumn]{aastex63}
\usepackage{amsmath}
\usepackage{natbib}
\usepackage{float}
\usepackage{CJKutf8} 
\usepackage{placeins}

\newcommand{\teff}{$T_\mathrm{eff}$}
\newcommand{\logg}{$\log{g}$}

\newcommand{\vsini}{$v\sin{i}$}
\newcommand{\kms}{km~s$^{-1}$}

\newcommand{\caltech}{Department of Astronomy, California Institute of Technology, Pasadena, CA 91125, USA}
\newcommand{\gps}{Division of Geological \& Planetary Sciences, California Institute of Technology, Pasadena, CA 91125, USA}
\newcommand{\ucsc}{Department of Astronomy \& Astrophysics, University of California, Santa Cruz, CA95064, USA}
\newcommand{\keck}{W. M. Keck Observatory, 65-1120 Mamalahoa Hwy, Kamuela, HI, USA}
\newcommand{\ucla}{Department of Physics \& Astronomy, 430 Portola Plaza, University of California, Los Angeles, CA 90095, USA}
\newcommand{\jpl}{Jet Propulsion Laboratory, California Institute of Technology, 4800 Oak Grove Dr.,Pasadena, CA 91109, USA}
\newcommand{\ucsd}{Center for Astrophysics and Space Sciences, University of California, San Diego, La Jolla, CA 92093}

\newcommand{\osu}{Department of Astronomy, The Ohio State University, 100 W 18th Ave, Columbus, OH 43210 USA}

\newcommand{\arizona}{James C. Wyant College of Optical Sciences, University of Arizona, Meinel Building 1630 E. University Blvd., Tucson, AZ. 85721}

\submitjournal{ApJ}

\shorttitle{High-Resolution Spectra of HD~33632 Ab}
\shortauthors{Hsu et al.}

\graphicspath{{./}{figures/}}

\begin{document}

\title{Rotation and Abundances of the Benchmark Brown Dwarf HD~33632 Ab from Keck/KPIC High-resolution Spectroscopy}

\correspondingauthor{Chih-Chun Hsu}
\email{chsu@northwestern.edu}

\author[0000-0002-5370-7494]{Chih-Chun Hsu}
\affil{Center for Interdisciplinary Exploration and Research in Astrophysics (CIERA), Northwestern University,
1800 Sherman Ave, Evanston, IL, 60201, USA}

\author[0000-0003-0774-6502]{Jason J. Wang
\begin{CJK*}{UTF8}{gbsn}(王劲飞)\end{CJK*}}
\affil{Center for Interdisciplinary Exploration and Research in Astrophysics (CIERA), Northwestern University,
1800 Sherman Ave, Evanston, IL, 60201, USA}
\affil{Department of Physics and Astronomy, Northwestern University, 2145 Sheridan Rd, Evanston, IL 60208, USA}

\author[0000-0002-6618-1137]{Jerry W. Xuan}
\affiliation{\caltech}

\author[0000-0003-2233-4821]{Jean-Baptiste Ruffio}
\affiliation{\ucsd}
\affiliation{\caltech}

\author[0000-0003-3165-0922]{Evan Morris}
\affiliation{\ucsc}

\author[0000-0002-1583-2040]{Daniel Echeverri}
\affiliation{\caltech}

\author[0000-0002-6171-9081]{Yinzi Xin}
\affiliation{\caltech}

\author[0000-0002-4934-3042]{Joshua Liberman}
\affiliation{\caltech}
\affiliation{\arizona}

\author[0000-0002-1392-0768]{Luke Finnerty}
\affiliation{\ucla}

\author[0000-0001-9708-8667]{Katelyn Horstman}
\affiliation{\caltech}

\author[0000-0003-1399-3593]{Ben Sappey}
\affiliation{\ucsd}


\author{Gregory W. Doppmann} 
\affiliation{\keck}

\author[0000-0002-8895-4735]{Dimitri Mawet}
\affiliation{\caltech}
\affiliation{\jpl}

\author[0000-0001-5213-6207]{Nemanja Jovanovic}
\affiliation{\caltech}

\author[0000-0002-0176-8973]{Michael P. Fitzgerald}
\affiliation{\ucla}

\author[0000-0001-8953-1008]{Jacques-Robert Delorme}
\affiliation{\keck}

\author[0000-0001-5299-6899]{J. Kent Wallace}
\affiliation{\jpl}

\author[0000-0002-6525-7013]{Ashley Baker}
\affiliation{\caltech}

\author{Randall Bartos} 
\affiliation{\jpl}

\author[0000-0003-0787-1610]{Geoffrey A. Blake}
\affiliation{\gps}

\author[0000-0003-4737-5486]{Benjamin Calvin}
\affiliation{\caltech}
\affiliation{\ucla}

\author{Sylvain Cetre} 
\affiliation{\keck}

\author[0000-0002-2019-4995]{Ronald A. L\'opez}
\affiliation{\ucla}

\author{Jacklyn Pezzato} 
\affiliation{\caltech}

\author{Tobias Schofield}
\affiliation{\caltech}

\author[0000-0001-6098-3924]{Andrew Skemer}
\affiliation{\ucsc}

\author[0000-0002-4361-8885]{Ji Wang}
\affiliation{\osu}


\begin{abstract}

We present the projected rotational velocity and molecular abundances for HD~33632 Ab obtained via Keck Planet Imager and Characterizer high-resolution spectroscopy.
HD~33632 Ab is a nearby benchmark brown dwarf companion at a separation of $\sim$20~au that straddles the L/T transition.
Using a forward-modeling framework with \added{on-axis host star spectra,} self-consistent substellar atmospheric and retrieval models for HD 33632 Ab, we derive a projected rotational velocity of 53 $\pm$ 3~{\kms} and
\replaced{water plus carbon}{carbon/water} monoxide mass fractions of $\log{\mathrm{CO}}$ = $-2.3 \pm 0.3$ and $\log{\mathrm{H_2O}}$ = $-2.7 \pm 0.2$. 
The inferred carbon-to-oxygen ratio (C/O = $0.58 \pm 0.14$), molecular abundances, and metallicity ([C/H] = $0.0 \pm 0.2$~dex) of HD 33632 Ab are consistent with its host star. 
Although detectable methane opacities are expected in L/T transition objects, we did not recover methane in our KPIC spectra, partly due to the high $v$sin $i$ and to disequilibrium chemistry at the pressures we are sensitive to.
We parameterize the spin as the ratio of rotation over break-up velocity, 
and compare HD~33632 Ab to a compilation of $>$200 very low-mass objects ($M$$\lesssim$0.1 M$_{\odot}$) that have spin measurements in the literature. 
There appears to be no clear trend for the isolated field low-mass objects versus mass, but a tentative trend is identified for low-mass companions and directly imaged exoplanets, similar to previous findings.
A larger sample of close-in gas giant exoplanets and brown dwarfs will critically examine our understanding of their formation and evolution through rotation and chemical abundance measurements.

\end{abstract}


\section{Introduction} \label{sec:intro}

More than 5500 exoplanets have been discovered since the discovery of 51 Peg b, the first exoplanet around a main sequence star \citep{Mayor:1995aa}, with $<$30 directly imaged (gas giant) exoplanets \citep[e.g.][]{Marois:2008aa, Macintosh:2015aa, Franson:2023ab}\footnote{The number of the directly imaged exoplanets is based on of \cite{Currie:2023ab} and the NASA Exoplanet Archive.}.
In contrast, tens of thousands of brown dwarfs have been found within a klioparsec (kpc) of the Sun. 
Brown dwarfs share similar temperatures and atmospheric properties with gas giant exoplanets \citep{Burrows:1997aa, Burrows:2001aa, Baraffe:2003aa}, but are more massive and brighter at the same age, allowing us to characterize their atmospheric properties more easily.

Measuring the rotation rates and atmospheric content of brown dwarfs and gas giant exoplanets via molecular spectroscopy enables constraining the theories of their formation and evolution, because
the rotation of these low-mass objects imprints their angular momentum evolution.
As brown dwarfs evolve into field age ($>$1~Gyr), they lose little angular momenta, and spin up until their radii contract to $\sim$1~R$_\mathrm{Jup}$, with the size governed by electron degeneracy pressure \citep{Zapatero-Osorio:2006aa, Vos:2017aa, Hsu:2021aa, Vos:2022aa}.
In the planetary-mass regime, objects also spin up and roughly follow angular momentum conservation as they age after disk clearance \citep{Bryan:2020ab}.
During their earlier phases, the magnetic fields of low-mass companions can interact with the magnetized circumplanetary disk and release their angular momenta \citep{Batygin:2018aa}. 
It is predicted that the angular momentum loss is mass-dependent because gas giant exoplanets or brown dwarf companions have different levels of magnetic fields, regulated by their masses, temperatures, and therefore degrees of ionization \citep{Ginzburg:2020aa}.
Tentative evidence of anti-correlation has been found between the rotation and companion mass \citep{Bryan:2020ab, Wang:2021aa}, in a small sample size of $<$20 low-mass companions.

For chemical abundances, the carbon-to-oxygen (C/O) ratio has been widely used to infer the location of planet formation, as the ice lines of H$_2$O, CO$_2$ and CO are different which lead to different observed (gas-phase) C/O ratios for planets assembled at different semi-major axes in the disk \citep{Oberg:2011ab, Madhusudhan:2012ab, Konopacky:2013aa, Lavie:2017aa, Nowak:2020aa, Molliere:2020aa, Molliere:2022aa, Whiteford:2023aa, Hoch:2023aa, Nasedkin:2024ab}.
Brown dwarf companions, either formed through gravitational core collapse or disk instability, are expected to share the same C/O ratios with their host stars because their formation is star-like \citep{Bate:2005aa}.
Detailed characterization of brown dwarf abundances not only facilitates our understanding of the star formation at the very low mass end but also serves as an independent calibration of the methodology used to characterize gas giant exoplanets \citep[e.g.,][]{Burningham:2017aa, Wang:2022aa, Adams:2023aa}.

However, a major challenge of studying brown dwarfs as exoplanet analogs is the observed degeneracy between age, mass, and bolometric luminosity.
Because brown dwarfs are unable to fuse hydrogen and are supported by electron degeneracy pressure \citep{Kumar:1962aa, Kumar:1963aa}, they constantly cool and evolve under hydrostatic equilibrium.
A given spectral type, which corresponds to a given range of effective temperature, could represent either a low-mass star, a brown dwarf, or a very young, hot exoplanet at $\sim$3000~K \citep{Burrows:2001aa, Baraffe:2003aa}.
One way to break the age degeneracy is through identifications of brown dwarfs in young star clusters or moving groups \citep{Gagne:2017aa, Gagne:2018aa, Gagne:2018ab, Schneider:2023aa}.
Another route is through the discovery of brown dwarfs in FGK binary systems because these solar-like stars have independent estimates of age, mass, and metallicity.
While wide binary FGK systems with brown dwarf companions have been identified (e.g. \citealp{Faherty:2021aa, Zhou:2023aa}), such widely separated companions typically have long orbital periods -- meaning that the dynamical masses are not well constrained over a timescale of a few decades.
Close companions, on the other hand, provide reliable dynamical masses through direct imaging and absolute and relative astrometry with \textit{Gaia} \citep{Gaia-Collaboration:2016aa, Gaia-Collaboration:2018ab, Gaia-Collaboration:2021ac, Gaia-Collaboration:2023aa} and \textit{Hipparcos} \citep{van-Leeuwen:1997aa, Perryman:1997ab, ESA:1997aa} as they provide better orbital phase coverage.

Thus, close-in, low-mass companions can serve as benchmark objects with independent ages and dynamical masses, but the precise characterization of their rotation and abundances are difficult due to being so close to the bright host star and the high contrasts (5--15~mag) that must be achieved to isolate them. 
Designed to overcome these challenges, the Keck Planet Imager and Characterizer (KPIC; \citealp{Mawet:2016aa, Mawet:2017aa, Mawet:2018aa, Delorme:2021aa}) is a fiber injection unit connecting the Keck/NIRSPEC spectrometer (R$\sim$35,000; \citealp{McLean:1998aa, McLean:2000aa, Martin:2018aa}) to the Keck II AO system via a single-mode fiber to provide high-resolution $K$-band spectroscopy at high-angular resolution.
With its high spectral resolution, rotation and chemical abundances can be reliably measured.
Several benchmark brown dwarfs and gas giant exoplanets have been characterized and reported with KPIC, providing robust abundances and rotation, including HR~8799 cde planets \citep{Wang:2021aa,Wang:2023aa}, HR~7672B \citep{Wang:2022aa} and HD~4747 B \citep{Xuan:2022aa}. 

HD~33632~Ab is a brown dwarf companion of mass 46$\pm$8~M$_\mathrm{Jup}$ \citep{Currie:2020aa} that straddles the L/T transition (L9.5$^{+1.0}_{-3.0}$) around the F8V star HD~33632A at $\sim$20~au separation.
The system was initially identified by a \textit{Gaia}/\textit{Hipparcos} acceleration trend and confirmed with SCExAO/CHARIS and Keck/NIRC2 imaging from \cite{Currie:2020aa}.
The host star HD~33632A has a field age of 1.0--2.5~Gyr \citep{Currie:2020aa} and slightly subsolar metallicity ([Fe/H] = $-0.15 \pm 0.03$~dex; \citealp{Rice:2020aa}).
With independent age, metallicity (abundances), and dynamical mass,  
HD~33632~Ab serves as a benchmark brown dwarf to break the observational degeneracy among the population of brown dwarfs and gas giant exoplanets.
The system properties are summarized in Table~\ref{table:system_properties}.

In this work, we present the follow-up Keck/KPIC observation of the HD~33632~Ab system to derive its companion rotation rate, radial velocity, and abundances.
Our manuscript is organized as follows.
In Section~\ref{sec:observe} we describe our KPIC observations.
In Section~\ref{sec:ccf} we show our detection of CO and H$_2$O using the cross-correlation method.
In Section~\ref{sec:forward_model} we forward-model our KPIC spectra using self-consistent substellar atmosphere models to derive radial and projected rotational velocities.
In Section~\ref{sec:retrieval} we employ our retrieval modeling framework to extract the CO and H$_2$O abundances and validate the non-detection of methane in the KPIC data.
In Section~\ref{sec:orbit} we update the orbital solutions of the HD~33632 Ab system using our measured companion RV and updated astrometry.
We compare the rotation of HD~33632 Ab with other low-mass objects to examine if there exists a population-level trend in Section~\ref{sec:rotation}.
We summarize our findings in Section~\ref{sec:sum}.


\begin{deluxetable}{lcc}
\tablecaption{HD~33632 System Properties \label{table:system_properties}}
\tablecolumns{3}
\tablehead{
\colhead{Property (unit)} &  \colhead{Value}  & \colhead{Ref.}
}
\startdata
\multicolumn{3}{c}{HD33632~A}\\
\hline
R.A. (J2000) & 05:13:17.45 & (1) \\
Dec. (J2000) & +37:20:14.32 & (1) \\
$\mu_{\alpha}$ (mas yr$^{-1}$) & $-144.922 \pm 0.031$ & (1) \\
$\mu_{\delta}$ (mas yr$^{-1}$) & $-136.772 \pm 0.022$ & (1) \\
Mass (M$_{\odot}$) & 1.1 $\pm$ 0.1 & (2) \\
Age (Gyr) & 1.0--2.5 & (2) \\
SpT & F8V & (3) \\
\textit{Gaia} $G$ & 6.351$\pm$0.003 & (1) \\
$J_\mathrm{\, MKO}$ (mag) & 5.43 $\pm$ 0.02 & (4) \\
$H_\mathrm{\, MKO}$ (mag) & 5.193 $\pm$ 0.015 & (4) \\
$K_{\rm S, \, MKO}$ (mag) & 5.17 $\pm$ 0.02 & (4) \\
$\pi$ (mas) & 37.895 $\pm$ 0.026 & (1) \\
distance (pc) & 26.388 $\pm$ 0.018 & (1) \\
RV ({\kms}) & $-1.75 \pm 0.12$ & (1) \\
{\vsini} ({\kms}) & $\lesssim$4 & (5)\tablenotemark{a}; (6) \\
{[}Fe\slash H{]} & $-0.15 \pm 0.03$ & (6) \\
{[}C\slash H{]} & $-0.13 \pm 0.05$ & (6) \\
C\slash O & 0.39$^{+0.12}_{-0.09}$ & (6); (7) \\
\hline
\multicolumn{3}{c}{HD33632~Ab}\\
\hline
Mass (M$_\mathrm{Jup}$) & 46 $\pm$ 8 & (2) \\
\nodata & 37 $\pm$ 7 & (7) \\
SpT & L9.5$^{+1.0}_{-3.0}$ & (2) \\
$J_\mathrm{\, MKO}$ (mag) & 16.91 $\pm$ 0.11 & (2) \\
$H_\mathrm{\, MKO}$ (mag) & 16.00 $\pm$ 0.09 & (2) \\
$K_{\rm S, \, MKO}$ (mag) & 15.37 $\pm$ 0.09 & (2) \\
{\vsini} ({\kms}) & $53 \pm 3$ & (7) \\
RV ({\kms})\tablenotemark{b} & $-8\pm3$ & (7) \\
{[}C\slash H{]} (dex) & 0.0$\pm$0.2 & (7) \\
C\slash O & 0.58$\pm$0.14 & (7) \\
$a$ (au) & 18$^{+5}_{-3}$ & (7) \\
$e$ & 0.25$^{+0.17}_{-0.18}$ & (7) \\
$i$ (deg) & 33$^{+11}_{-21}$ & (7) \\
$P$ (yr) & 74$^{+34}_{-16}$ & (7) \\
\enddata
\tablenotetext{a}{Note that the measured {\vsini} = 0.44~{\kms} in \cite{Rice:2020aa} using the Keck/HIRES data has a resolution limit of $\sim$4~{\kms}. }
\tablenotetext{b}{Barycentric RV measured on MJD 59538.436}
\tablerefs{(1) \cite{Gaia-Collaboration:2023aa}; (2) \cite{Currie:2020aa}; (3) \cite{Gray:2003aa}; (4) \cite{Cutri:2003aa}; (5) \cite{Nordstrom:2004aa}; (6) \cite{Rice:2020aa}; (7) This work}
\end{deluxetable}

\section{Observations and Data Reduction} \label{sec:observe}
We obtained our high-resolution near-infrared spectra of HD~33632A and HD~33632 Ab on 2021 September 18 (UT) and 2021 November 20 (UT), using the Keck Planet Imager and Characterizer (KPIC; \citealp{Mawet:2016aa, Mawet:2017aa, Mawet:2018aa, Delorme:2021aa}).
KPIC is the adaptive optics (AO) system mounted on the Keck II telescope that is optimized for high-resolution spectroscopy and high-contrast coronagraphic imaging, which couples the light using a single-mode fiber injection unit to the Near-InfraRed SPECtrometer (NIRSPEC; R$\sim$35,000; \citealt{McLean:1998aa, McLean:2000aa, Martin:2018aa, Lopez:2020aa}) and NIRC2 \citep{van-Dam:2006aa, Wizinowich:2006aa, Vargas-Catalan:2016aa, Serabyn:2017aa}.
The KPIC instrumental design is detailed in \cite{Mawet:2017aa,Delorme:2021aa}.
The ``Kband-new'' filter on NIRSPEC covers a wavelength range of 1.91--2.55~$\micron$.
KPIC has four single-mode fibers (fluoride 6.5/125~$\micron$) to place the science targets, and two fibers were typically used to enable nod subtraction.
On 2021 September 18, four spectra of HD~33632A were observed, each of 60~s integration time, and nine spectra of HD~33632 Ab were acquired, each 600~s integration time. The fibers were either 2 or 3.
For our calibration data, the M2.5III giant HIP 81497 was observed for wavelength calibration, while A0 HD 33704 served to obtain the spectral traces and the instrumental response function.
On 2021 November 20, fourteen spectra of HD~33632A were observed, each of 30~s integration time, with twelve spectra of HD~33632 Ab obtained, each of 600~s integration time. The fibers were either 1 or 2. 
Here, the M0.5III giant HIP 95771 served as the wavelength calibrator, while A1V $^*$ ome Aur \citep{Zuckerman:2011aa} was observed for spectral order tracing and spectral response.
On both nights, the NIRSPEC backgrounds with the same integration were also obtained, so that the sky backgrounds could be either removed using another nodding position or the thermal background measured under the same integration time (see below).

The data were reduced using the \texttt{KPIC Data Reduction Pipeline}\footnote{\url{https://github.com/kpicteam/kpic_pipeline}}, detailed in Section~3 of \cite{Wang:2021aa}.
The reduction steps include instrumental thermal background subtraction, trace-finding using the telluric standard star (typically A0V stars)\footnote{The telluric standard stars are used to identify the trace and derive the spectral response. Our reduced spectra include the telluric absorption and will be forward modeled using the observed star and telluric spectra. See Section~\ref{sec:forward_model} for details.}, the standard spatial and spectral rectification and optimal extraction \citep{Horne:1986aa}, and finally wavelength calibration using the early M giant star spectra. 
The wavelength calibration precision of Keck/NIRSPEC spectrograph is typically around 0.1--0.5 {\kms}, using the earth absorption lines from telluric standard star spectra \citep{Blake:2010aa, Burgasser:2016aa, Hsu:2021aa, Theissen:2022aa, Hsu:2023aa}.

We assessed the data quality for each night based on the end-to-end throughput based on the associated M giant spectra using HIP 81497 and HIP 95771, respectively.
Using their $K$-band photometries (0.467~mag and 0.711~mag) from \cite{Cutri:2003aa} and effective temperatures (3774~K and 3972~K) from \cite{Stassun:2019aa}, the 95$^\mathrm{th}$-percentile peak throughputs of HIP 81497 on 2021 September 18 and HIP 95771 on 2021 November 20 are 1.3\% (poor KPIC performance) and 2.4\% (typical KPIC performance), respectively, as a result of a higher and more unstable seeing on 2021 September 18 compared to the 2021 November 20 night\footnote{The S/N, defined as the 99$^\mathrm{th}$ percentile of observed flux, $\sim$ 13 on 2021 September 18; the S/N $\sim$ 20 on 2021 November 20. The companion fluxes have S/Ns of $\sim$1.8 and $\sim$3.4, respectively, inferred from our best-fit companion fluxes in Section~\ref{sec:forward_model}.}.

Various subtraction methods for data reduction are used under different circumstances, to maximize the detection of the companion flux and minimize the speckle from the host star.
The first subtraction method uses the thermal background of the same integration as the science file. 
The second subtraction method, nod-subtraction, uses the science file of the same integration time on another fiber, similar to the traditional pair subtraction on slit-fed spectroscopy. 
For our HD~33632 Ab data on 2021 November 20, we found that the nod-subtracted flux on star fiber 1 and background-subtracted flux on star fiber 2 provided the highest signal-to-noise, as the speckle flux on fiber 2 caused poor subtraction(s).
For data on the same fiber, the bad pixels were removed using a 3-sigma-clipping outlier rejection, and the resulting statistical uncertainties were then computed.
We focus on NIRSPEC orders 33 (2.29-2.34~{\micron}), 32 (2.36-2.41~{\micron}), and 31 (2.43-2.49~{\micron}), in which we could obtain robust wavelength calibrations with sufficient companion fluxes.

\section{Cross-Correlation Method to Detect Molecules} \label{sec:ccf}
The cross-correlation method using spectral templates is a powerful tool for detecting possible species in low-mass companions (e.g., \citealp{Konopacky:2013aa, Ruffio:2021aa, Wang:2021aa, Xuan:2022aa, Patapis:2022aa, Malin:2023aa}).
Specifically, we employed a least-squares cross-correlation function (CCF) to cross-correlate the observed companion flux and incorporate the star flux contribution, telluric and instrument response.
Our method has been detailed in \cite{Wang:2021aa} and \cite{Xuan:2022aa}, so we briefly summarize it here.

To cross-correlate the observed spectra (orders 31--33; 2.29--2.49~$\micron$) with the forward-model molecular templates\replaced{,}{.}
The forward model includes the companion and star flux contributions. For the companion flux component, we used the molecular templates (CO, H$_2$O, and CH$_4$) from the Sonora-Bobcat model sets (\citealp{Marley:2021aa}; see the justification in \citealp{Wang:2021aa}).
\replaced{they}{These templates} were first multiplied by the telluric absorption and the spectral response function.
The specific molecular templates were derived from the cloudless Sonora-Bobcat model set \citep{Marley:2018aa, Marley:2021aa} by turning on specific molecular species including carbon monoxide (CO), water (H$_2$O), methane (CH$_4$), and carbon monoxide and water (CO $+$ H$_2$O).
The telluric spectra were directly taken from our observed spectra of $^*$ ome Aur.
The spectral response function, defined as the sensitivity of a given spectral order, was measured by the observed flux (A1V $^*$ ome Aur) over the expected flux using the telluric spectra, normalized at the 99$^\mathrm{th}$ percentile flux ratio to reduce the effects from the outlier fluxes. 
For the theoretical atmospheric model of the telluric spectra, we used the \texttt{PHOENIX ACES AGSS COND} models \citep{Husser:2013aa}, assuming the effective temperature $T_\mathrm{eff}$ = 9400~K, surface gravity $\log{g}$ = 4.0~cm s$^{-2}$ dex, and solar metallicity, which is consistent with the empirical $T_\mathrm{eff}$ and $\log{g}$ relations from \cite{Pecaut:2013aa}.
For the star flux contribution, we used the observed on-axis star spectra as an empirical template. 
The cross-correlation signal of the companion flux can then be estimated through the maximum likelihood. We used the least-squares method by optimizing the observed companion spectra and the forward-model, as a function of radial velocity shift. Interested readers are referred to \cite{Wang:2021aa} for this mathematical formulation.
Note that our result might still be biased by imperfect modeling the star flux contribution (see Section~\ref{sec:forward_model} for details).

\begin{figure*}[ht]
    \centering
    \replaced{\includegraphics[width=0.49\textwidth]{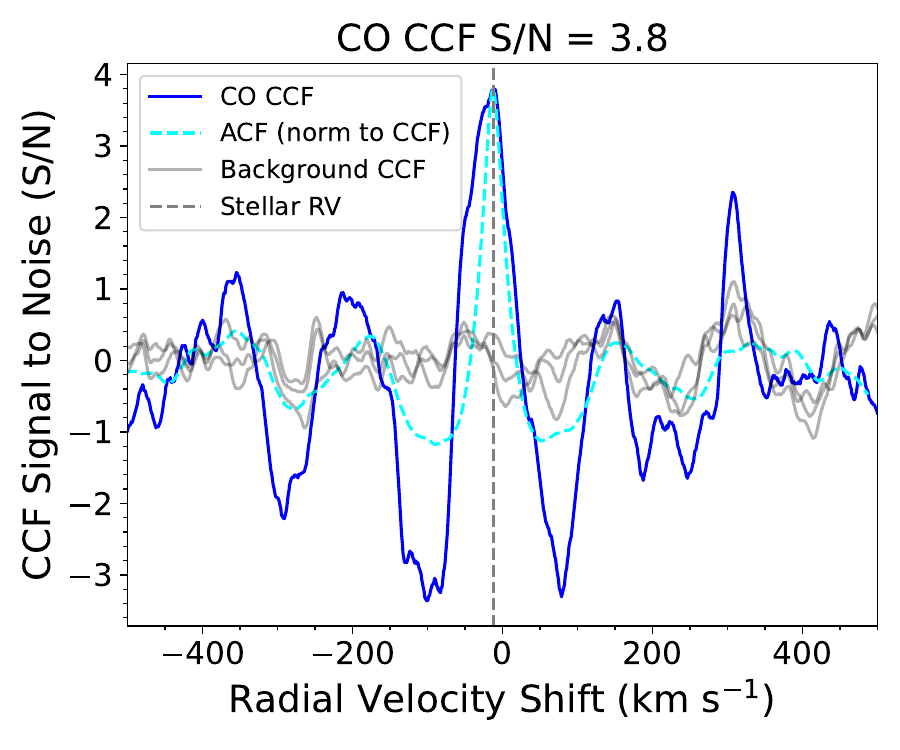}}
    {\includegraphics[width=0.49\textwidth, trim={0 1.3cm 0 0}]{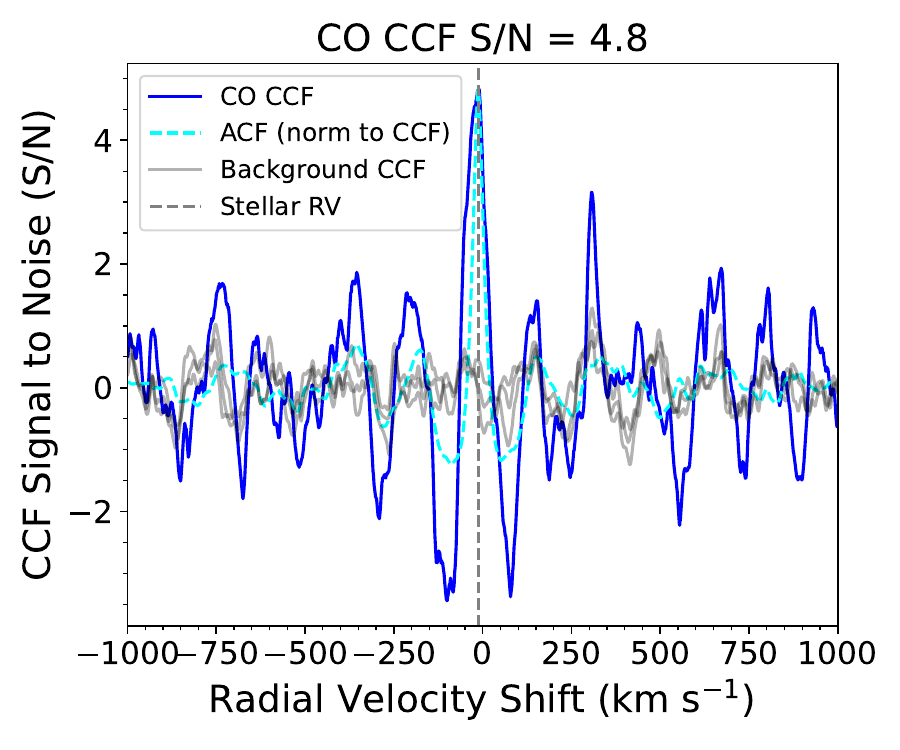}}
    \replaced{\includegraphics[width=0.49\textwidth]{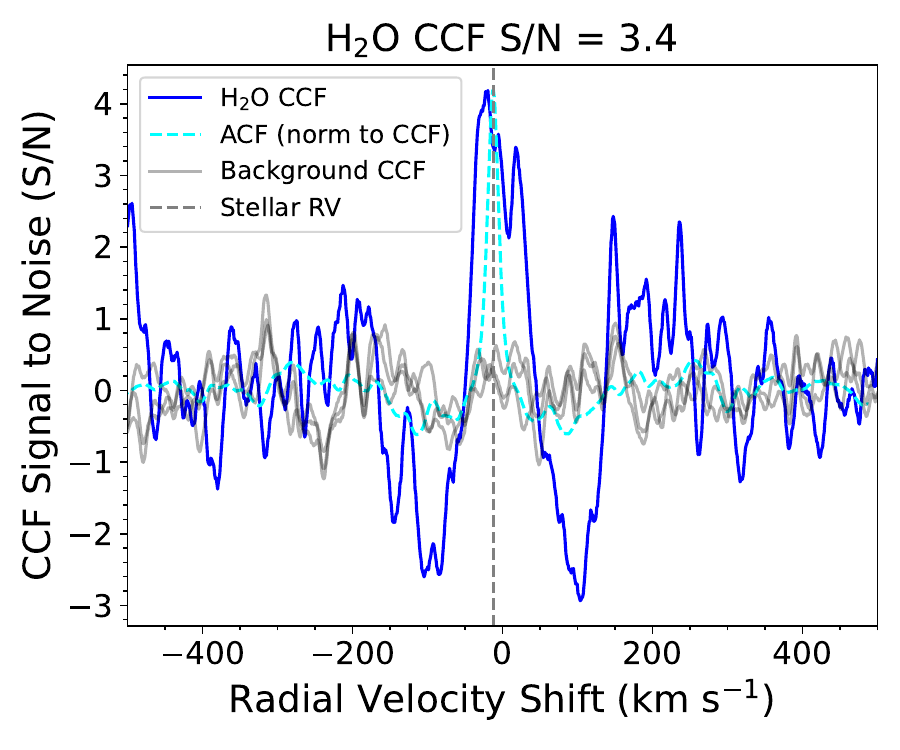}}
    {\includegraphics[width=0.49\textwidth, trim={0 1.3cm 0 0}]{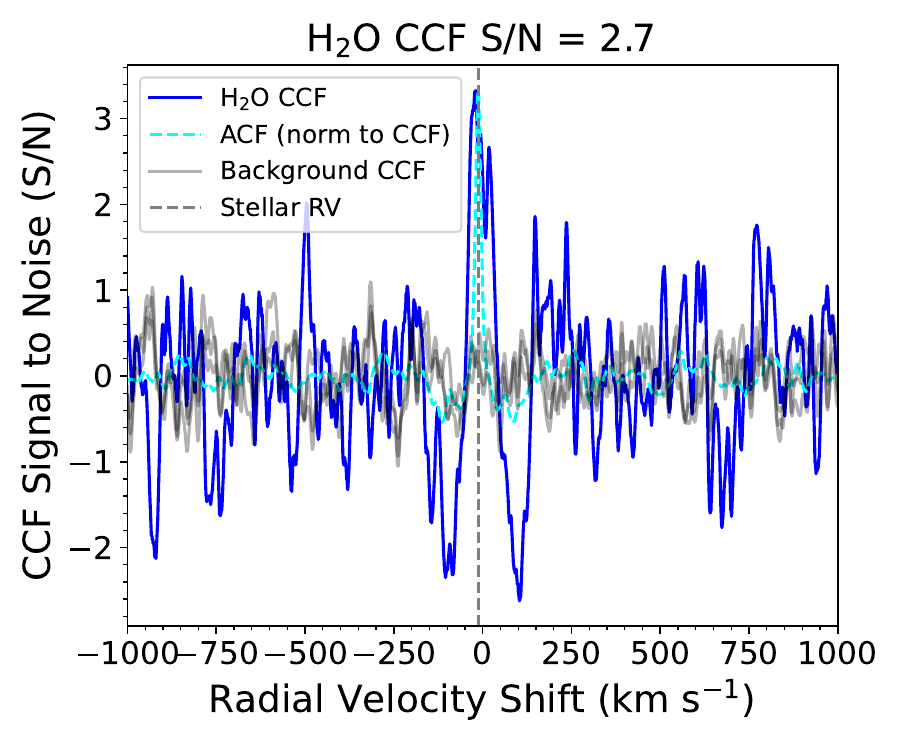}}
    \replaced{\includegraphics[width=0.49\textwidth]{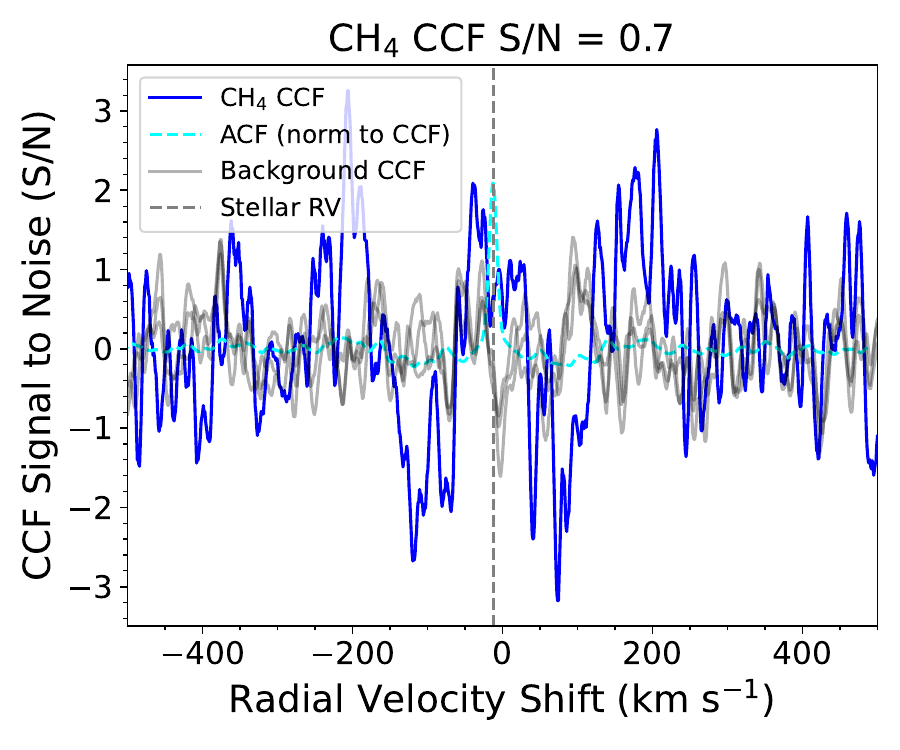}}
    {\includegraphics[width=0.49\textwidth, trim={0 1.3cm 0 0}]{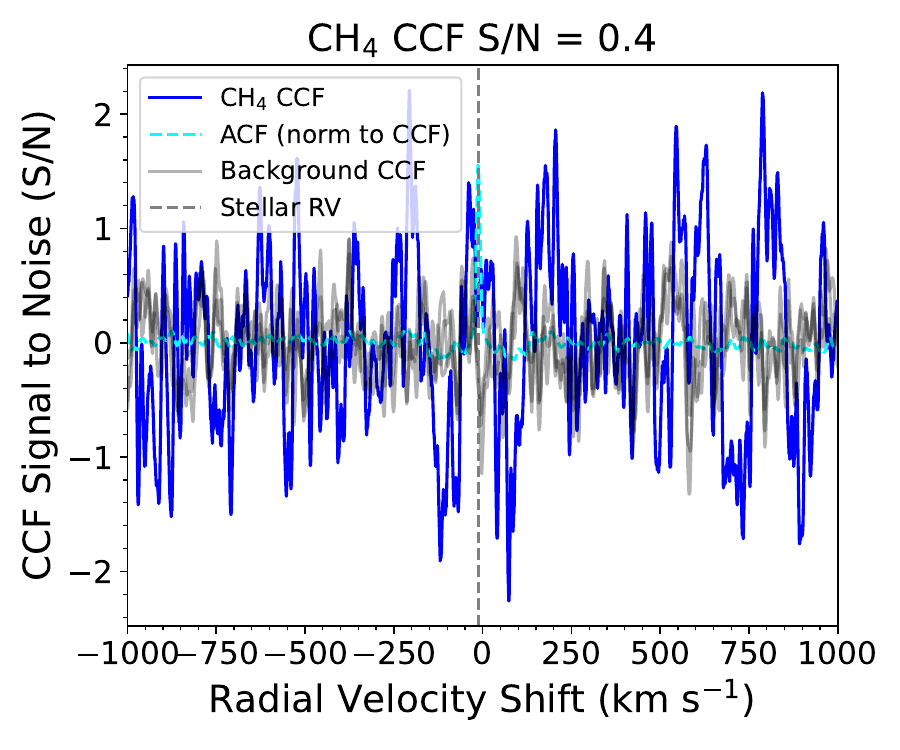}}
    \replaced{\includegraphics[width=0.49\textwidth]{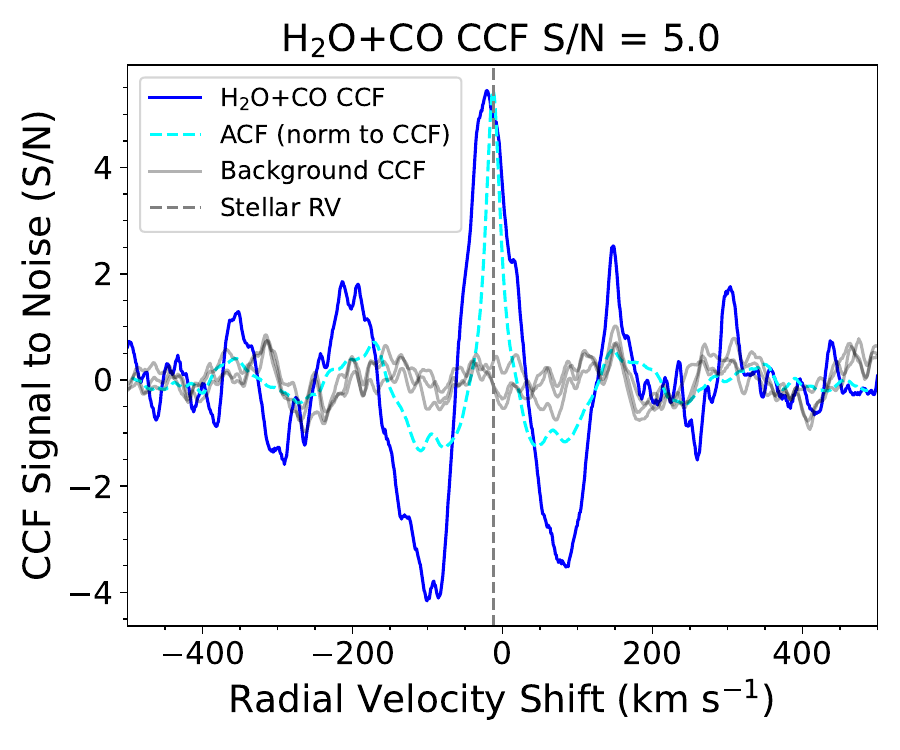}}
    {\includegraphics[width=0.49\textwidth, trim={0 1.3cm 0 0}]{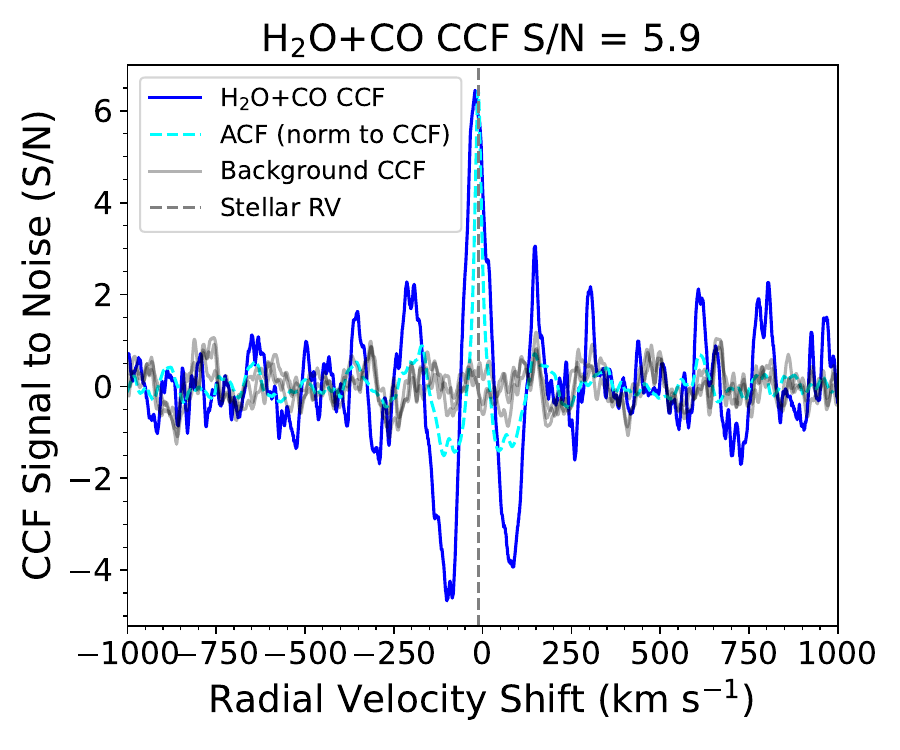}}
    \includegraphics[width=0.49\textwidth]{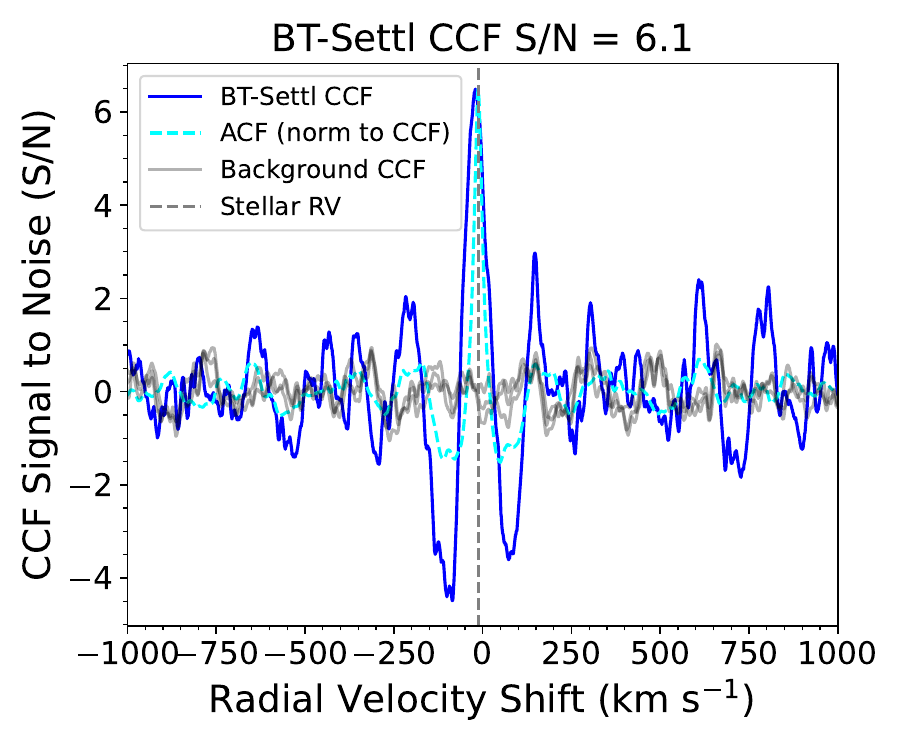}
    \includegraphics[width=0.49\textwidth]{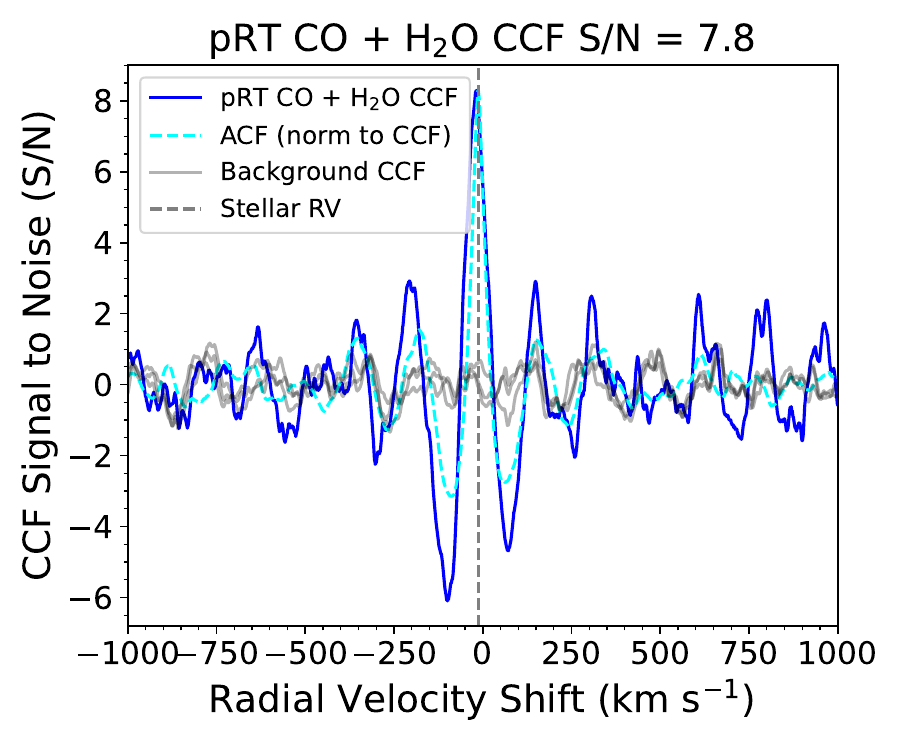}
    \caption{Cross-correlation function (CCF) of our KPIC spectrum of HD~33632 Ab with the molecular templates derived from the Sonora-Bobcat models \citep{Marley:2018aa, Marley:2021aa}.
    \textit{Upper-left}: CCF (blue solid line) of our KPIC spectra with respect to the CO molecular templates. 
    The background CCFs of three locations are shown in grey lines.
    The stellar barycentric-included RV ($-$12.65~{\kms}) is depicted in the vertical dashed grey line.
    The auto-correlation function (ACF) of the CO molecular templates, normalized to the peak of CCF, is plotted in dashed cyan line.
    \textit{Upper-right}: Same as the upper-left panel for the H$_2$O molecular templates.
    \replaced{\textit{Lower-left}}{\textit{Middle-left}}: Same as the upper-left panel for the CH$_4$ molecular templates.
    \replaced{\textit{Lower-right}}{\textit{Middle-right}}: Same as the upper-left panel for the CO and H$_2$O combined molecular templates. \textit{Lower-left}: Same as the upper-left panel for the best-fit BT-Settl models (Section~\ref{sec:forward_model}).
    \textit{Lower-right}: Same as the upper-left panel for the CO and H$_2$O combined molecular templates from the best-fit forward retrieval model (Section~\ref{sec:retrieval}).}
    \label{fig:ccf}
\end{figure*}

Our confidence in the detection(s) of molecular species was computed based on the signal-to-noise ratio (S/N), where the signal of molecular \replaced{cross-correlation function (CCF)}{CCF} was first normalized to the baseline using the CCF wings of the last $+/-$\replaced{200}{500} km s$^{-1}$, and then compared with the 
standard deviation of CCF wings of the same last $+/-$\replaced{200}{500} km s$^{-1}$ as the noise.
Under this formulation, the CCF S/N could be negative, and using the CCF wings to estimate allows us to quantify the instrument systematics in addition to the background noise.
The radial velocity (RV) shifts of CCF range from \replaced{$-500$~{\kms} to $+500$~{\kms}}{$-1000$~{\kms} to $+1000$~{\kms}} in the step of 1~{\kms}.
Figure~\ref{fig:ccf} illustrates the CCFs of HD~33632 Ab (with both stellar and brown dwarf fluxes), compared to the RV of the primary HD~33632A at the time of observation ($-$12.65~{\kms}), which includes its systematic RV = 
$-$1.75$\pm$0.12~{\kms} \citep{Gaia-Collaboration:2023aa} and its barycentric velocity at the time of the observation ($+$10.903~{\kms}).
We detected water (peak S/N = \replaced{3.4}{4.8}) and carbon monoxide (S/N = \replaced{3.8}{2.7}), with a combined CO $+$ H$_2$O S/N of \replaced{5.0}{5.9} in our NIRSPEC spectra using the cross-correlation method.
The CCF wings for CO and H$_2$O are largely attributed to the CO band and H$_2$O, roughly separated by their line widths, which can also be seen in the corresponding auto-correlation functions.
We were unable to detect CH$_4$ in our NIRSPEC spectra (S/N = 0.7) due to the relatively large CCF wings off the expected companion velocity as well as more comparable noise amplitude. The structure in the CCFs for the primary species can be used to further enhance the detection confidence, in the context of forward models, and so we re-examine if CH$_4$ is present in our NIRSPEC spectra of HD~33632Ab using the retrieval method in Section~\ref{sec:retrieval}. 
We examined telluric variations that cause false positive detection by excluding telluric strong regions (transmission $<$ 0.4) and found similar SNRs for all of the species examined above.
To further illustrate our confidence in detection of HD~33632Ab in our KPIC spectra, we also showed CCFs of the best-fit BT-Settl model and baseline forward retrieval CO and H$_2$O model and found CCF SNRs of 6.1 and 7.8, respectively (See Sections~\ref{sec:forward_model} and \ref{sec:retrieval} for details of deriving the best-fit models).

\section{Forward-Modeling Method} \label{sec:forward_model}

To extract the physical parameters of the brown dwarf companion, we used a forward-modeling framework to joint-fit the stellar speckle and companion fluxes with the earth's atmospheric and NIRSPEC instrumental broadening profiles due to the star companion flux contrast $\Delta K_S = 10.2$~mag.
This forward-modeling method has been adopted in several high-resolution and medium-resolution spectrometers for close brown dwarf companions and exoplanets, including Keck/KPIC \citep{Wang:2021aa, Wang:2022aa, Xuan:2022aa, Wang:2023aa, Ruffio:2023aa, Xuan:2024aa}, VLT/CRIRES+ \citep{Landman:2023aa}, Keck/OSIRIS \citep{Ruffio:2019aa, Wilcomb:2020aa, Ruffio:2021aa, Hoch:2023aa}, VLT/SINFONI \citep{Petrus:2021aa}, and VLT/HiRISE \citep{Vigan:2023aa}.
This method is particularly useful for modeling low signal-to-noise ratio data as our companion flux is lower than the speckle flux (see below).
We analyzed the NIRSPEC spectral orders 31 (2.43--2.49~{\micron}), 32 (2.36--2.41~{\micron}), and 33 (2.29--2.34~{\micron}). 
Following \cite{Wang:2021aa}, our forward model 
is shown in the equation below:

\begin{equation}
\begin{split}
F[p] & = \Bigg[ \alpha_{p} \times \bigg(M \Big[p^* \big(\lambda \big[ 1 + \frac{\mathrm{RV}^*}{c}\big] \big) , T_{\text{eff}}, \log \, g \Big]  \\ 
& \otimes \kappa_R (v\sin{i}) \bigg) \times T \big[ p^*(\lambda) \big] \Bigg] \otimes \kappa_G (\Delta \nu_\text{inst} \times \sigma_\mathrm{scale}) \\ 
& + \alpha_{s, f} \times D_s[p^*(\lambda)] \, .    
\end{split}
\label{eqn:forward_model_science}
\end{equation}

Here, $p^*(\lambda)$ is the pixel as a function of wavelength $\lambda$ from our wavelength calibration, $M$ is the self-consistent stellar atmosphere model grids as a function of effective temperature $T_{\text{eff}}$ (K) and surface gravity $\log{g}$ (cgs dex), which is then convolved with the projected rotational velocity $v\sin{i}$ ($\otimes$ denotes convolution, with $\kappa_R$ as the rotational broadening kernel), corrected for radial velocity $\mathrm{RV}^*$\footnote{The $\mathrm{RV}^*$ here is uncorrected for the barycentric velocity, which we corrected after the forward modeling routine and reported in Table~\ref{table:forward_model_result}.}.
The companion model is then corrected for the companion scale factor $\alpha_{p}$ to fit its observed flux contribution, and multiplied by the spectra response $T \big[ p^*(\lambda) \big]$ (derived from the observed A0V telluric standard star spectra).
We then model the stellar flux scale factor $\alpha_{s, f}$ for each fiber $f$ to model the speckle flux contribution $D_s[p^*(\lambda)]$ in the data (i.e. two speckle flux scale factors because there are two fibers observed\footnote{We typically observe the star and companion using two fibers to enable pair subtraction.}), instrumental profile scale term $\sigma_\mathrm{scale}$ for the instrumental profile $\Delta \nu_\text{inst}$ ($\kappa_G$ is the Gaussian broadening kernel), and noise jitter term in each order.
The speckle flux $D_s[p^*(\lambda)]$ is directly drawn from our observed on-axis star flux, which also includes the telluric absorption.
The term $\sigma_\mathrm{scale}$ is to account for the non-symmetric line spread function in the spatial and spectral directions \citep{Trujillo:2001aa}. We measured the line spread function in the spatial direction in the trace profile during our data reduction, and found that the spectral direction is typically larger by $\sim 10\%$ \citep{Wang:2021aa}.
Additionally, we include the noise jitter term $\sigma_\mathrm{jitter}$, sampled in log space, and combined it with the extracted noise in quadrature for each order.
Therefore, there are seventeen parameters in our full forward model, including five parameters for the companion, and twelve parameters for the star and nuisance parameters, since we model three NIRSPEC orders (orders 31--33). 

Our log-likelihood function $\mathcal{L}$ is defined as 
\begin{equation}
\ln \, \mathcal{L} = -0.5 \times \left[ \sum \chi^2 + \sum{\ln ( 2 \pi (\sigma^{*})^2 ) } \right],
\end{equation}
where $\chi^2$ is the chi-square defined as the square of data minus the full forward model over the inflated noise $\sigma^{*}$. The second term is the normalization constant with the inflated noise.

The brown dwarf atmosphere models are drawn from the BT-Settl \citep{Allard:2012ab} and Sonora \citep{Marley:2021aa} models \added{with resolutions of $\sim$235000 and $\sim$200000, respectively}, both assuming solar metallicity. We linearly interpolated the model grid points across different {\teff} and {\logg},
and used the nested sampling method \citep{Skilling:2004aa, Skilling:2006aa} with the \texttt{dynesty} package \citep{Speagle:2020aa} to derive our best-fit parameters.
\deleted{We prefer nested sampling over the Markov chain Monte Carlo sampling method because our model dimension is high (17 parameters), as nested sampling better samples the multi-modal distribution for each parameter, and allows us to compute the Bayes factor which can be used to compare different models.}
We used 1,000 live points to sample and followed the default stopping criteria when the difference of log evidence between iterations is below $\epsilon$ = 1.009\footnote{$\epsilon = 10^{-3} \times (K-1) + 0.01 = 1.009$, where $K = 1000$ is the number of nested sampling live points \citep{Speagle:2020aa}.}.

We used uniform distributions for our nested sampling priors.
Basically, we covered the reasonable parameter range for the {\teff} from 800~K to 2400~K and {\logg} from 3.5 to 5.5 dex cgs, RV from $-100$ to $+100$~{\kms}, and {\vsini} from 0 to 100~{\kms}.
The companion scale factor $\alpha_{p}$ ranges from 0 to 100 in data numbers (DN), the speckle scale factors $\alpha_{s, f}$ range from 0 to 6180 DN (for each fiber and each order; 6 parameters in total), the instrumental profile scale factor goes from 1.0 to 1.2 (for each order; 3 parameters in total), and the noise jitter is drawn from 0.1 to 30 DN (for each order; 3 parameters in total).

We modeled the data observed on 2021 September 18 (UT) and 2021 November 20 (UT), with results shown in Table~\ref{tab:nirspec}.
As noted in Section~\ref{sec:observe}, the data on 2021 September 18 were much worse (lower throughput; S/N = 13) than those from 2021 November 20 (S/N = 20), so the measurements of 2021 September 18 listed in Table~\ref{tab:nirspec} are presented for largely for completeness, and we adopt our results from 2021 November 20 measurements for the corresponding analysis.

\begin{figure*}
    \centering
    \includegraphics[width=\textwidth]{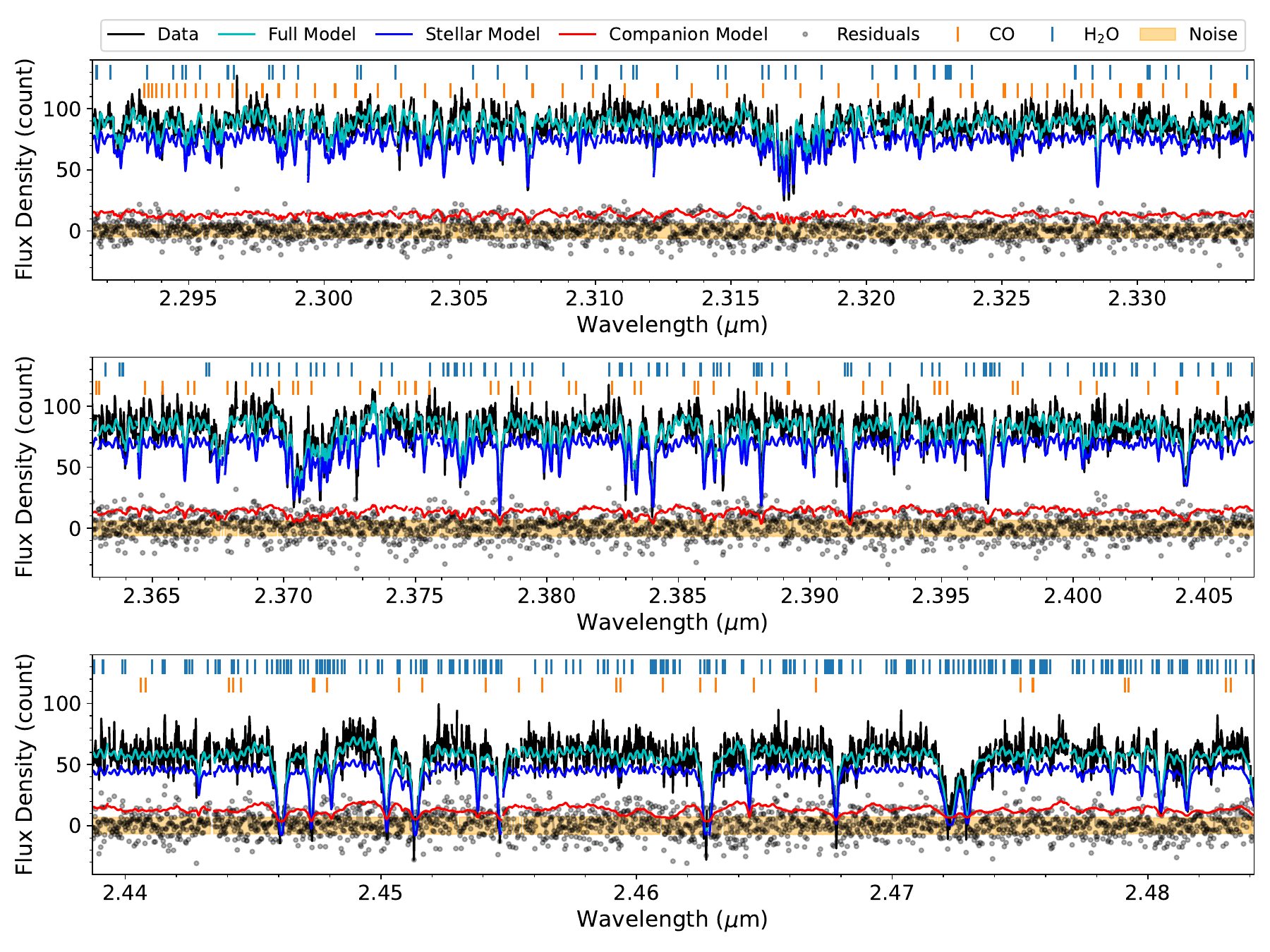}
    \caption{KPIC spectrum of HD~33632 Ab and its best-fit forward model derived from the BT-Settl model grids.
    The data were taken on 2021 November 20 (UT).
    The spectral orders 31--33 on fiber 1 are shown in black lines.
    The full forward model, including the stellar speckle and \replaced{planet}{companion} fluxes, is in cyan lines.
    The stellar model pulled directly from the observed on-axis KPIC spectrum of HD~33632A, is shown in blue lines. 
    The planet model, which is the best-fit BT-Settl model with the observed telluric profile, is illustrated in red lines. 
    The residual (data $-$ full forward model) is plotted in grey dots, consistent with uncorrelated noise, whereas the data noise is shown in the orange-shaded region. The CO and H$_2$O features are labeled in vertical orange and light blue lines, respectively.}
    \label{fig:kpic_spectrum_btsettl}
\end{figure*}

\begin{figure}
    \centering
    \includegraphics[width=0.45\textwidth]{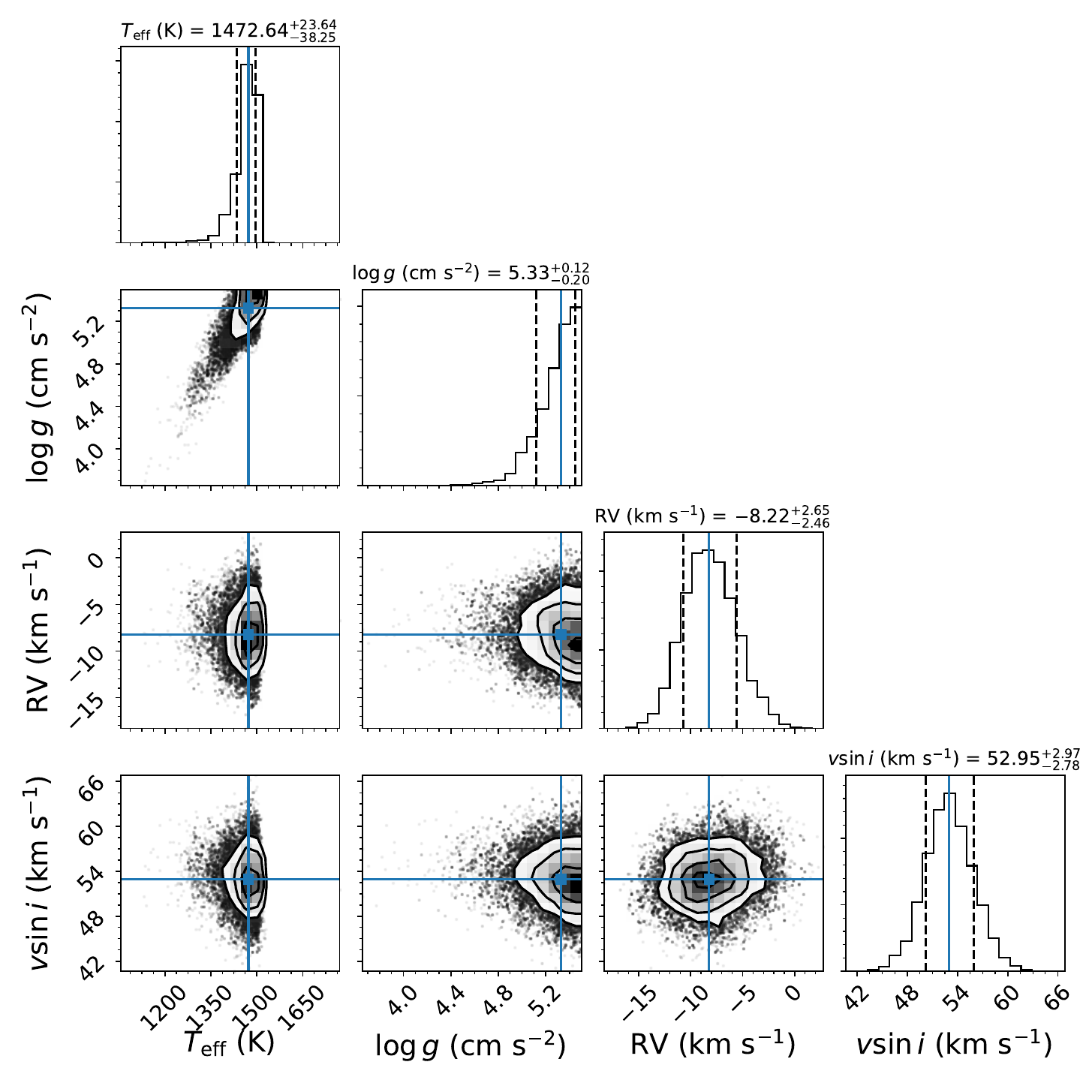}
    \includegraphics[width=0.45\textwidth]{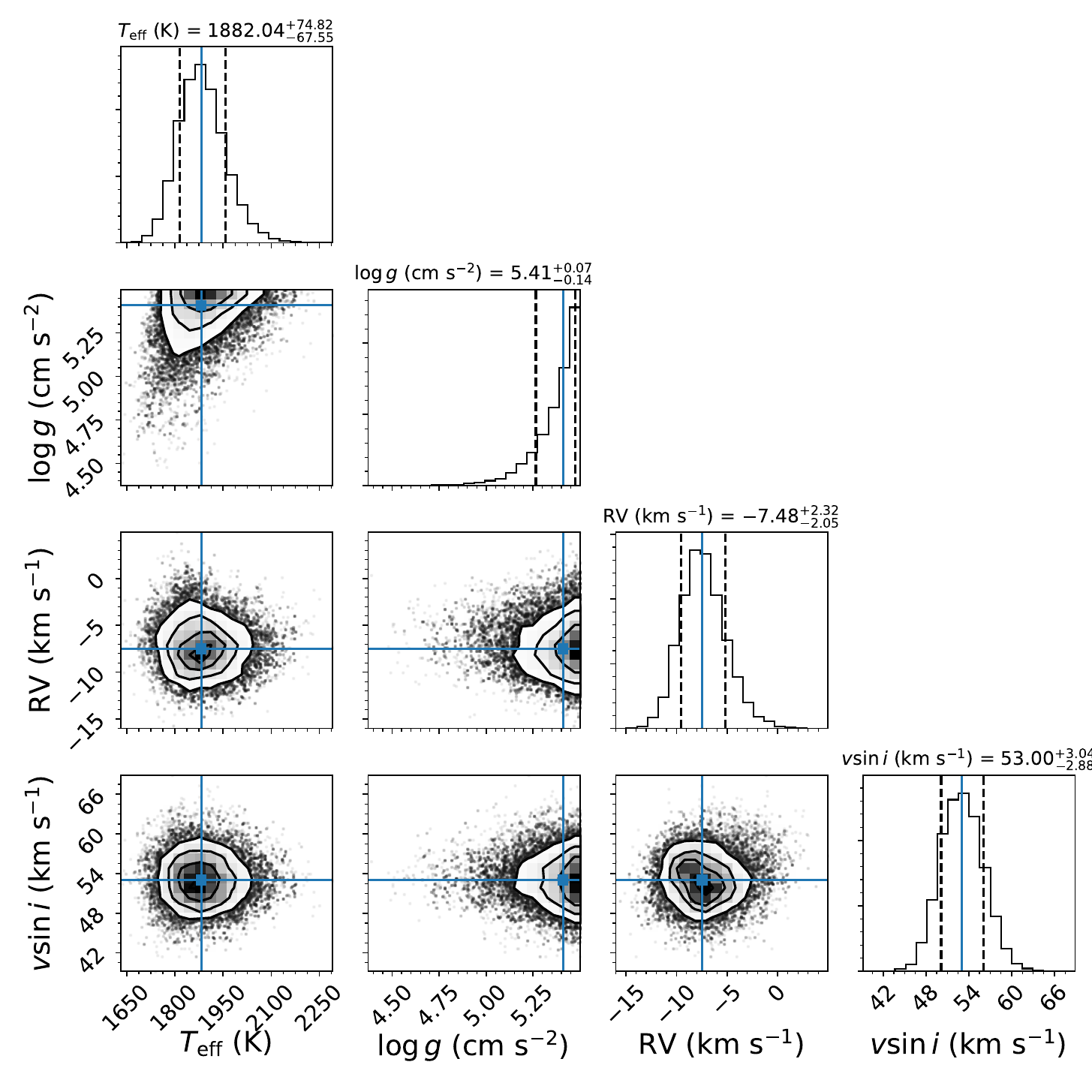}
    \caption{Posterior probability distribution for the substellar parameters of HD~33632 Ab using the BT-Settl (top) and Sonora (bottom) models.
    The KPIC data were taken on 2021 November 20 (UT).
    The spectral orders 31--33 on fiber 1 are shown in black lines, while
    the blue line indicates the median values and the dashed lines denote the 16$^\mathrm{th}$ and 84$^\mathrm{th}$ percentiles.}
    \label{fig:kpic_corner_btsettl}
\end{figure}

Our forward-modeling best-fit model and posterior probability distributions with the BT-Settl are shown in Figures~\ref{fig:kpic_spectrum_btsettl}--\ref{fig:kpic_corner_btsettl}\footnote{For completeness, the best-fit spectra with the Sonora model are shown in Figures~\ref{fig:kpic_spectrum_sonora}.}
Figure~\ref{fig:kpic_spectrum_btsettl} shows that our best-fit forward models match the data well, and our residual (data$-$model) auto-correlation functions are consistent with uncorrelated noise.
Our companion flux is lower than the speckle flux by 3.5--5.4$\times$, depending on the order, but the companion model clearly shows the CO ($\nu$2--0) bandhead in order 33 around 2.3~$\micron$, which validates the CO detection in Section~\ref{sec:ccf}.
The modeled {\vsini} of $53 \pm 3$~{\kms} indicates that HD~33632 Ab is a relatively fast rotator, which we discuss in detail in Section~\ref{sec:rotation}.
The best-fit companion RV ($-8 \pm 3$~{\kms} shows evidence of orbital motion compared to its primary RV ($-1.75 \pm 0.12$~{\kms}; \citealp{Gaia-Collaboration:2023aa})\footnote{While we observed on-axis KPIC spectra for the host star, measuring its RV and {\vsini} is not possible because the stellar absorption lines are very shallow for F8V HD~33632 in the wavelength used in our analysis.}, which we will later constrain the associated orbital parameters in Section~\ref{sec:orbit}.
Our {\logg}=$5.33^{+0.12}_{-0.20}$~cgs dex for the BT-Settl model and {\logg}=$5.41^{+0.07}_{-0.14}$ for the Sonora model hit the surface gravity ceiling of the model grids, which are the typical fitting issue \replaced{for}{using} self-consistent model grids for field late-M, L, and T dwarfs in $K$-band high-resolution spectra \citep{Del-Burgo:2009aa, Burgasser:2016aa, Hsu:2021aa, Hsu:2024aa}.
The high surface gravity, at a constant mass, represents the small radius issue in the retrievals in Section~\ref{sec:retrieval}. 
The modeled {\teff} values are very different, {\teff}=$1473^{+24}_{-38}$~K for the BT-Settl model and {\teff}=$1882^{+75}_{-68}$~K for the Sonora model.
The BT-Settl models incorporate clouds, while Sonora models are cloudless, and the difference in {\teff} among these two models is similar to the findings from T dwarf high-resolution $K$-band spectra in \cite{Hsu:2021aa}.
Notably, the BT-Settl {\teff} estimates have an abrupt drop in the posterior, due to the model treatment of chemistry change at the L/T transition, and its {\teff} is also consistent with the spectral type estimate of L9.5$^{+1.0}_{-3.0}$ from \cite{Currie:2020aa}.

One advantage of the nested sampling method is that we can use the Bayes factor to compare which model is more favored.
If the Bayes factor, defined by the ratio of the evidence of two competing models, is smaller by 100 times, the model in the denominator (smaller evidence) is decisively rejected \citep{Jeffreys:1961aa, Kass:1995aa}. We found the BT-Settls are decisively favored over the Sonora models ($\Delta$ Bayes factor = 1.3 $\times$ $10^{-6}$, and the $\chi^2$ (19784 versus 19812) also supported the same conclusion).
Therefore, we adopted the best-fit parameters from the BT-Settl models using the 2021 November 20 data, for which RV = $-8 \pm 3$~{\kms} and {\vsini} = $53 \pm 3$~{\kms}. 

While we forward-modeled the star light contribution using the observed on-axis star flux, there might be residual star light that was not completely taken into account in our modeling frameworks, such as the temporal variations of the slightly different airmass between observing the companion and the host star, humidity variations within the full observation, and the difference in optical paths between these frames. 
Our host star is an F8V star \citep{Gray:2003aa}, with an effective temperature of $\sim$6180~K \citep{Pecaut:2013aa}. Such temperature is not expected to present any strong molecular features with only a few weak absorption lines ($\lesssim$10$\%$ absorption) in the wavelength range used in our analysis, which we also validated in modeling the observed on-axis star spectra. The observed on-axis star spectra contain mostly telluric absorption features with the fringing modulation.


\begin{deluxetable*}{llcccccccccc}\label{table:forward_model_result}
\tablecaption{HD~33632 Ab NIRSPEC Measurements \label{tab:nirspec}} 
\tabletypesize{\scriptsize} 
\tablehead{ 
\colhead{UT Date} & 
\colhead{MJD\tablenotemark{a}} & 
\colhead{exposure} & 
\colhead{throughput\tablenotemark{b}} & 
\colhead{orders} &
\colhead{Model\tablenotemark{c}} & 
\colhead{{\teff}} & 
\colhead{{\logg}} & 
\colhead{RV\tablenotemark{d}}  & 
\colhead{{\vsini}}  &
\colhead{{Bayes factor\tablenotemark{e}}}  &
\colhead{$\chi^2$} \\ 
& & (s) & ($\%$) & &
\colhead{(K)} & 
\colhead{(cm s$^{-2}$)} & 
\colhead{(km s$^{-1}$)}  & 
\colhead{(km s$^{-1}$)}  & 
} 
\startdata
2021 Sep 18 & 59475.620 & 5400 & 1.3 & 31--33 & B & $1169^{+130}_{-95}$ & $4.0^{+0.6}_{-0.3}$ & $-10 \pm 4$ & $63^{+8}_{-7}$ & 1.0 & 21289 \\
\nodata & \nodata & \nodata & \nodata & \nodata & S & $1800^{+118}_{-104}$ & $5.2^{+0.2}_{-0.4}$ & $-12^{+4}_{-5}$ & $55^{+6}_{-7}$ & 0.01 & 21317 \\
2021 Nov 20 & 59538.436 & 14400 & 2.4 & 31--33 & B & $1473^{+24}_{-38}$ & $5.33^{+0.12}_{-0.20}$ & $-8 \pm 3$ & $53 \pm 3$ & 1.0 & 19784 \\
\nodata & \nodata & \nodata & \nodata & \nodata & S & $1882^{+75}_{-68}$ & $5.41^{+0.07}_{-0.14}$ & $-7 \pm 2$ & $53 \pm 3$ & 1.3 $\times$ 10$^{-6}$ & 19812 \\
\enddata 
\tablenotetext{a}{Modified Julian Dates (MJD) at the middle of the observing sequence}
\tablenotetext{b}{The end-to-end throughput. Our typical KPIC performance is 2--3\%, and $<$2\% is considered poor performance.}
\tablenotetext{c}{The self-consistent substellar atmosphere model grid. ``B'' is the BT-Settl CIFIST model \citep{Allard:2012ab}; ``S'' is the Sonora model \citep{Marley:2021aa}.}
\tablenotetext{d}{Barycentric radial velocity}
\tablenotetext{e}{We used the BT-Settl CIFIST model \citep{Allard:2012ab} as our baseline model for each night to compute the Bayes factor.}
\end{deluxetable*} 

\section{Retrieval Method} \label{sec:retrieval}

While the cross-correlation method provides evidence of CO and H$_2$O in the atmosphere of HD 33632 Ab, their abundances cannot be constrained (Section~\ref{sec:ccf}).
Furthermore, it is unclear whether methane is present, or not, in our KPIC data of this L/T transition object.
These goals can be achieved with a forward-modeling retrieval method.
Here, the package \texttt{petitRADTRANS} \citep{Molliere:2019ab, Nasedkin:2024aa} is used to perform our retrieval analysis, \deleted{mostly }following the analysis outlined in \cite{Xuan:2022aa}, with slightly different priors listed in Table~\ref{table:retrieval_priors}.
We measured the CO and H$_2$O abundances and placed an upper limit on the non-detection of CH$_4$.
Three combinations of data were considered for our retrievals, including (1) retrievals with the high-resolution $K$-band spectra, (2) joint-fitting the high-resolution $K$-band spectra and $JHK$-band photometry, and (3) joint-fitting the high-resolution $K$-band spectra and low-resolution near-infrared spectra.

\subsection{Retrieval Setup for High-resolution Spectra}


Our HD~33632 Ab spectra are retrieved with a forward retrieval modeling framework similar to Section~\ref{sec:forward_model}.
We fit the same NIRSPEC orders 31--33 using the data from 2021 November 20 (UT) by fitting an atmosphere model generated from retrievals, for which
we adopted the pressure-temperature ($P$-$T$) profile from \cite{Molliere:2020aa}. 
The optical depth $\tau$ as $\tau = \delta P^{\alpha}$, where $P$ is the pressure, and $\alpha$ and $\delta$ are free parameters. 
At the middle altitude, the internal temperature $T_\mathrm{int}$, using the Eddington approximation, follows $T(\tau)^4 = \frac{3}{4} T_\mathrm{int}^4 (\frac{2}{3} + \tau)$.
At high altitude between $P = 0.1$~bar and $\tau = 0.1$, a cubic spline interpolation is used to sample in equidistant location $\log{P}$ space using three parameters $T_1$, $T_2$, $T_3$.
We log-sample sampled our pressure grid from 10$^{-5}$--10$^{2}$~bar.

The major opacity species of L/T transition objects in $K$-band are CO, H$_2$O, and CH$_4$, so we included line-by-line species for all CO isotopologues (`CO\_all\_iso'), the principal H$_2$O isotopologue (`H2O\_main\_iso') from the HITEMP database \citep{Rothman:2010aa}, and the parent $^{12}$CH$_4$ species from \cite{Hargreaves:2020aa} (`CH4\_hargreaves\_main\_iso').
We also included the Rayleigh scattering species for hydrogen and helium \citep{Dalgarno:1962aa, Chan:1965aa}, as well as continuum collision-induced absorption (CIA) opacities for H$_2$-H$_2$ and H$_2$-He\deleted{$_2$} \citep{Gray:2008aa}.
The opacities have a resolution of $R = 10^6$, so we down-sampled the opacity table by a factor of four to speed up computations, as this resolution remains sufficiently high with respect to our KPIC spectra, for which $R \sim 35,000$.
The sampled wavelength range for emission retrieval spectra is 2.28--2.50~{\micron} to cover all three NIRSPEC orders.

Four chemistry scenarios were considered.
One is a chemical equilibrium model.
The second is a chemical non-equilibrium model, parameterized by the quench pressure $P_\mathrm{quench}$ \citep{Molliere:2020aa}.
$P_\mathrm{quench}$ treats the abundances of CO, H$_2$O, and CH$_4$ as constant above a given pressure level as the time scale of the chemical reaction is longer than the mixing time scale \citep{Zahnle:2014aa}.
The last two retrievals are free chemistry retrievals, which fit each species as a constant across all pressure levels.
To examine potential CH$_4$ present in our high-resolution spectra, we ran two sets of free retrievals, including: (1) CO, H$_2$O, and CH$_4$ and (2) CO and H$_2$O. 
To compute the [C/H] metallicity under free retrievals, we adopted the solar value of carbon (8.43 $\pm$ 0.05) from \cite{Asplund:2009aa}.

For our cloud assumptions, we explored cloudless, and the EddySed cloud models from \cite{Ackerman:2001aa} with two particle shapes of MgSiO$_3$. 
Our choice of MgSiO$_3$ is motivated by the investigation of the cloudy L dwarf HD~4747 B in \cite{Luna:2021aa}, who found that MgSiO$_3$ is the dominant species in L/T transition objects \citep{Gao:2020aa, Marley:2021aa}.
The first cloud model assumes amorphous spherical particles under Mie scattering (am), while the second cloud model has crystalline irregular shape particles using the distribution of hollow spheres method (DHS; \citealp{Min:2005aa, Min:2016aa}) (cd).
Four parameters are fitted in the EddySed cloud model, including the cloud mass fraction of MgSiO$_3$ ($\tilde{X}_\mathrm{MgSiO_3}$), the vertical Eddy diffusion coefficient $K_{zz}$, the sedimentation efficiency factor ($f_\mathrm{sed}$), the log-normal size distribution of the particle $\sigma_g$. 

The mass, radius, and parallax of HD~33632 Ab are required to compute emission spectra.
We assume the mass $46\pm8$~M$_\mathrm{Jup}$ from \cite{Currie:2020aa} as a Gaussian prior.
For our radius, we adopted a uniform prior between 1.0 $\pm$ 0.4~R$_\mathrm{Jup}$.
We used the \textit{Gaia} DR3 parallax 37.895 $\pm$ 0.026~mas \citep{Gaia-Collaboration:2023aa}, but this has minimal effect on our retrievals as our NIRSPEC spectra are normalized and not flux-calibrated.

Finally, we produce our substellar emission spectral model $M$ from \texttt{petitRADTRANS}.
The full forward retrieval model is simply generated by replacing the companion substellar atmosphere term $M$ in Equation~\ref{eqn:forward_model_science}.
We considered a total of 24 models for high-resolution spectral retrievals, including four chemistry models, three cloud assumptions, and two radius priors (Section~\ref{subsec:highres_retrieval}).
For joint retrievals, we consider 11 scenarios which are a subset of the high-resolution retrievals (Section~\ref{subsec:joint_retrieval}).
Table~\ref{table:retrieval_priors} lists the priors of our forward retrieval modeling method.
As there are 25--30 parameters in our forward retrieval model, the nested sampling with \texttt{dynesty} is used to obtain our best-fit models and posteriors, which is suitable to sample multi-modal and high-number parameter models.
We used $K = 200$ live points to run our nested sampling,
with the multi-ellipsoids bounding method to account for multi-model posteriors, and the random walk sampling method.
The stopping criteria are the default from \texttt{dynesty}, using 1\% of the unaccounted evidence remaining and parameterized by the difference of log evidence $\Delta \ln{z}$ under 0.209 (i.e. $10^{-3} \times (K-1) + 0.01 = 0.209$) for all of our fitting routines \citep{Speagle:2020aa}.

\begin{deluxetable}{lcc}
\tablecaption{Forward Retrieval Modeling Priors \label{table:retrieval_priors}}
\tablecolumns{3}
\tablehead{
\colhead{Description} &  \colhead{Symbol (unit)}  & \colhead{Priors\tablenotemark{a}}
}
\startdata
Mass & $M$ (M$_\mathrm{Jup}$) & $\mathcal{N}$(46.4, 8)\tablenotemark{b} \\
Radius\tablenotemark{c} & $R$ (R$_\mathrm{Jup}$) & $\mathcal{U}$(0.6, 1.4) \\
\nodata & \nodata & $N$(0.86, 0.03) \\
Projected Rot. Vel. & {\vsini} ({\kms}) & $\mathcal{U}$(0, 100) \\
Radial Velocity & RV ({\kms}) & $\mathcal{N}$($-100$, 100) \\
Carbon/Oxygen\tablenotemark{d} & C/O & $\mathcal{U}$(0.1, 1.7) \\
Metallicity\tablenotemark{d} & [C/H] & $\mathcal{U}$($-1.5$, 1.5) \\
Mass Fractions of CO & m$_\mathrm{CO}$ & $\mathcal{U}$(10$^{-7}$, 10$^{-1}$)\tablenotemark{e} \\
Mass Fractions of H$_2$O & m$_\mathrm{H_2O}$ & $\mathcal{U}$(10$^{-7}$, 10$^{-1}$)\tablenotemark{e} \\
Mass Fractions of CH$_4$ &  m$_\mathrm{CH_4}$ & $\mathcal{U}$(10$^{-10}$, 10$^{-1}$)\tablenotemark{e} \\
Internal Temperature & T$_\mathrm{int}$ (K) & $\mathcal{U}$(500, 3000) \\
$T_3$ & $T_3$ (K) & $\mathcal{U}$(0, $T_c$)\tablenotemark{f} \\
$T_2$ & $T_2$ (K) & $\mathcal{U}$(0, $T_3$) \\
$T_1$ & $T_1$ (K) & $\mathcal{U}$(0, $T_2$) \\
$\alpha$ & $\alpha$ (mas) & $\mathcal{U}$(1, 2) \\
$\log{\delta}$\tablenotemark{g} & $\log{\delta}$ & $P_\mathrm{phot} \in $ $\mathcal{U}$[10$^{-3}$, 100] \\
Quench Pressure\tablenotemark{h} & $P_\mathrm{quench}$ (bar) & $\mathcal{U}$($10^{-4}$, $10^{3}$) \\
Mass Frac. of MgSiO$_3$ & $\log{\tilde{X}_\mathrm{MgSiO_3}}$ & $\mathcal{U}$($-2.3$, 1.0) \\
Vert. Eddy Diff. Coeff. & $\log{K_{zz}}$ (cm s$^{-2}$) & $\mathcal{U}$(5, 13) \\
Sediment Fraction & $f_\mathrm{sed}$ & $\mathcal{U}$(0.0, 10.0) \\
Scatter of Particle Size & $\sigma_g$ & $\mathcal{N}$(37.895, 0.026) \\
\replaced{Planet}{Companion} Flux & \replaced{Planet}{Companion} Flux (DN) & $\mathcal{U}$(0, 100) \\
Speckle Flux for Fiber 1\tablenotemark{i} & Speckle Flux (DN) & $\mathcal{U}$(0, 6180) \\
Speckle Flux for Fiber 2\tablenotemark{i} & Speckle Flux (DN) & $\mathcal{U}$(0, 6180) \\
Scale Term for LSF\tablenotemark{i} & LSF $\sigma$ & $\mathcal{U}$(1.0, 1.2) \\
Error Jitter\tablenotemark{i} & Error Jitter (DN) & $\mathcal{U}$(0.1, 30.0) \\
\enddata
\tablenotetext{a}{$\mathcal{N}$ stands for normal priors with the mean and standard deviation in the parentheses, while $\mathcal{U}$ represents uniform priors with the lower and upper bounds in the parentheses.}
\tablenotetext{b}{Mass measurements from \cite{Currie:2020aa}}
\tablenotetext{c}{We have two different radius priors, one with a wider range of uniform priors and the other with a narrow range of normal priors from \cite{Baraffe:2003aa} evolutionary models. See Section~\ref{sec:retrieval} for details.}
\tablenotetext{d}{Parameters are fitted in chemical equilibrium and in chemical disequilibrium with quench pressure.}
\tablenotetext{e}{Sampled in the log space}
\tablenotetext{f}{$T_c$ is the connection temperature used to connect the photosphere boundary with the Eddington approximation $T_3 = (\frac{3}{4} T_\mathrm{int}^4 \times (0.1 + \frac{2}{3})^\frac{1}{4}$.}
\tablenotetext{g}{Log optical depth $\log{\delta}$ was sampled from a uniform distribution of the photospheric pressure $P_\mathrm{phot}$ where the optical depth $\tau = 1$ and $\delta = P_\mathrm{phot}^{-\alpha}$. See the discussion and justification of these parameters in \cite{Molliere:2020aa}.}
\tablenotetext{h}{Only used when the quench pressure is used to model the disequilibrium chemistry}
\tablenotetext{i}{A separate parameter was used for each order (for orders 31--33).}
\end{deluxetable}

\subsection{High-resolution Retrieval Results} \label{subsec:highres_retrieval}

\begin{figure}
    \centering
    \includegraphics[width=0.48\textwidth]{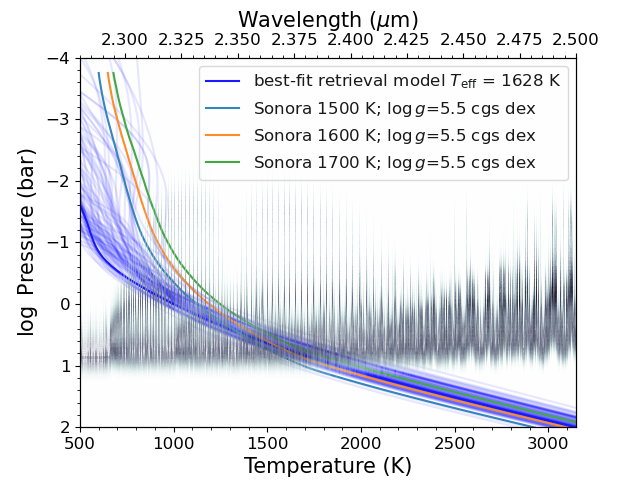}
    \includegraphics[width=0.48\textwidth]{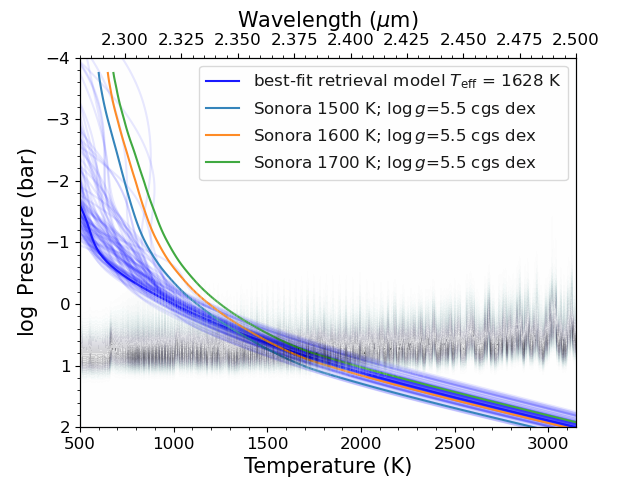}
    \caption{Pressure-temperature ($P$-$T$) profile and emission contributions of our baseline forward retrieval model for HD~33632 Ab.
    Our $P$-$T$ profiles are randomly drawn from 100 best-fit baseline forward retrieval models (blue), and the Sonora models of 1500, 1600, and 1700~K under $\log{g}$ = 5.5 dex cgs in light blue, orange, and green lines, respectively.
    The emission contribution function (pressure vs. wavelength) without/with rotation included are plotted in the left and right panels, respectively.
    The fast rotation of HD~33632 Ab limits our ability to constrain the $P$-$T$ profiles in the upper atmosphere.
    }
\label{fig:emission_contribution}
\end{figure}

\begin{figure*}
    \centering
    \replaced{\includegraphics[width=0.49\textwidth]{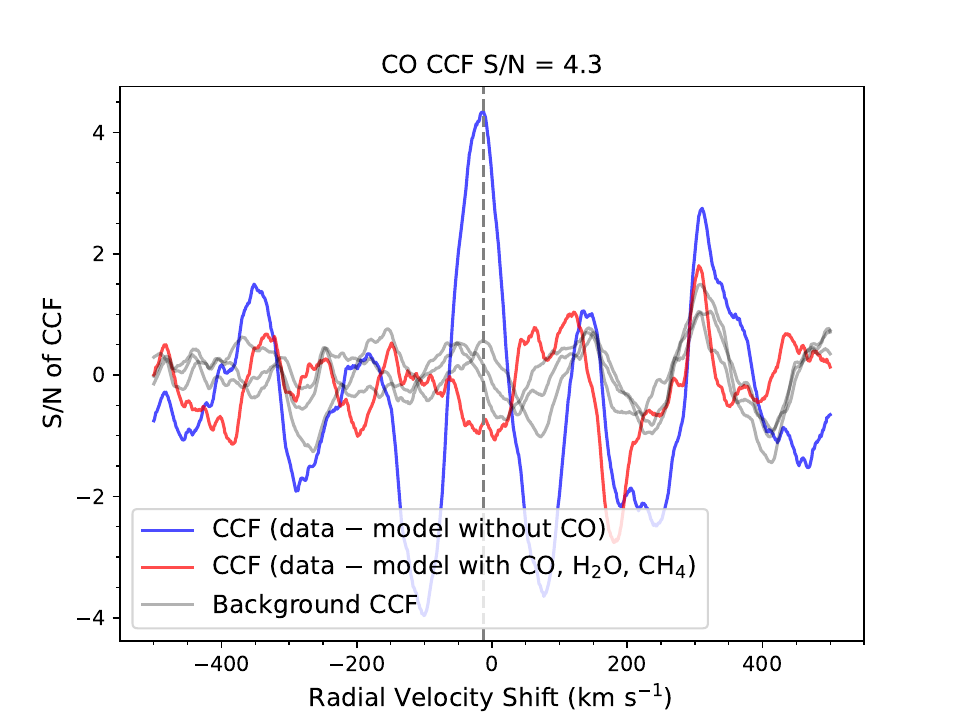}}
    {\includegraphics[width=0.49\textwidth]{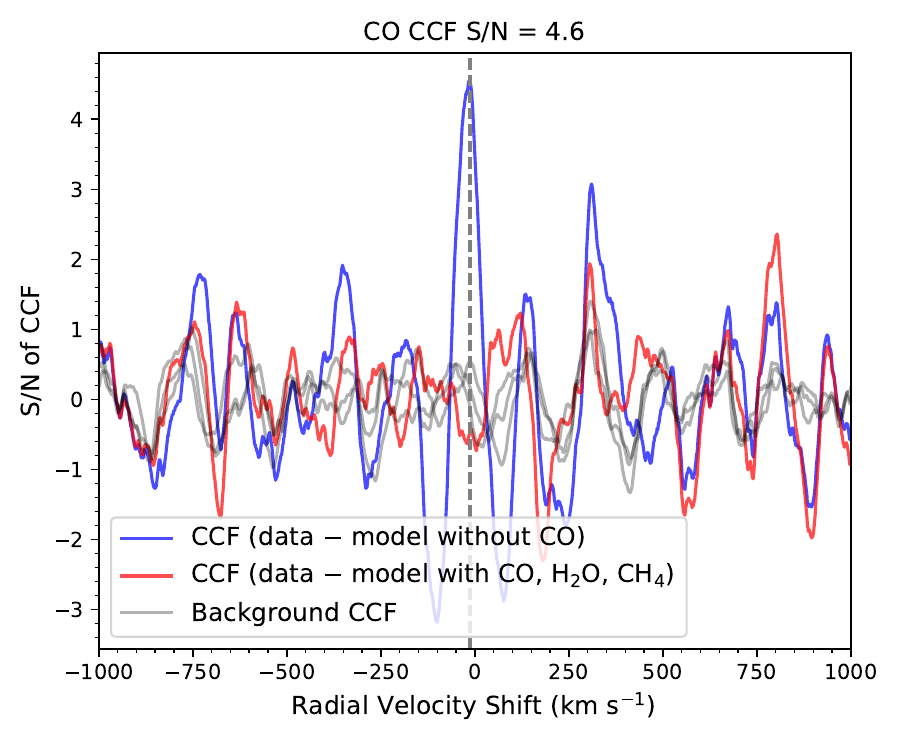}}
    \replaced{\includegraphics[width=0.49\textwidth]{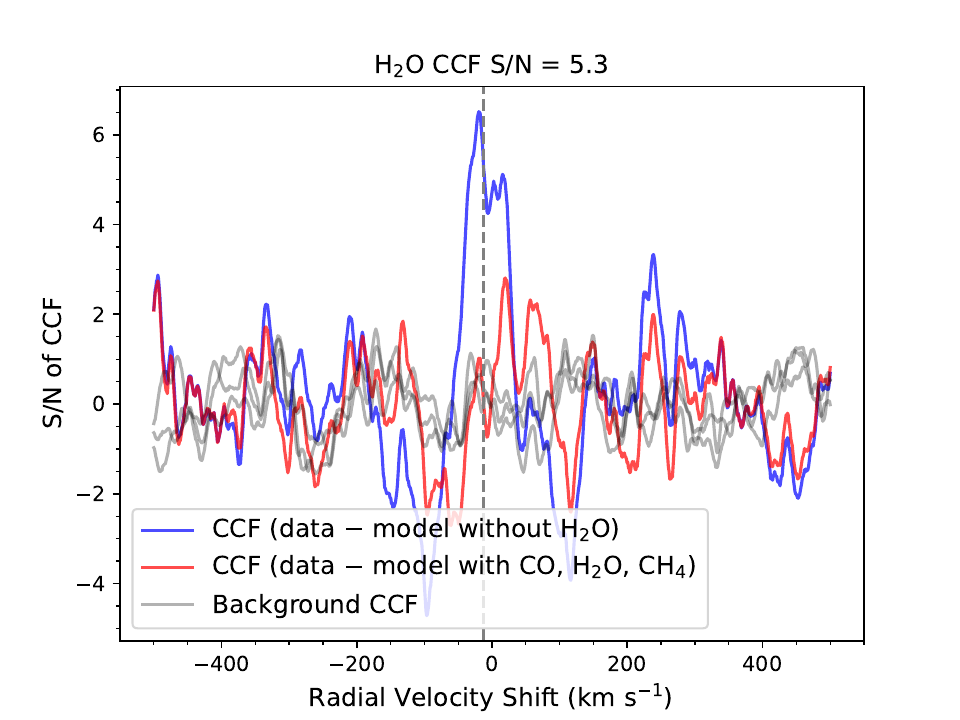}}
    {\includegraphics[width=0.49\textwidth]{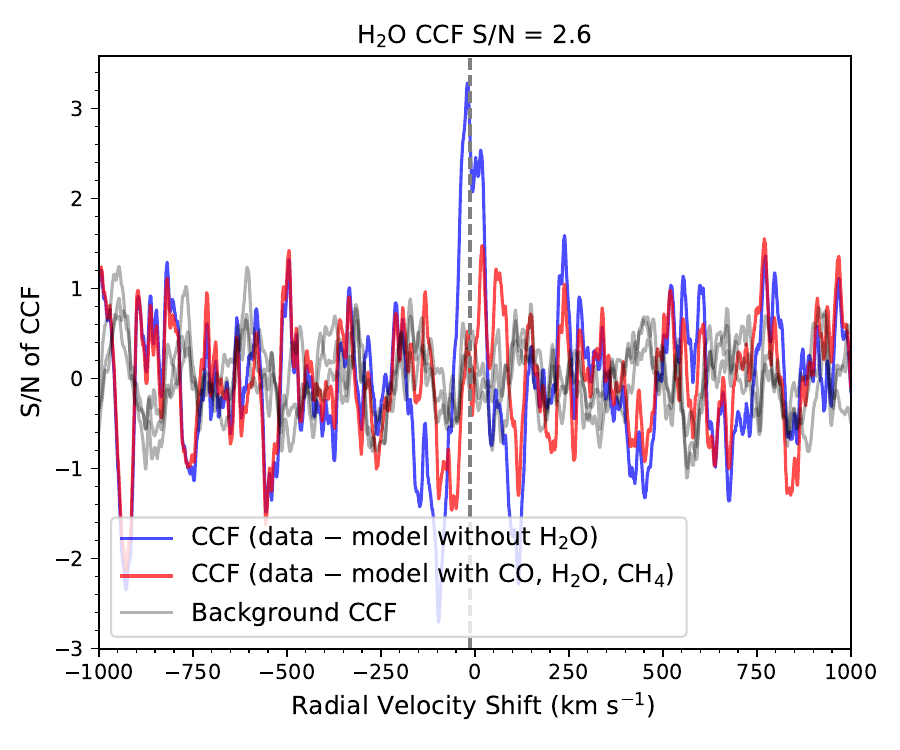}}
    \replaced{\includegraphics[width=0.49\textwidth]{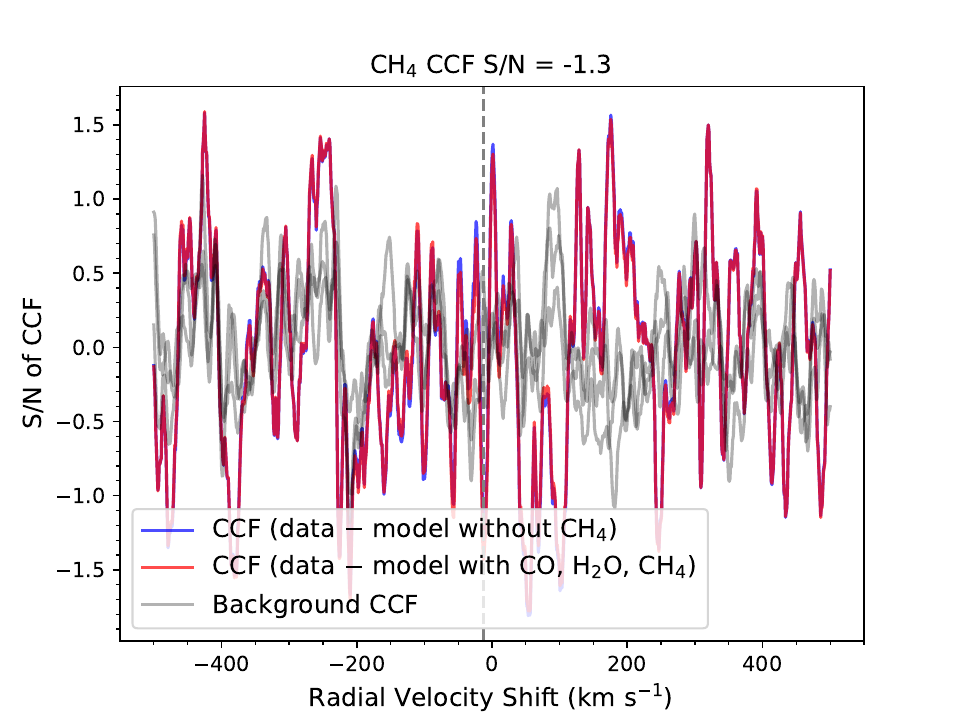}}
    {\includegraphics[width=0.49\textwidth]{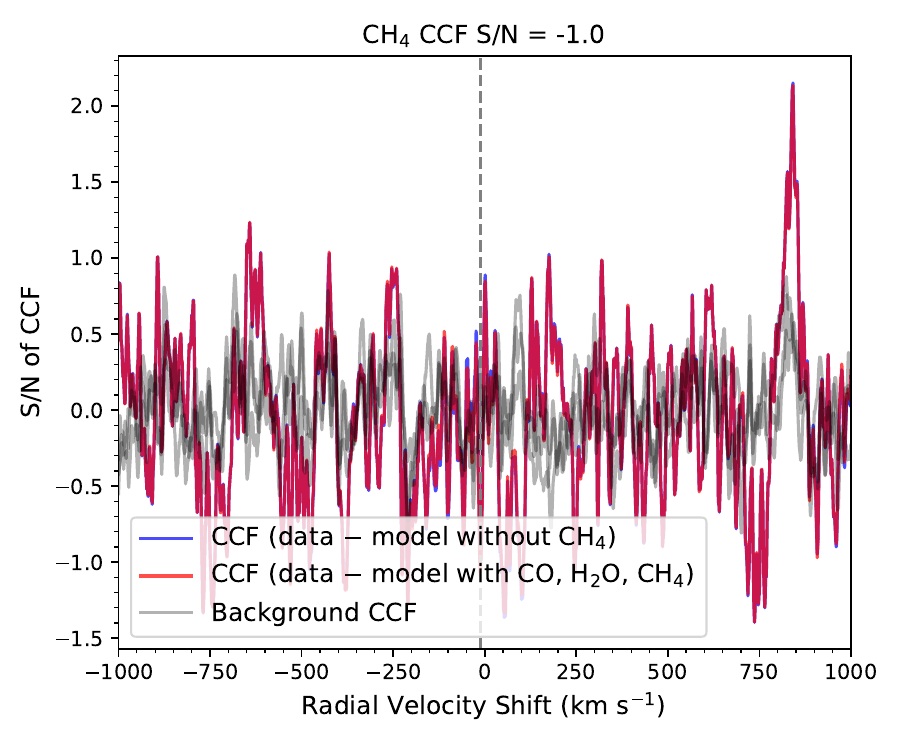}}
    \replaced{\includegraphics[width=0.49\textwidth]{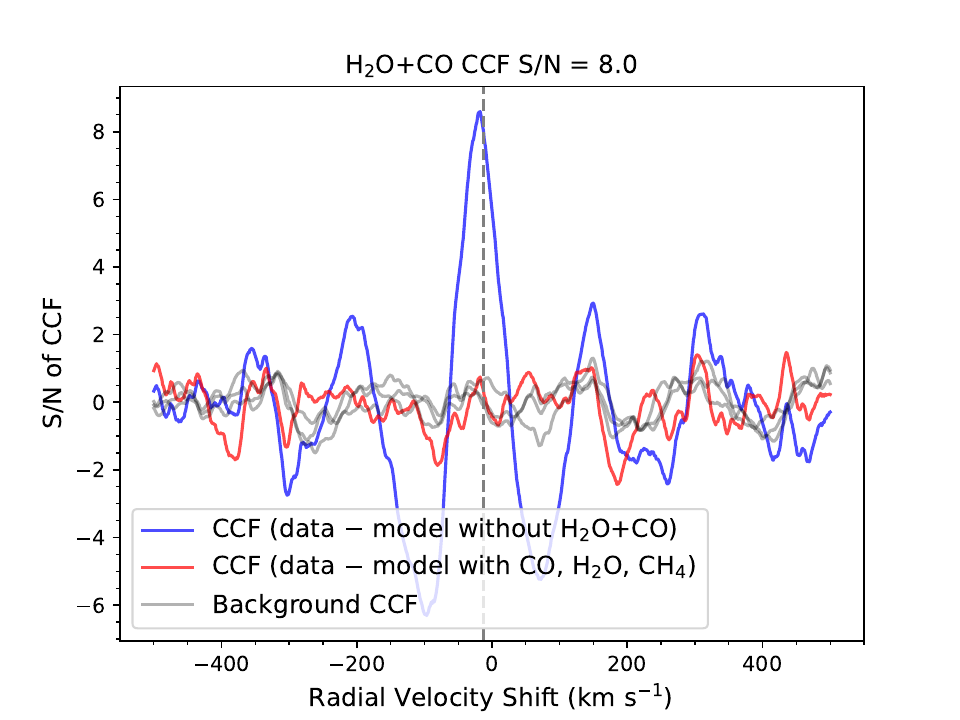}}
    {\includegraphics[width=0.49\textwidth]{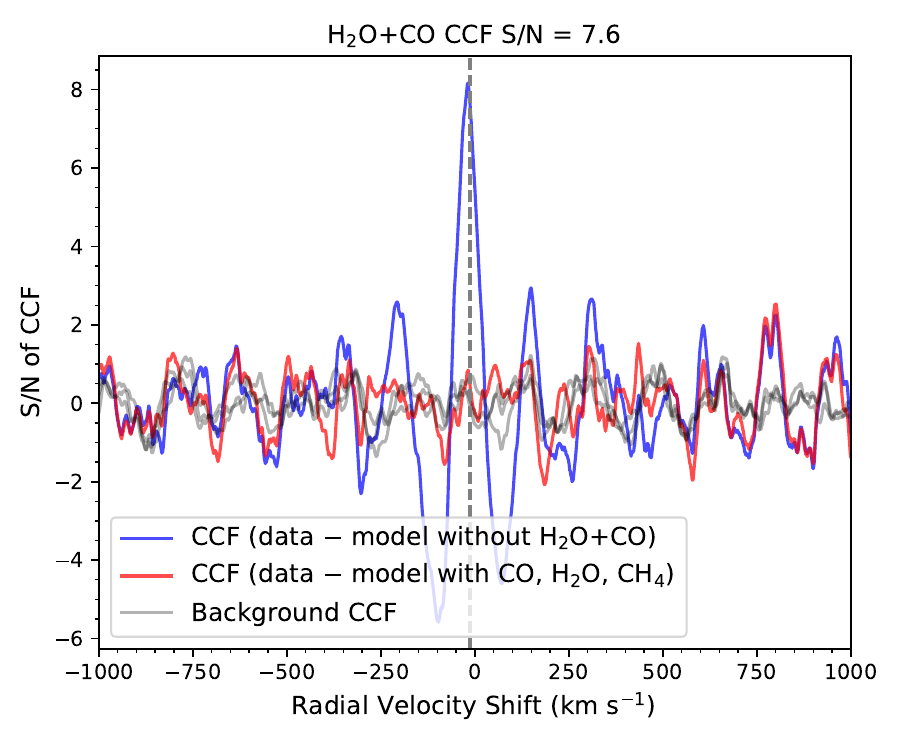}}
    \caption{Cross-correlation function (CCF) of the residuals of our KPIC spectrum for HD~33632 Ab using the molecular templates derived from our baseline forward retrieval model.
    \textit{Upper-left}: 
    CCFs of the residuals against the CO template.
    The residuals are defined as the difference between our KPIC data and the best-fit model with all (CO, H$_2$O, and CH$_4$) species and without the CO molecules (i.e. H$_2$O and CH$_4$ only), which are labeled in red and blue lines, respectively. 
    The background CCFs in three locations on the detector are shown in grey lines.
    The stellar barycentric-included RV ($-$12.65~{\kms}) is depicted in the vertical dashed grey line.
    \textit{Upper-right}: Same as the upper-left panel, but for H$_2$O.
    \textit{Lower-left}: Same as the upper-left panel, but for CH$_4$.
    \textit{Lower-right}: Same as the upper-left panel, for the combined CO $+$ H$_2$O molecular templates.}
    \label{fig:ccf_retrievals}
\end{figure*}

\replaced{To ensure we retrieve unbiased estimates of}{In an attempt to minimize bias in our estimates of} C/O and [C/H] for HD~33632 Ab, several retrievals were conducted to ensure the robustness of our retrieved parameters, including different chemistry, clouds, radius priors, and data combinations. 
The full retrieval results are summarized in Table~\ref{table:retrieval_result}.

Our baseline forward retrieval model assumes cloudless, disequilibrium chemistry parameterized with the quench pressure $P_\mathrm{quench}$ with a wide uniform radius prior between 0.6 and 1.4~R$_\mathrm{Jup}$.
We found that disequilibrium chemistry is highly favored over the chemical equilibrium model using the Bayes factor (strong $\Delta \ln{B}$ = 2.3--4.6 and decisive $\Delta \ln{B}$ $>4.6$; \citealp{Kass:1995aa}). 
The quench pressure $\log{P_\mathrm{quench}} = 2.0\pm0.6$~bar indicates a high degree of chemical disequilibrium\footnote{For completeness, our best-fit spectra and posteriors of our baseline forward retrieval models are shown in Figures~\ref{fig:retrieval_spectrum_baseline} and \ref{fig:retrieval_posterior_baseline}, respectively.}.
The best-fit companion emission spectra are similar to our forward-modeling fits.
Our RV ($-8.8^{+2.0}_{-1.9}$~{\kms}) and {\vsini} ($50^{+2}_{-3}$~{\kms}) are consistent within 1$\sigma$ with our forward-modeling results (RV = $-8 \pm 3$~{\kms} and {\vsini} = $53 \pm 3$~{\kms}), demonstrating the robustness of our retrieval routine fit.
However, our retrieved radius is extremely small, stacked at the lower end of our priors ($\lesssim 0.7$~R$_\mathrm{Jup}$).
The small radius issue has been widely reported in the literature (e.g. \citealp{Gonzales:2018aa, Gonzales:2020ab, Burningham:2021aa, Hood:2023aa}), which cannot be resolved with sophisticated cloud models \citep{Burningham:2021aa}.
As will be shown later in this section, our retrieved abundances, C/O and [C/H] are robust, validated by paralleled sets of retrievals using the restricted Gaussian priors for the radii inferred from the \cite{Baraffe:2003aa} evolutionary models.

Another way to validate our retrieval results is through inspection of our best-fit $P$-$T$ profiles.
Figure~\ref{fig:emission_contribution} shows 100 randomly drawn best-fit $P$-$T$ profiles from our baseline forward retrieval models.
The inferred effective temperatures of our baseline retrieval model {\teff} = $1628^{+109}_{-97}$~K, which are consistent with Sonora models between 1600--1800~K near the photosphere and inner atmosphere.
However, the $P$-$T$ profiles in the upper atmosphere of our baseline model are not well constrained compared to those expected from the Sonora models.
While in theory the high-resolution spectra would allow us to resolve the line cores of species (but more difficult than the wings due to lower fluxes, thus higher fractional errors) such as CO bandhead which can reach into the upper atmosphere (upper panel of Figure~\ref{fig:emission_contribution}), the fast rotation of HD~33632 Ab $\sim$50~{\kms} smears out the emission contribution of CO lines and thus hinders our ability to constrain the upper atmosphere using our KPIC $K$-band spectra.

One major goal of using the forward retrieval modeling method is to possibly detect methane in the L/T transition object HD~33632 Ab.
The baseline retrieval result indicates a CH$_4$ mass fraction of $\log$ CH$_4$ = $-3.82^{+0.4}_{-1.0}$, almost \replaced{100}{30} times lower than the measured CO mass fraction of $\log$ CO = $-2.33 \pm 0.17$, or 15 times lower than the H$_2$O mass fraction ($\log$ H$_2$O = $-2.62^{+0.19}_{-0.18}$)\footnote{The corresponding volume mixing ratios for CO, H$_2$O, and CH$_4$ are $-3.34 \pm 0.18$, $-3.54^{+0.17}_{-0.18}$, and $-5.2^{+0.4}_{-0.6}$, respectively.}.
While the upper and lower bounds of CH$_4$ is constrained, this does not necessarily mean that we detect methane in our NIRSPEC spectra.
Indeed, the low mass fractions for methane in our free retrievals validate the non-detection of methane, with the mass fraction of methane $\log$ CH$_4$ = $-7.5^{+1.8}_{-1.6}$ to $-8.1^{+1.6}_{-1.2}$.
CCFs of a given molecular species against the residual of our data and model without a given species can potentially corroborate if there is indeed no detection of methane, following the technique in \cite{Xuan:2022aa}.
Therefore, we cross-correlate a given species (CO, H$_2$O, or CH$_4$) with the residuals of our KPIC data minus the best-fit retrieval models with/without a given species turned on, as shown in Figure~\ref{fig:ccf_retrievals}.
Similar to our findings in Section~\ref{sec:ccf}, we found clear detections of CO (S/N = \replaced{4.3}{4.6}), H$_2$O (S/N = \replaced{5.3}{2.6}), and CO $+$ H$_2$O (S/N = \replaced{8.0}{7.6}), but there is no detection in the CCF for CH$_4$ (S/N = \replaced{$-$1.3}{$-$1.0}).

To validate that our retrieved parameters are robust regardless of our radius priors, we ran the retrievals under the restricted Gaussian radius priors $R_\mathrm{prior} = 0.86\pm0.03$~R$_\mathrm{Jup}$, with the radii inferred from the \cite{Baraffe:2003aa} evolutionary models.
We found that the inferred $v\sin{i}$s under restricted priors are consistent but generally higher by $\sim$5~{\kms} compared to the wide uniform priors.
The higher radii from the restricted priors, under the same mass priors, imply a smaller {\logg}.
Therefore, higher {\vsini} values compensate for the smaller pressure broadening in the line profile.
The RVs, retrieved C/O and [C/H] are generally consistent under the wide and restricted radius priors.
The posteriors of {\logg} using the BT-Settl and Sonora models are stacked at the highest available {\logg} $=$5.5 dex cm s$^2$, which has been a common issue in fitting high-resolution spectra at $\sim$2.3~$\micron$ for field low-mass stars and brown dwarfs (e.g. \citealp{Hsu:2021aa, Hsu:2023aa, Hsu:2024aa}), also implying small radii within the self-consistent modeling grids.
It is noted that our forward-model framework that incorporates the on-axis star spectra might still have residual star light, as discussed in Section~\ref{sec:forward_model}.



\begin{figure*}
    \centering
    \includegraphics[width=\textwidth]{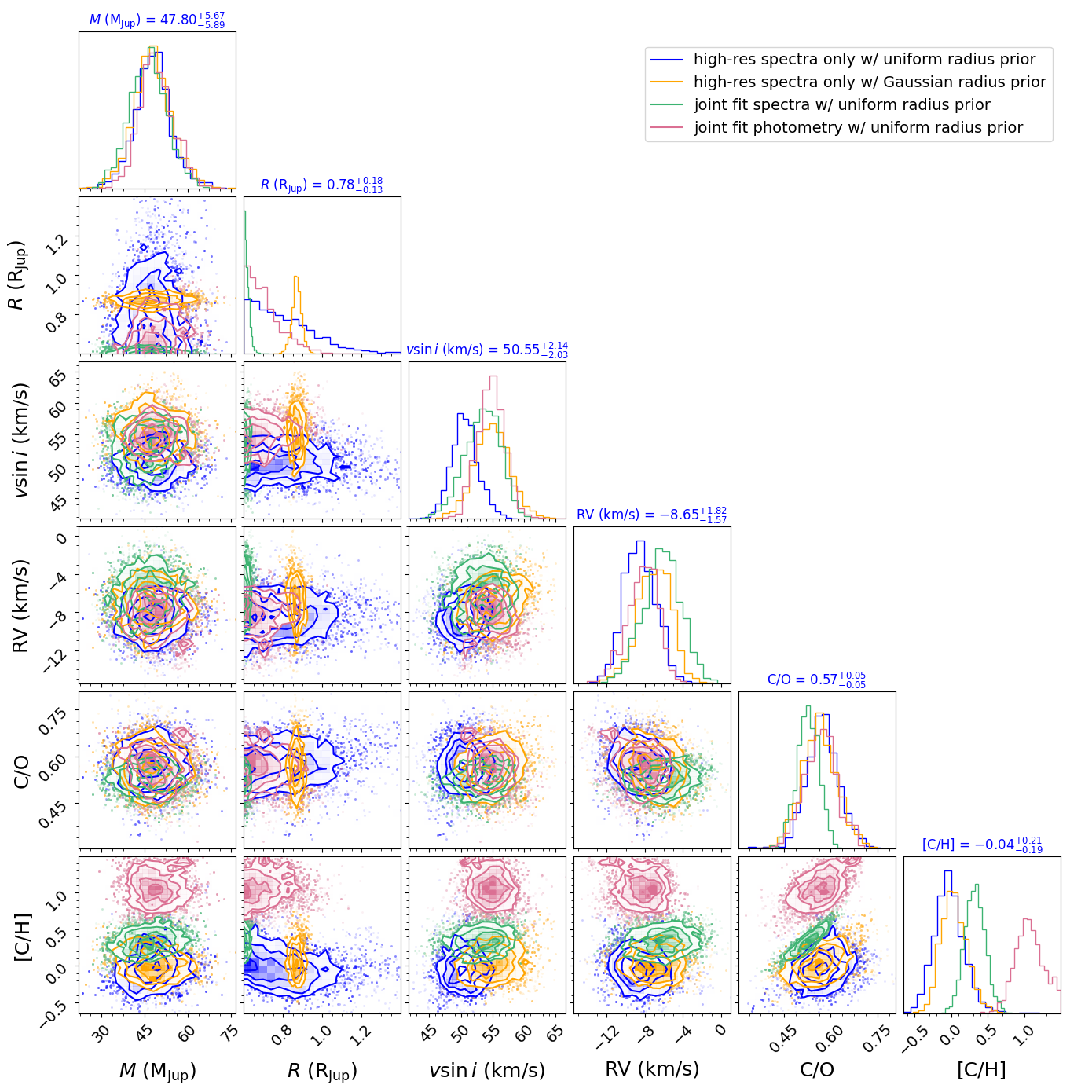}
    \caption{
    Posterior probability distributions of our retrieval models of HD~33632 Ab for the selected substellar parameters.
    We compared the retrievals using a combination of high-resolution, and low-resolution spectra, as well as photometry.
    The median values are labeled in solid lines, and the 16$^\mathrm{th}$ and 84$^\mathrm{th}$ percentiles are labeled in dashed lines.
    We compare the following scenarios: high-resolution spectra only for uniform/wide radius priors (blue) and for Gaussian/restricted radius priors (orange); 
    joint-fitting high-resolution spectra and low-resolution spectra for uniform/wide radius priors (bluish green); and
    joint-fitting high-resolution spectra and $JHK$ photometry for uniform/wide radius priors (reddish purple).
    }
    \label{fig:retrieval_posterior_compare}
\end{figure*}

\begin{longrotatetable}
\begin{deluxetable*}{lccccccccccccccc}
\tablecaption{HD~33632 Ab Forward Retrieval Modeling Result \label{table:retrieval_result}}
\tabletypesize{\scriptsize}
\tablehead{
\colhead{Chemistry\tablenotemark{a}} &  \colhead{Cloud\tablenotemark{b}}  & \colhead{Data\tablenotemark{c}} & \colhead{$R$ Prior\tablenotemark{d}} & 
\colhead{Mass} & \colhead{Radius} & \colhead{$v\sin{i}$} & \colhead{RV} &
\colhead{C/O} & \colhead{[C/H]} & \colhead{m$_\mathrm{CO}$\tablenotemark{e}} & \colhead{m$_\mathrm{H_20}$\tablenotemark{e}} & 
\colhead{m$_\mathrm{CH_4}$\tablenotemark{e}} & \colhead{$T_\mathrm{eff}$} & \colhead{$\log{g}$} & \colhead{$\log{B}$\tablenotemark{f}} \\
\colhead{} &  \colhead{}  & \colhead{} & \colhead{} &
\colhead{(M$_\mathrm{Jup}$)} & \colhead{(R$_\mathrm{Jup}$)} & \colhead{(km s$^{-1}$)} & \colhead{(km s$^{-1}$)} & 
\colhead{} & \colhead{} & \colhead{} & \colhead{} & \colhead{} &  \colhead{(K)} & \colhead{(dex cm s$^{-2}$)} & \colhead{} \\
}
\startdata
BT-Settl & cloudy & HI & $\mathcal{U}$ & \nodata & \nodata & $53 \pm 3$ & $-8 \pm 3$ & \nodata & \nodata & \nodata & \nodata & \nodata & $1473^{+24}_{-38}$ & $5.33^{+0.12}_{-0.20}$ & \nodata \\
Sonora & cloudless & HI & $\mathcal{U}$ & \nodata & \nodata & $53 \pm 3$ & $-7 \pm 2$ & \nodata & \nodata & \nodata & \nodata & \nodata & $1882^{+75}_{-68}$ & $5.41^{+0.07}_{-0.14}$ & \nodata \\
\hline
Chem EQ & cloudless & HI & $\mathcal{U}$ & $51.0^{+7.4}_{-7.3}$ & $0.67^{+0.12}_{-0.05}$ & $48.1^{+2.4}_{-2.2}$ & $-7.7^{+1.8}_{-2.1}$ & $0.51^{+0.04}_{-0.04}$ & $-0.04^{+0.17}_{-0.15}$ & $-2.33^{+0.17}_{-0.17}$ & $-2.62^{+0.19}_{-0.18}$ & $-3.82^{+0.42}_{-0.98}$ & $1919^{+90}_{-89}$ & $5.43^{+0.1}_{-0.14}$ & $-9.4$ \\
Chem EQ & cloudless & HI & $\mathcal{G}$ & $49.8^{+7.4}_{-6.3}$ & $0.85^{+0.03}_{-0.03}$ & $53.0^{+2.7}_{-2.7}$ & $-7.8^{+2.0}_{-2.3}$ & $0.52^{+0.04}_{-0.04}$ & $-0.12^{+0.16}_{-0.13}$ & $-2.41^{+0.18}_{-0.13}$ & $-2.74^{+0.18}_{-0.14}$ & $-4.02^{+0.44}_{-0.89}$ & $1901^{+98}_{-92}$ & $5.23^{+0.06}_{-0.06}$ & $-11.86$ \\
Chem EQ & am & HI & $\mathcal{U}$ & $51.9^{+6.2}_{-5.5}$ & $0.68^{+0.13}_{-0.06}$ & $49.1^{+2.0}_{-2.4}$ & $-7.7^{+1.9}_{-2.1}$ & $0.53^{+0.04}_{-0.04}$ & $0.18^{+0.32}_{-0.26}$ & $-2.12^{+0.3}_{-0.3}$ & $-2.5^{+0.27}_{-0.2}$ & $-3.92^{+0.53}_{-0.63}$ & $1764^{+157}_{-130}$ & $5.44^{+0.09}_{-0.19}$ & $-3.13$ \\
Chem EQ & am & HI & $\mathcal{G}$ & $49.9^{+7.7}_{-5.7}$ & $0.86^{+0.03}_{-0.03}$ & $52.8^{+2.6}_{-2.8}$ & $-8.1^{+2.2}_{-2.2}$ & $0.52^{+0.04}_{-0.04}$ & $-0.07^{+0.24}_{-0.16}$ & $-2.35^{+0.26}_{-0.17}$ & $-2.67^{+0.21}_{-0.17}$ & $-4.11^{+0.49}_{-0.67}$ & $1858^{+113}_{-156}$ & $5.23^{+0.06}_{-0.06}$ & $-7.27$ \\
Chem EQ & cd & HI & $\mathcal{U}$ & $48.8^{+8.2}_{-6.4}$ & $0.69^{+0.13}_{-0.07}$ & $47.7^{+2.3}_{-2.5}$ & $-8.1^{+1.9}_{-2.0}$ & $0.51^{+0.04}_{-0.04}$ & $-0.04^{+0.19}_{-0.16}$ & $-2.34^{+0.21}_{-0.16}$ & $-2.64^{+0.19}_{-0.18}$ & $-3.87^{+0.4}_{-1.02}$ & $1891^{+93}_{-85}$ & $5.4^{+0.11}_{-0.17}$ & $-5.95$ \\
Chem EQ & cd & HI & $\mathcal{G}$ & $49.6^{+6.6}_{-6.5}$ & $0.85^{+0.02}_{-0.03}$ & $51.8^{+2.8}_{-2.9}$ & $-6.7^{+1.8}_{-2.0}$ & $0.5^{+0.05}_{-0.05}$ & $-0.06^{+0.22}_{-0.16}$ & $-2.35^{+0.23}_{-0.2}$ & $-2.63^{+0.2}_{-0.14}$ & $-3.93^{+0.42}_{-0.7}$ & $1859^{+79}_{-141}$ & $5.23^{+0.06}_{-0.06}$ & $-5.76$ \\
Pquench & cloudless & HI & $\mathcal{U}$ & $47.5^{+6.9}_{-6.9}$ & $0.72^{+0.19}_{-0.09}$ & $49.6^{+2.2}_{-2.5}$ & $-8.8^{+2.0}_{-1.9}$ & $0.57^{+0.06}_{-0.05}$ & $0.01^{+0.18}_{-0.18}$ & $-2.25^{+0.18}_{-0.17}$ & $-2.63^{+0.16}_{-0.17}$ & $-4.55^{+0.18}_{-0.2}$ & $1628^{+109}_{-97}$ & $5.34^{+0.13}_{-0.18}$ & $0.0$ \\
Pquench & cloudless & HI & $\mathcal{G}$ & $49.1^{+7.5}_{-6.4}$ & $0.86^{+0.03}_{-0.03}$ & $55.8^{+2.6}_{-2.4}$ & $-8.6^{+1.9}_{-2.0}$ & $0.58^{+0.05}_{-0.05}$ & $0.0^{+0.19}_{-0.17}$ & $-2.26^{+0.19}_{-0.17}$ & $-2.64^{+0.17}_{-0.18}$ & $-4.59^{+0.19}_{-0.17}$ & $1614^{+102}_{-94}$ & $5.22^{+0.06}_{-0.07}$ & $11.22$ \\
Pquench & am & HI & $\mathcal{U}$ & $49.1^{+6.7}_{-6.7}$ & $0.72^{+0.17}_{-0.09}$ & $50.4^{+2.2}_{-2.5}$ & $-8.7^{+1.7}_{-1.8}$ & $0.57^{+0.05}_{-0.04}$ & $0.07^{+0.19}_{-0.16}$ & $-2.18^{+0.2}_{-0.16}$ & $-2.56^{+0.17}_{-0.15}$ & $-4.58^{+0.17}_{-0.18}$ & $1553^{+107}_{-119}$ & $5.37^{+0.12}_{-0.19}$ & $-4.37$ \\
Pquench & am & HI & $\mathcal{G}$ & $45.2^{+6.5}_{-6.0}$ & $0.86^{+0.03}_{-0.03}$ & $55.2^{+2.6}_{-2.6}$ & $-8.4^{+2.0}_{-2.3}$ & $0.59^{+0.04}_{-0.05}$ & $0.06^{+0.2}_{-0.18}$ & $-2.2^{+0.21}_{-0.18}$ & $-2.61^{+0.21}_{-0.21}$ & $-4.65^{+0.24}_{-0.29}$ & $1559^{+95}_{-97}$ & $5.19^{+0.07}_{-0.07}$ & $6.56$ \\
Pquench & cd & HI & $\mathcal{U}$ & $47.8^{+5.7}_{-5.9}$ & $0.78^{+0.18}_{-0.13}$ & $50.6^{+2.1}_{-2.0}$ & $-8.6^{+1.8}_{-1.6}$ & $0.57^{+0.05}_{-0.05}$ & $-0.04^{+0.21}_{-0.19}$ & $-2.3^{+0.19}_{-0.2}$ & $-2.68^{+0.2}_{-0.19}$ & $-4.71^{+0.24}_{-0.35}$ & $1621^{+170}_{-158}$ & $5.29^{+0.17}_{-0.19}$ & $6.48$ \\
Pquench & cd & HI & $\mathcal{G}$ & $47.5^{+6.8}_{-6.1}$ & $0.87^{+0.02}_{-0.02}$ & $54.9^{+2.7}_{-2.7}$ & $-7.0^{+1.7}_{-1.8}$ & $0.56^{+0.06}_{-0.05}$ & $0.03^{+0.19}_{-0.17}$ & $-2.24^{+0.2}_{-0.16}$ & $-2.58^{+0.14}_{-0.16}$ & $-4.65^{+0.22}_{-0.23}$ & $1535^{+111}_{-133}$ & $5.19^{+0.07}_{-0.06}$ & $16.33$ \\
C+H+C & cloudless & HI & $\mathcal{U}$ & $47.6^{+8.2}_{-7.1}$ & $0.76^{+0.19}_{-0.12}$ & $50.7^{+2.4}_{-2.6}$ & $-9.4^{+2.1}_{-1.9}$ & $0.63^{+0.06}_{-0.06}$ & $-0.06^{+0.16}_{-0.14}$ & $-2.3^{+0.16}_{-0.14}$ & $-2.71^{+0.18}_{-0.17}$ & $-7.52^{+1.75}_{-1.55}$ & $2584^{+156}_{-182}$ & $5.32^{+0.15}_{-0.21}$ & $8.59$ \\
C+H+C & cloudless & HI & $\mathcal{G}$ & $50.2^{+6.4}_{-6.1}$ & $0.86^{+0.03}_{-0.02}$ & $55.4^{+2.6}_{-2.4}$ & $-8.8^{+1.8}_{-1.9}$ & $0.6^{+0.06}_{-0.06}$ & $-0.07^{+0.18}_{-0.18}$ & $-2.31^{+0.18}_{-0.18}$ & $-2.69^{+0.17}_{-0.16}$ & $-7.71^{+1.54}_{-1.39}$ & $2417^{+269}_{-299}$ & $5.23^{+0.06}_{-0.06}$ & $-12.05$ \\
C+H+C & am & HI & $\mathcal{U}$ & $47.8^{+7.3}_{-6.2}$ & $0.76^{+0.22}_{-0.12}$ & $50.6^{+2.3}_{-2.4}$ & $-8.6^{+2.0}_{-1.9}$ & $0.6^{+0.07}_{-0.06}$ & $-0.05^{+0.21}_{-0.2}$ & $-2.29^{+0.21}_{-0.2}$ & $-2.65^{+0.19}_{-0.21}$ & $-7.81^{+1.62}_{-1.42}$ & $2136^{+448}_{-379}$ & $5.29^{+0.17}_{-0.22}$ & $9.66$ \\
C+H+C & am & HI & $\mathcal{G}$ & $47.1^{+6.5}_{-6.4}$ & $0.86^{+0.03}_{-0.03}$ & $56.8^{+3.0}_{-2.5}$ & $-8.7^{+2.5}_{-2.1}$ & $0.62^{+0.05}_{-0.05}$ & $-0.03^{+0.17}_{-0.17}$ & $-2.27^{+0.16}_{-0.17}$ & $-2.67^{+0.15}_{-0.16}$ & $-7.49^{+1.45}_{-1.48}$ & $1910^{+616}_{-215}$ & $5.2^{+0.07}_{-0.07}$ & $1.4$ \\
C+H+C & cd & HI & $\mathcal{U}$ & $48.4^{+7.3}_{-6.6}$ & $0.72^{+0.2}_{-0.09}$ & $50.1^{+2.5}_{-2.5}$ & $-8.8^{+1.8}_{-1.8}$ & $0.6^{+0.06}_{-0.06}$ & $-0.06^{+0.23}_{-0.18}$ & $-2.3^{+0.22}_{-0.18}$ & $-2.67^{+0.19}_{-0.17}$ & $-8.14^{+1.6}_{-1.19}$ & $1999^{+476}_{-231}$ & $5.34^{+0.14}_{-0.19}$ & $4.8$ \\
C+H+C & cd & HI & $\mathcal{G}$ & $51.4^{+6.1}_{-5.8}$ & $0.87^{+0.03}_{-0.03}$ & $56.2^{+2.7}_{-2.8}$ & $-8.5^{+2.2}_{-1.9}$ & $0.6^{+0.06}_{-0.05}$ & $-0.1^{+0.15}_{-0.15}$ & $-2.34^{+0.15}_{-0.15}$ & $-2.72^{+0.15}_{-0.14}$ & $-7.8^{+1.59}_{-1.46}$ & $1883^{+454}_{-214}$ & $5.23^{+0.06}_{-0.06}$ & $11.1$ \\
C+H & cloudless & HI & $\mathcal{U}$ & $43.7^{+5.4}_{-5.4}$ & $0.75^{+0.21}_{-0.11}$ & $50.5^{+2.0}_{-2.2}$ & $-9.4^{+2.2}_{-1.9}$ & $0.62^{+0.07}_{-0.06}$ & $-0.12^{+0.17}_{-0.17}$ & $-2.37^{+0.16}_{-0.17}$ & $-2.77^{+0.18}_{-0.17}$ & \nodata & $2530^{+196}_{-228}$ & $5.27^{+0.16}_{-0.21}$ & $9.85$ \\
C+H & cloudless & HI & $\mathcal{G}$ & $51.9^{+6.2}_{-5.3}$ & $0.85^{+0.03}_{-0.03}$ & $55.9^{+2.7}_{-2.9}$ & $-9.0^{+2.3}_{-2.0}$ & $0.62^{+0.06}_{-0.06}$ & $-0.06^{+0.17}_{-0.16}$ & $-2.3^{+0.16}_{-0.16}$ & $-2.69^{+0.16}_{-0.17}$ & \nodata & $2422^{+230}_{-262}$ & $5.25^{+0.06}_{-0.06}$ & $9.17$ \\
C+H & am & HI & $\mathcal{U}$ & $49.2^{+6.6}_{-6.7}$ & $0.71^{+0.17}_{-0.08}$ & $51.3^{+2.1}_{-2.1}$ & $-9.3^{+1.8}_{-1.6}$ & $0.62^{+0.06}_{-0.05}$ & $-0.01^{+0.19}_{-0.18}$ & $-2.25^{+0.19}_{-0.17}$ & $-2.65^{+0.19}_{-0.18}$ & \nodata & $1832^{+520}_{-172}$ & $5.36^{+0.13}_{-0.19}$ & $7.39$ \\
C+H & am & HI & $\mathcal{G}$ & $53.3^{+4.6}_{-3.7}$ & $0.85^{+0.02}_{-0.03}$ & $55.8^{+3.1}_{-3.0}$ & $-8.4^{+2.3}_{-1.8}$ & $0.62^{+0.05}_{-0.06}$ & $-0.01^{+0.18}_{-0.19}$ & $-2.25^{+0.17}_{-0.19}$ & $-2.66^{+0.17}_{-0.16}$ & \nodata & $1874^{+444}_{-153}$ & $5.27^{+0.04}_{-0.04}$ & $10.96$ \\
C+H & cd & HI & $\mathcal{U}$ & $49.4^{+6.9}_{-6.7}$ & $0.75^{+0.21}_{-0.11}$ & $50.5^{+2.4}_{-2.6}$ & $-8.5^{+2.0}_{-1.8}$ & $0.6^{+0.06}_{-0.06}$ & $-0.04^{+0.2}_{-0.17}$ & $-2.28^{+0.2}_{-0.17}$ & $-2.64^{+0.16}_{-0.15}$ & \nodata & $2169^{+436}_{-399}$ & $5.32^{+0.14}_{-0.21}$ & $12.82$ \\
C+H & cd & HI & $\mathcal{G}$ & $51.1^{+6.2}_{-5.5}$ & $0.86^{+0.03}_{-0.03}$ & $55.9^{+2.4}_{-2.6}$ & $-9.2^{+2.1}_{-1.7}$ & $0.63^{+0.05}_{-0.06}$ & $0.0^{+0.18}_{-0.16}$ & $-2.24^{+0.18}_{-0.16}$ & $-2.66^{+0.19}_{-0.17}$ & \nodata & $1834^{+603}_{-209}$ & $5.23^{+0.05}_{-0.05}$ & $1.93$ \\
\hline
Chem EQ & cd & HI + PH & $\mathcal{G}$ & $48.5^{+7.3}_{-6.3}$ & $0.83^{+0.03}_{-0.03}$ & $53.5^{+2.6}_{-2.7}$ & $-8.3^{+1.9}_{-2.1}$ & $0.53^{+0.03}_{-0.03}$ & $1.08^{+0.23}_{-0.22}$ & $-1.25^{+0.22}_{-0.18}$ & $-1.71^{+0.16}_{-0.15}$ & $-3.43^{+0.23}_{-0.26}$ & $1316^{+26}_{-27}$ & $5.24^{+0.07}_{-0.07}$ & $-22.02$ \\
Pquench & cloudless & HI + PH & $\mathcal{U}$ & $46.7^{+7.8}_{-6.8}$ & $0.65^{+0.07}_{-0.04}$ & $55.5^{+2.3}_{-2.5}$ & $-8.0^{+1.5}_{-1.6}$ & $0.53^{+0.06}_{-0.05}$ & $0.86^{+0.24}_{-0.2}$ & $-1.43^{+0.21}_{-0.18}$ & $-1.82^{+0.19}_{-0.16}$ & $-3.92^{+0.23}_{-0.24}$ & $1391^{+51}_{-68}$ & $5.41^{+0.1}_{-0.11}$ & $0.0$ \\
Pquench & am & HI + PH & $\mathcal{U}$ & $45.9^{+6.8}_{-6.4}$ & $0.67^{+0.1}_{-0.05}$ & $55.9^{+2.5}_{-2.6}$ & $-7.9^{+1.4}_{-2.1}$ & $0.54^{+0.05}_{-0.05}$ & $0.86^{+0.24}_{-0.29}$ & $-1.44^{+0.22}_{-0.26}$ & $-1.81^{+0.19}_{-0.22}$ & $-3.95^{+0.23}_{-0.22}$ & $1361^{+76}_{-111}$ & $5.41^{+0.09}_{-0.14}$ & $-3.81$ \\
Pquench & cd & HI + PH & $\mathcal{U}$ & $48.7^{+6.6}_{-5.3}$ & $0.69^{+0.11}_{-0.07}$ & $54.8^{+2.1}_{-2.2}$ & $-8.1^{+1.4}_{-1.8}$ & $0.57^{+0.05}_{-0.05}$ & $1.06^{+0.22}_{-0.18}$ & $-1.27^{+0.19}_{-0.17}$ & $-1.66^{+0.13}_{-0.15}$ & $-4.03^{+0.28}_{-0.23}$ & $1296^{+103}_{-160}$ & $5.4^{+0.08}_{-0.14}$ & $-3.44$ \\
Pquench & cd & HI + PH & $\mathcal{G}$ & $47.0^{+6.8}_{-6.9}$ & $0.85^{+0.03}_{-0.03}$ & $56.1^{+2.6}_{-2.5}$ & $-7.9^{+1.3}_{-1.7}$ & $0.57^{+0.05}_{-0.05}$ & $1.07^{+0.18}_{-0.17}$ & $-1.24^{+0.15}_{-0.14}$ & $-1.67^{+0.14}_{-0.11}$ & $-3.94^{+0.21}_{-0.23}$ & $1217^{+36}_{-62}$ & $5.21^{+0.07}_{-0.07}$ & $-22.4$ \\
\hline
Chem EQ & cd & HI + LO & $\mathcal{G}$ & $47.4^{+6.7}_{-6.4}$ & $0.75^{+0.02}_{-0.02}$ & $51.3^{+3.1}_{-2.9}$ & $-6.9^{+1.6}_{-1.8}$ & $0.55^{+0.03}_{-0.04}$ & $0.57^{+0.15}_{-0.14}$ & $-1.73^{+0.15}_{-0.13}$ & $-2.19^{+0.06}_{-0.06}$ & $-3.21^{+0.21}_{-0.27}$ & $1373^{+17}_{-18}$ & $5.32^{+0.06}_{-0.07}$ & $-16.25$ \\
Pquench & cloudless & HI + LO & $\mathcal{U}$ & $43.7^{+6.4}_{-6.9}$ & $0.61^{+0.02}_{-0.01}$ & $54.8^{+2.5}_{-2.9}$ & $-7.6^{+1.9}_{-2.0}$ & $0.53^{+0.03}_{-0.03}$ & $0.31^{+0.14}_{-0.12}$ & $-1.97^{+0.13}_{-0.13}$ & $-2.39^{+0.07}_{-0.07}$ & $-3.57^{+0.13}_{-0.11}$ & $1484^{+13}_{-18}$ & $5.45^{+0.06}_{-0.08}$ & $0.0$ \\
Pquench & am & HI + LO & $\mathcal{U}$ & $46.1^{+6.6}_{-7.2}$ & $0.61^{+0.02}_{-0.01}$ & $54.0^{+3.1}_{-3.5}$ & $-5.6^{+1.6}_{-1.8}$ & $0.53^{+0.04}_{-0.04}$ & $0.35^{+0.15}_{-0.14}$ & $-1.94^{+0.16}_{-0.14}$ & $-2.36^{+0.07}_{-0.07}$ & $-3.55^{+0.12}_{-0.12}$ & $1478^{+14}_{-21}$ & $5.48^{+0.06}_{-0.07}$ & $8.98$ \\
Pquench & cd & HI + LO & $\mathcal{U}$ & $46.4^{+6.4}_{-6.6}$ & $0.61^{+0.02}_{-0.01}$ & $53.5^{+2.6}_{-3.1}$ & $-6.2^{+1.9}_{-2.1}$ & $0.53^{+0.04}_{-0.04}$ & $0.32^{+0.14}_{-0.14}$ & $-1.96^{+0.14}_{-0.14}$ & $-2.38^{+0.07}_{-0.06}$ & $-3.55^{+0.1}_{-0.11}$ & $1475^{+13}_{-16}$ & $5.48^{+0.06}_{-0.07}$ & $2.74$ \\
Pquench & cd & HI + LO & $\mathcal{G}$ & $42.9^{+5.7}_{-6.3}$ & $0.76^{+0.02}_{-0.02}$ & $53.1^{+2.7}_{-2.5}$ & $-7.7^{+1.9}_{-1.7}$ & $0.55^{+0.04}_{-0.04}$ & $0.54^{+0.16}_{-0.14}$ & $-1.76^{+0.17}_{-0.15}$ & $-2.21^{+0.06}_{-0.07}$ & $-3.3^{+0.09}_{-0.1}$ & $1360^{+21}_{-21}$ & $5.27^{+0.05}_{-0.07}$ & $-24.17$ \\
\enddata
\tablenotetext{a}{Our chemistry settings are (1) `Chem EQ': chemical equilibrium; (2) `Pquench': chemical non-equilibrium using quench pressure from \cite{Molliere:2020aa}; (3) `C+H+C': chemical non-equilibrium using free retrievals of the CO + H$_2$O + CH$_4$ molecules; (4) `C+H': chemical non-equilibrium using free retrievals of the CO + H$_2$O molecules.}
\tablenotetext{b}{Our cloud settings are (1) `cloudless': no cloud; (2) `am': cloud model from \cite{Ackerman:2001aa} with the MgSiO$_3$ amorphous; (3) `cd': cloud model from \cite{Ackerman:2001aa} with the MgSiO$_3$ condensate.}
\tablenotetext{c}{Our data types are (1) `HI': high-resolution spectra; (2) `HI + PH': high-resolution spectra and photometric measurements from \cite{Currie:2020aa}; (3) `HI + LO': high-resolution spectra and low-resolution spectra from \cite{Currie:2020aa}.}
\tablenotetext{d}{Our radius priors include (1) $\mathcal{U}$: the wide uniform priors and (2) $\mathcal{G}$: the restricted Gaussian priors using the \cite{Baraffe:2003aa} brown dwarf evolutionary models.}
\tablenotetext{e}{Mass fractions for CO (m$_\mathrm{CO}$), H$_2$O (m$_\mathrm{H_2O}$), and CH$_4$ (m$_\mathrm{CH_4}$)}
\tablenotetext{f}{The Bayes factor of log base 10. 
Because the Bayes factor depends on the data used in the fit, we set our baseline models using the cloudless disequilibrium `Pquench' retrievals under the wide uniform radius priors for each data combination (HI; HI + PH; HI + LO).}
\end{deluxetable*}
\end{longrotatetable}

\subsection{Retrievals with High-resolution and Joint-fitting Low-resolution Spectra or Broad-band Photometry}\label{subsec:joint_retrieval}

While we showed that our retrievals using high-resolution spectra are insensitive to the radius priors, we explored whether joint-fitting low-resolution spectra or broad-band photometry could resolve the small radius issue.
Our retrieval setting is similar to Section~\ref{subsec:highres_retrieval}, following the prescription in \cite{Xuan:2022aa}.
The low-resolution spectra and broadband photometry are compiled from the SCExAO/CHARIS data in \cite{Currie:2020aa} (i.e. $J$ = 16.91 $\pm$ 0.11, $H$ = 16.00 $\pm$ 0.09, $K_s$ 15.37 $\pm$ 0.09 and their Table 2).
While our joint-fitting forward retrieval modeling is similar to that for the high-resolution spectra alone, we included the major molecular absorbers of CO, H$_2$O, CH$_4$ from HITEMP database \citep{Hargreaves:2020aa}, and NH$_3$, CO$_2$, FeH, VO (B. Plez), TiO (B. Plez), K (F. Allard), Na (F. Allard) from \citep{Molliere:2019ab}.
The sampled wavelength is between 0.8 to 2.5~{\micron} using the correlated-k sampling method.
For broadband photometry, the MKO $JHK_s$ profiles are multiplied with our broadband, low-resolution retrieval model before computing log-likelihood\footnote{\url{http://irtfweb.ifa.hawaii.edu/~nsfcam/filters.html}}.
The log-likelihood is the sum of the high-resolution spectra and low-resolution spectra or the broadband photometry.
As running these retrievals roughly doubles the computation time, we only examined our baseline combinations of retrievals, which are chemical non-equilibrium with the quench pressure.

\begin{figure}
    \centering
    \includegraphics[width=0.48\textwidth]{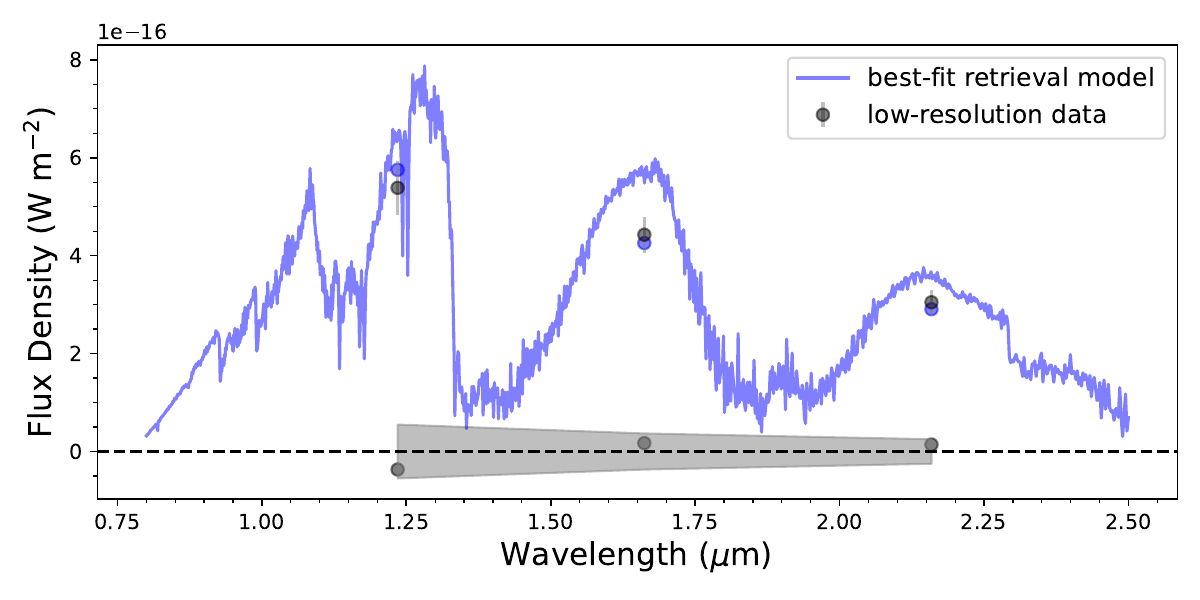}
    \includegraphics[width=0.48\textwidth]{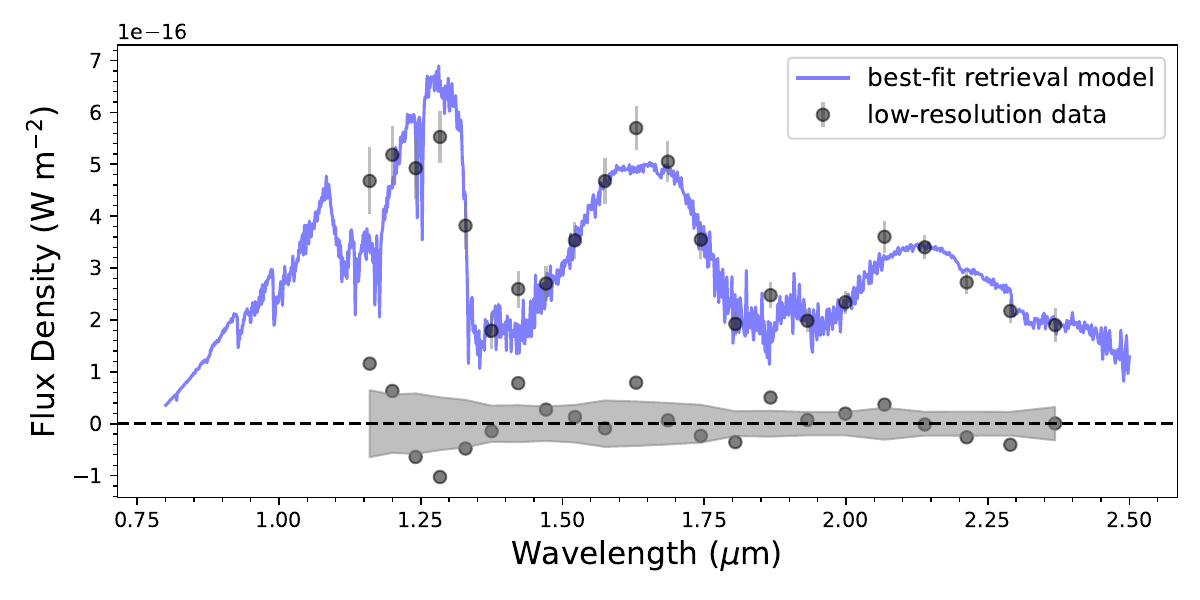}
    \caption{Joint-fitting retrieval results for our baseline models. \textit{Top}: Joint-fitting $JHK$ photometry. \textit{Bottom}: Joint-fitting near-infrared low-resolution spectra. The low-resolution data or photometry were compiled from \cite{Currie:2020aa} (black dots). The best-fit model is shown in blue, with the residual (data $-$ model) in black dots near zero and data noise in the grey-shaded region.
    }
    \label{fig:joint_fit_retrievals}
\end{figure}

Figure~\ref{fig:retrieval_posterior_compare} and Table~\ref{table:retrieval_result} compare our forward retrieval models under a select set of data combinations.
Figure~\ref{fig:joint_fit_retrievals} shows our baseline joint-fitting retrieval results with photometry and low-resolution spectra.
The continuum from low-resolution spectra or photometry is difficult to reproduce, partly due to residual stellar light contribution (e.g., AF Lep b low-resolution spectra reported in \citealp{Mesa:2023aa, De-Rosa:2023aa}).
Joint-fitting low-resolution spectra or broadband near-infrared photometry \deleted{(under the wide uniform radius priors) }with our wide, uniform priors did not resolve the radius issue.
The retrieved mass, $v\sin{i}$, RV, and C/O are consistent.
However, the measured metallicities are dependent on different data combinations.
Retrievals of joint-fitting the broadband photometry give [C/H]=0.85--1.1, while Retrievals of joint-fitting low-resolution spectra return [C/H]=0.3--0.6, much higher than our measured metallicity of [C/H]$\sim$0.0 from high-resolution retrievals.
The main reason is that the low-resolution spectra or photometry are more sensitive to either clouds or the deep atmosphere if it is clear. The effective optical depth ties directly to the inferred metallicity, which reflects the [C/H] trend. This effect mirrors the trend for [O/H], via water, which cancels out the C/O estimate. This means that our retrieved C/O is more robust than [C/H], which holds even for a relatively fast rotator such as HD~33632 Ab.
As our high-resolution spectra with CO and H$_2$O molecules anchor our metallicities, we regard the inferred [C/H]s from joint-fitting low-resolution or broadband photometry as not robust.
Better characterization or suppression of host star light for low-resolution spectra in the future might provide us with more consistent results, to precisely model the continuum and thus clouds of low-temperature objects (e.g., clouds might affect the inferred metallicity; \citealp{Calamari:2024aa}).

\subsection{Effective Temperatures}

Comparisons of the effective temperatures inferred in our retrievals allow us to compare these results against self-consistent modeling grids.
Our effective temperatures are computed by integrating between the wavelength range of 0.5 to 30.0~$\micron$.
We used the correlated-k method to sample the molecular species used in our low-resolution spectral retrievals (Section~\ref{subsec:joint_retrieval}), with the uncertainties computed from the Monte Carlo method.

For {\teff} inferred from our high-resolution spectral retrievals, the cloudless free retrievals have unphysically high temperatures ($\sim$2400--2500~K). Clouds could mitigate the issue.
The {\teff}s under the chemical equilibrium ($\sim$1720--1920~K) are similar to those {\teff} using Sonora models ($\sim 1880$~K).
Among our high-resolution retrievals, the lowest {\teff}s are the retrievals parameterized by the quench pressure ($\sim$1550--1630~K), consistent with the {\teff} from the BT-Settl model fit ($\sim 1470$~K).
While our best-fit effective temperatures are generally higher than the L9.5 spectral type ($\sim$1300~K; \citealp{Mamajek:2013aa}), this could indicate that HD~33632 Ab could be an earlier spectral type than the reported L9.5$^{+1.0}_{-3.0}$ from \cite{Curiel:2020aa}, where methane is not yet significantly converted from carbon monoxide.
The literature spectral type is uncertain due to relatively low S/N spectra ($\sim$20--40) with SCExAO/CHARIS from \cite{Currie:2020aa}, and the literature spectral type to effective temperature has a scatter $\sim$113~K \citep{Filippazzo:2015aa}.
However, we disregard our {\teff} measured through the high-resolution spectra because our high-resolution spectra are normalized in the fitting routine and not flux-calibrated, limited in a narrow wavelength where the cloud effect is more evident from broad-band photometry and wide wavelength coverage. Additionally, the {\teff}s inferred from high-resolution near-infrared spectra using theoretical stellar/substellar atmosphere models could deviate by 500~K from the expected effective temperatures measured with photometry or low-resolution spectra \citep{Del-Burgo:2009aa, Hsu:2021aa, Hsu:2024aa}.
Our joint-fit retrievals with low-resolution spectra and photometry generally imply {\teff} = 1200--1500~K, consistent with the literature spectral type. 

\subsection{Comparison to Host Star Metallicity and Abundances}\label{subsec:host_abundances}

Our final goal of the retrieval modeling is to validate whether our inferred metallicity and abundances (C/O ratio) are consistent with the host star. 
HD~33632A has several metallicity measurements in the literature.
The wide range of metallicity determinations ($-0.01$ to $-0.33$~dex) in the literature highlights systematics in the photometric and spectroscopic methods and the corresponding data \citep{Ammons:2006aa, Valenti:2005aa, Chen:2000aa, Soubiran:2016aa, Anderson:2012aa, Soubiran:2010aa, Taylor:2005aa, Marsakov:1995aa, Mishenina:2004aa, Suchkov:2003aa, Holmberg:2009aa}.
Spectroscopic metallicity is generally preferred, as it suffers less from the imprecise parallax measurements compared to other methods (esp. pre-\textit{Gaia} era),
and these measurements are in good agreement ($-0.18$ to $-0.25$) with the reported uncertainties.
Among the spectroscopic constraints, we adopted the host star abundance measurements from \cite{Rice:2020aa} which reported the only available C and O abundances for HD~33632A.
The abundances with Keck/HIRES spectra using a data-driven method from \cite{Rice:2020aa}, which trained their sample based on the SPOCS measurements \citep{Valenti:2005aa}, indicated that [Fe/H] = $-0.15\pm0.03$, [C/H] = $-0.13\pm0.05$ and [O/H] = $0.01\pm0.07$.
The inferred C/O ratio = 0.39$^{+0.12}_{-0.09}$ with uncertainty propagated by the Monte Carlo method.

In our retrievals with high-resolution spectra, the cloud assumptions typically lower our inferred {\teff}, which can be significant by a few hundred Kelvins, but we found consistent C/O and [C/H] regardless of clouds.
While the cloudy models are generally favored over the cloudless models, with our high-resolution spectra, clouds signatures are difficult to recover in our high-resolution spectra, as we are fitting only a narrow range of wavelengths near $\sim$2.3~{\micron}.
The low-resolution or photometry joint-fit retrievals do provide more consistent {\teff}s compared to the expected spectral type, but the measured metallicities and abundances are much higher compared to the retrievals using only high-resolution spectra (Section~\ref{subsec:joint_retrieval}).
The high-resolution spectra can resolve the molecular lines which enable more accurate abundances, so our adopted C/O and [C/H] are only based on our high-resolution spectral retrievals.

The brown dwarf companion is expected to have the same metallicity and abundances as its host star.
In our baseline retrieval, the measured [C/H] = $-0.01\pm0.18$ and C/O = 0.57$\pm$0.06 for HD~33632 Ab are consistent within 1$\sigma$ and 1.5$\sigma$ (or 0.16 dex and 0.18 difference), respectively.
Other settings using the high-resolution spectra are consistent with the uncertainties.
To realistically quantify the systematics of C/O and [C/H] in our retrieval models, we compared four brown dwarf companions with available host star metallicities observed with KPIC (\citealp{Wang:2022aa, Xuan:2022aa, Xuan:2024aa}.
We found the average differences of C/O and [C/H] (or [Fe/H]) compared to the host stars of these systems are 0.14 and 0.17, respectively, so we determined the C/O and [C/H] and associated uncertainties for HD~33632 Ab as 0.58$\pm$0.14 and 0.0$\pm$0.2~dex, respectively.

\section{Orbital Analysis} \label{sec:orbit}

In this section, we present the updated orbital solutions for the HD~33632 Ab system. 
The companion RV, the new astrometry from \textit{Gaia} DR3, and the EDR3 edition of the Hipparcos-Gaia Catalog of Accelerations (HGCA) from \cite{Brandt:2021aa}\footnote{We used the recently updated \texttt{orbitize!} available on \url{https://github.com/sblunt/orbitize}; with the commit 0fb5435. The current \texttt{orbitize!} is capable of fitting relative astrometry, companion (relative) RV, Hipparcos IAD data, as well as the HGCA catalog.} are available to refine the orbital solutions since the discovery by \cite{Currie:2020aa}.
We incorporated the stellar RV from \cite{Gaia-Collaboration:2023aa} ($-$1.75$\pm$0.12~{\kms})\footnote{The stellar RV is consistent in the literature since 2006 to 2022, including $-$1.7 $\pm$ 0.2~{\kms} \citep{Gontcharov:2006aa}, $-$1.8 $\pm$ 0.3~{\kms} \citep{Kharchenko:2007aa, Casagrande:2011aa}, $-$1.7 $\pm$ 0.1~{\kms} \citep{de-Bruijne:2012aa}, $-$1.88 $\pm$ 0.15~{\kms} \citep{Brandt:2021aa}, and $-$1.9 $\pm$ 0.3~{\kms}\citep{Tsantaki:2022aa}. While we also observed stellar spectra with KPIC, the stellar features in $K$-band are not useful to derive RV.} and companion RV from our KPIC spectra ($-$8$\pm$3~{\kms}), relative astrometry from \cite{Currie:2020aa} (three epochs), absolute astrometry, and the Lick RV measurements from \cite{Fischer:2014aa}.
For absolute astrometry, we fit \textit{Gaia} DR3 \citep{Gaia-Collaboration:2023aa} and \textit{Hipparcos} Intermediate Astrometric Data (IAD) data \citep{Perryman:1997ab,van-Leeuwen:2007aa}, or the Gaia EDR3 Edition of Hipparcos–Gaia Catalog of Accelerations from \cite{Brandt:2021aa}.
The \textit{Gaia} RV of the host star HD~33632A is used to compute the relative RV for the \texttt{orbitize!}. 
We also fit the overall systematic RV for Lick RV measurements because their RVs are measured relative to a given epoch (JD=2451206.78125).

For completeness, we also fit the orbital solution without the absolute astrometry and one with the Hipparcos-Gaia (EDR3) Catalog of Accelerations (HGCA) from \cite{Brandt:2021aa}, instead of the \textit{Gaia} and \textit{Hipparcos} IAD data directly, to examine any possible systematic uncertainties in our approach.

We used the \texttt{orbitize!} package \citep{Blunt:2020aa} to fit orbital solutions for the HD~33632 Ab system.
Our best-fit orbital parameters are obtained using the Markov chain Monte Carlo (MCMC) method using the \texttt{emcee} \citep{Foreman-Mackey:2013aa} package.
There are eleven parameters in the orbital modeling, including the semi-major axis (SMA; $a$), eccentricity ($e$), inclination ($i$), argument of periastron ($\omega$), position angle of nodes ($\Omega$), epoch of periastron passage ($\tau$) (as a fraction of orbital period), parallax ($\pi$), RV offset ($\gamma_0$) for host star RV (to model the systematic RV for the host star), RV error jitters\footnote{Note that the RV jitter term is to inflate the RV measurements for un-accounted noise, which is a common practice for radial velocity-related orbital fitting \citep{Howard:2014aa}.} ($\sigma_0$ and $\sigma_1$) for host and companion RVs, and the host star mass ($m_0$).

To fit the orbital solutions, we assumed Gaussian priors for the host star mass (1.11$\pm$0.09~M$_{\odot}$; \citealp{Currie:2020aa}) and the parallax ($\pi$=37.8953$\pm$0.0263~mas from \textit{Gaia} DR3; \citealp{Gaia-Collaboration:2023aa}). 
We assumed uniform priors for companion mass between 0.0 to 0.1~M$_{\odot}$.
The priors for the remaining parameters used the default setting in \texttt{orbitize!}:
$a$ with log-uniform priors from 0.001 to 10,000~au; $e$ with uniform priors from 0 to 1; $i$ with sine distribution (i.e. uniform inclinations); $\omega$ with uniform priors from 0 to 2$\pi$~rad; $\Omega$ with uniform priors from 0 to $\pi$~rad; $\tau$ with uniform priors from 0 to 1; $\gamma$ for the host star with uniform priors from $-5$ to $+5$~{\kms}; RV jitter for host star and companion with log-uniform priors from 0.0001 to 0.05~{\kms}.
Our MCMC sampler was set with 1,000 walkers, 50,000,000 steps (for all walkers), and the first 90$\%$ steps were discarded as burn-ins.

Under these assumptions, we adopted the best-fit orbital parameters using the HGCA (EDR3) with both the host and companion RVs, with our updated orbital parameters as $a$ = 18.4$^{+5.2}_{-2.8}$~au, 
$e$ = 0.25$^{+0.17}_{-0.18}$, 
$i$ = 32.94$^{+10.98}_{-20.56}$~deg, $\pi$ = 37.9$\pm$0.03~mas, companion mass $m_1$ = 36.63$^{6.69}_{4.17}$~M$_{\odot}$, 
primary mass $m_0$ = $1.11^{+0.09}_{-0.08}$~M$_{
\odot}$.
Our adopted best-fit orbit is shown in Figure~\ref{fig:hgca_orbit}.
Figure~\ref{fig:orbit_fit_compare} and Table~\ref{tab:orbit} compare the best-fit orbital solutions for the semi-major axis, eccentricity, parallax, inclination, and primary and secondary masses under various assumptions.
For the semi-major axis (SMA), we found that the \textit{Hipparcos} Intermediate Astrometric Data (IAD) data returned very different measurements compared to other methods, which were attributed to the underestimated uncertainties from the \textit{Hipparcos} IAD data. Therefore, we did not adopt any orbital solutions from the \textit{Hipparcos} IAD data.
The HGCA methods provided consistent SMAs compared to the measurements from \cite{Currie:2020aa}. Among the HGCA fits, the posteriors of SMAs are consistent but slightly bimodal, and we found that adding the host and companion RVs made our SMAs toward the lower valley $\sim$15~au.
The eccentricity is not well constrained, with its posteriors consistent with a wide range (0.0--0.6 within 95\% confidence intervals) due to the sparse orbital coverage.
The posteriors of parallaxes are consistent with our Gaussian priors, and as described are significantly different ($\sim$3.6$\sigma$) compared to the \textit{Gaia} DR2 parallax (37.6462$\pm$0.0643~mas) used in \cite{Currie:2020aa}.
The posteriors of inclinations and primary masses of our HGCA fit and relative astrometry with primary and companion RVs are consistent with the \cite{Currie:2020aa} measurements, which are reasonable given that we started with the same mass priors as in \cite{Currie:2020aa}. However, we found that the primary and companion masses from our \textit{Gaia} \textit{Hipparcos} IAD fits are significantly different than our other settings, due to the aforementioned underestimated \textit{Hipparcos} IAD uncertainties.
For the companion masses, the posteriors are consistent with our HGCA fits and \cite{Currie:2020aa} measurements.
The fits with relative astrometry and RVs are unable to constrain the companion masses well because the mass and SMA measurements are more sensitive to better orbit phase coverage from the absolute astrometry.
We, therefore, adopted the HGCA fit with both primary and companion RVs as it has all of the information available and is consistent with the literature values from \cite{Currie:2020aa}.


\section{Placing HD~33632 Ab to the Rotational Demographics of Very Low-mass Objects} \label{sec:rotation}

In this section, we compare our measured {\vsini} of HD~33632 Ab to the population of the very low-mass (VLM) objects reported in the literature, by comparing its {\vsini} against spectral type, and by rotational velocity against mass.

\subsection{Overall Projected Rotational Velocity Distribution}
The rotational evolution of very-low-mass objects ($M<0.1$~M$_{\odot}$) has been examined in several studies, for the ultracool dwarf ({\teff} $\leq$3000~K; spectral type of M6 or later; \citealp{Kirkpatrick:2005aa}) regime \citep{Zapatero-Osorio:2006aa, Radigan:2012aa, Metchev:2015aa, Prato:2015aa, Vos:2017aa, Vos:2018aa, Vos:2019aa, Hsu:2021aa, Popinchalk:2021aa, Hsu:2024aa} and for low-mass companions and gas giant exoplanets \citep{Bryan:2018aa, Bryan:2020ab, Xuan:2020aa, Wang:2021aa}.

\begin{figure}
    \centering
    \includegraphics[width=0.48\textwidth]{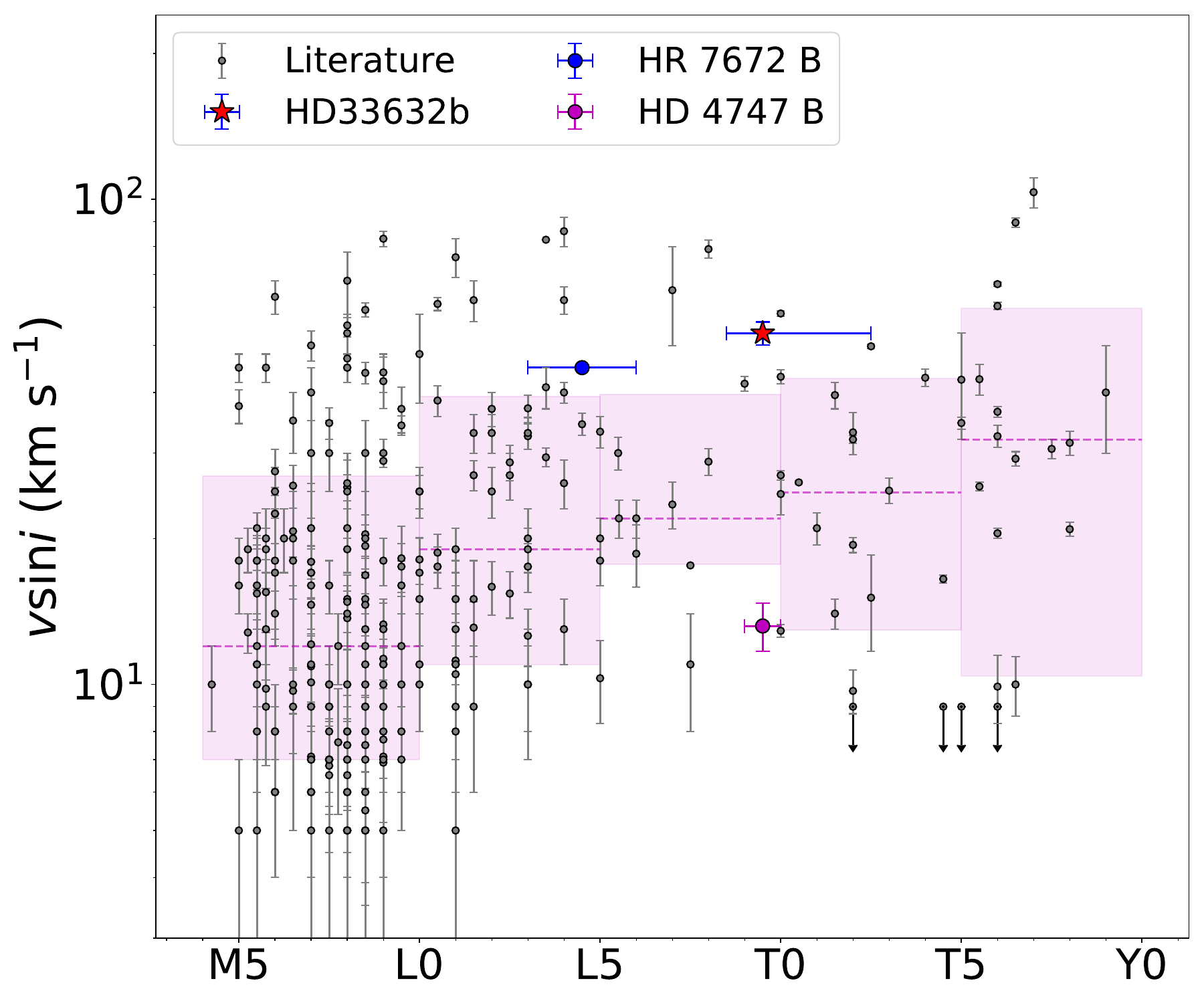}
    \caption{Projected rotational velocities versus spectral types of the very-low-mass objects (M5--T9).
    The $v\sin{i}$ measurements are labeled in grey dots, while the companions published with KPIC are labeled as the red star for the magenta dot for HR~7672 B \citep{Wang:2022aa}, the blue dot for HD~4747 B \citep{Xuan:2022aa}, and the red star for HD~33632 Ab (this work), respectively.
    The median values of the sample binned per five subtypes are labeled in dotted magenta lines, with their standard deviations and magenta-shaded regions.
    The majority of this sample is compiled from \cite{Crossfield:2014aa} and \cite{Hsu:2021aa, Hsu:2024aa}. 
    See Section~\ref{sec:rotation} for details.
    }
    \label{fig:vsini_rotation_population}
\end{figure}

Figure~\ref{fig:vsini_rotation_population} illustrates the current consensus of $v\sin{i}$ distributions over spectral types for VLM objects, ranging from M5 to T9.
The $v\sin{i}$ sample has 457 objects, compiled from \cite{Crossfield:2014aa, Hsu:2021aa, Tannock:2021aa, Hsu:2024aa}\deleted{; and Hsu et al. in prep}.
These high-resolution spectroscopic measurements are limited at $v\sin{i}$ $\sim$5--9~{\kms} for mid-M to T dwarfs. 
The overall increase of {\vsini} toward later spectral types can be explained by the spin-up mechanism under constant angular momentum evolution \citep{Zapatero-Osorio:2006aa, Hsu:2021aa}.
HD~33632 Ab, while being a relatively fast rotator ({\vsini} = 53$\pm$3~{\kms}), falls within the {\vsini} trend of the L dwarfs.
Interestingly, HD 4747 B, an L9.5 object, has a much lower $v\sin{i}$ (13.2$^{+1.4}_{-1.5}$~{\kms}; \citealp{Xuan:2022aa}) and methane detection, compared to HD~33632 Ab of the same spectral type.


\subsection{Comparison of Rotation with Mass}
Our comparisons of projected rotational velocity versus spectral type suffer from observational degeneracy between age and mass.
To possibly break this degeneracy, we attempted to parameterize the spin, evolved to the same age, versus mass, which might show evidence of a mass-dependent trend for rotation \citep{Batygin:2018aa, Ginzburg:2020aa}.

We examined the companion spin versus the mass of known low-mass companions ($M<0.1$~M$_{\odot}$), gas giant exoplanets, and solar system planets, following the analysis outlined in \cite{Wang:2021aa}.
The spins are parameterized as rotational velocity over break-up velocity.
Our prescription is detailed as follows.
We constructed a sample that has $v\sin{i}$ or rotational period measurements for companions, free-floating VLM objects, and solar system planets (i.e. Jupiter and Saturn). 
Our VLM rotational sample is mostly constructed from \cite{Bryan:2020ab, Tannock:2021aa, Popinchalk:2021aa, Wang:2021aa, Hsu:2021aa, Vos:2022aa, Hsu:2024aa}, for a total of 237 objects.
Among these catalogs, the majority of the sources lack direct mass measurements, so we cross-matched with the 2MASS catalog \citep{Cutri:2003aa} to obtain their $H-$band magnitudes and parallaxes from \textit{Gaia} DR3 \citep{Gaia-Collaboration:2023aa}.
Using absolute magnitudes and reasonable assumptions of their ages, their mass and radii can be inferred from interpolation of theoretical VLM evolutionary models \citep{Baraffe:2003aa}.
Next, we need to convert these measurements into rotational velocities.
There are two types of spins in our sample, including $v\sin{i}$ and rotational periods ($P$) from the photometric time series.
For $v\sin{i}$, the geometric sine distribution is assumed to compute the rotational velocity.
For rotational periods, radii are needed to infer the rotational velocity. 
We used the \cite{Baraffe:2003aa} models with age and another parameter to linearly interpolate and infer the model radius, which includes the mass (preferred), or absolute $H$-band magnitude.
We first use the dynamical masses, then the literature masses, and then the absolute $H$-band magnitude.
For ages, the known clusters are used \citep{Gagne:2018ab} (sampled as uniform distributions within their 1-$\sigma$ uncertainties), or a field age is assumed otherwise (sampled as uniform distributions between 0.5 to 10~Gyr).
We also removed the sources younger than 10~Myr to avoid possible disk interaction that slow down the spin and bias our analysis.
Using these inferred parameters (age, mass, radii), we can compute their rotational velocities and breakup velocities.
The breakup velocity is defined as $v_\text{breakup} = \sqrt{\frac{GM}{R}}$, where mass $M$ and radius $R$ are derived from the \cite{Baraffe:2003aa} evolutionary models, and $G$ is the gravitational constant.
Finally, to minimize the age effect on our analysis, we computed their ``final`` rotational velocities and ``final`` breakup velocities by evolving their rotation to 5~Gyr assuming conservation of angular momentum.
For the derived parameters, the associated uncertainties were propagated using the Monte Carlo method with 1000 draws.

\begin{figure*}
    \centering
    \includegraphics[width=0.45\textwidth]{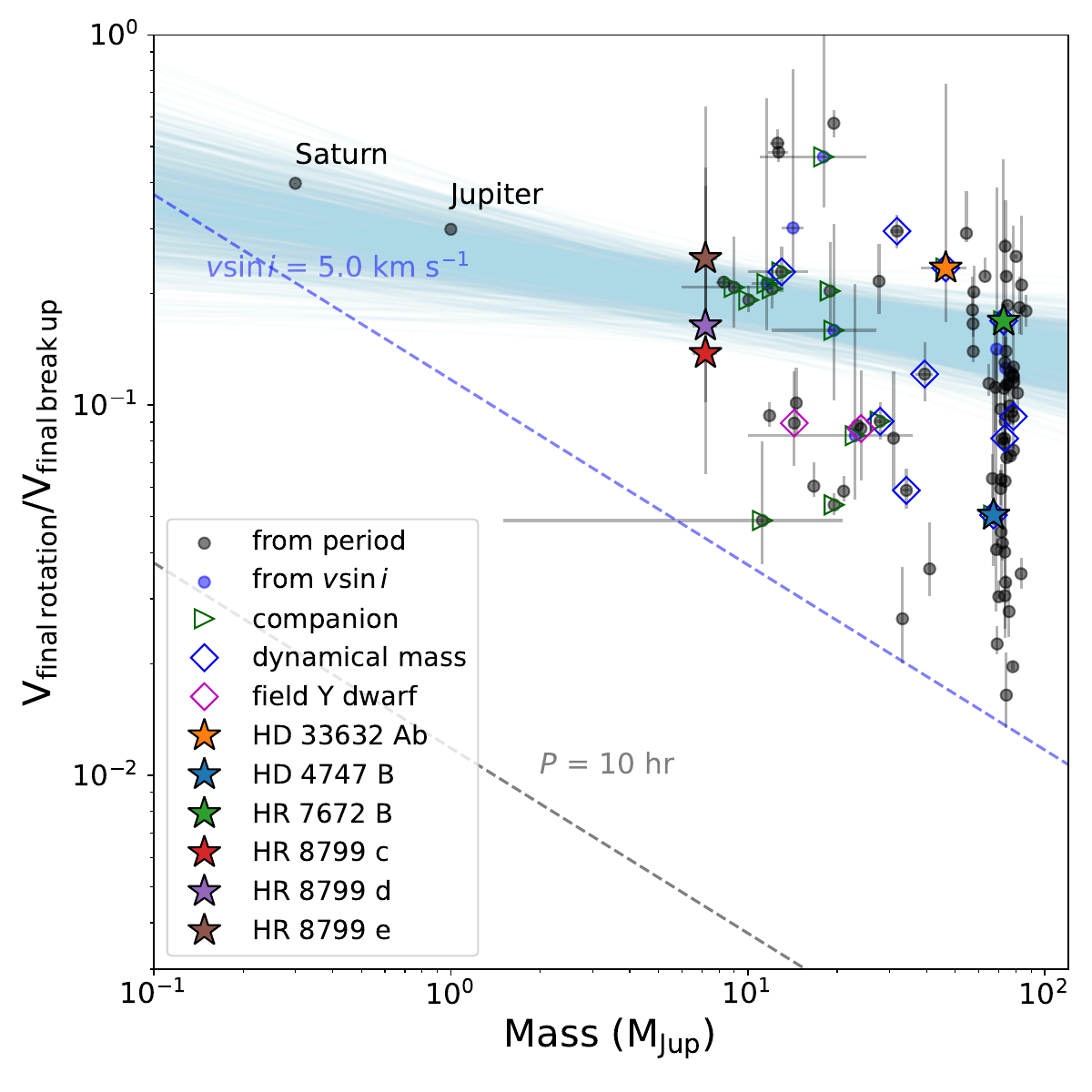}
    \includegraphics[width=0.45\textwidth]{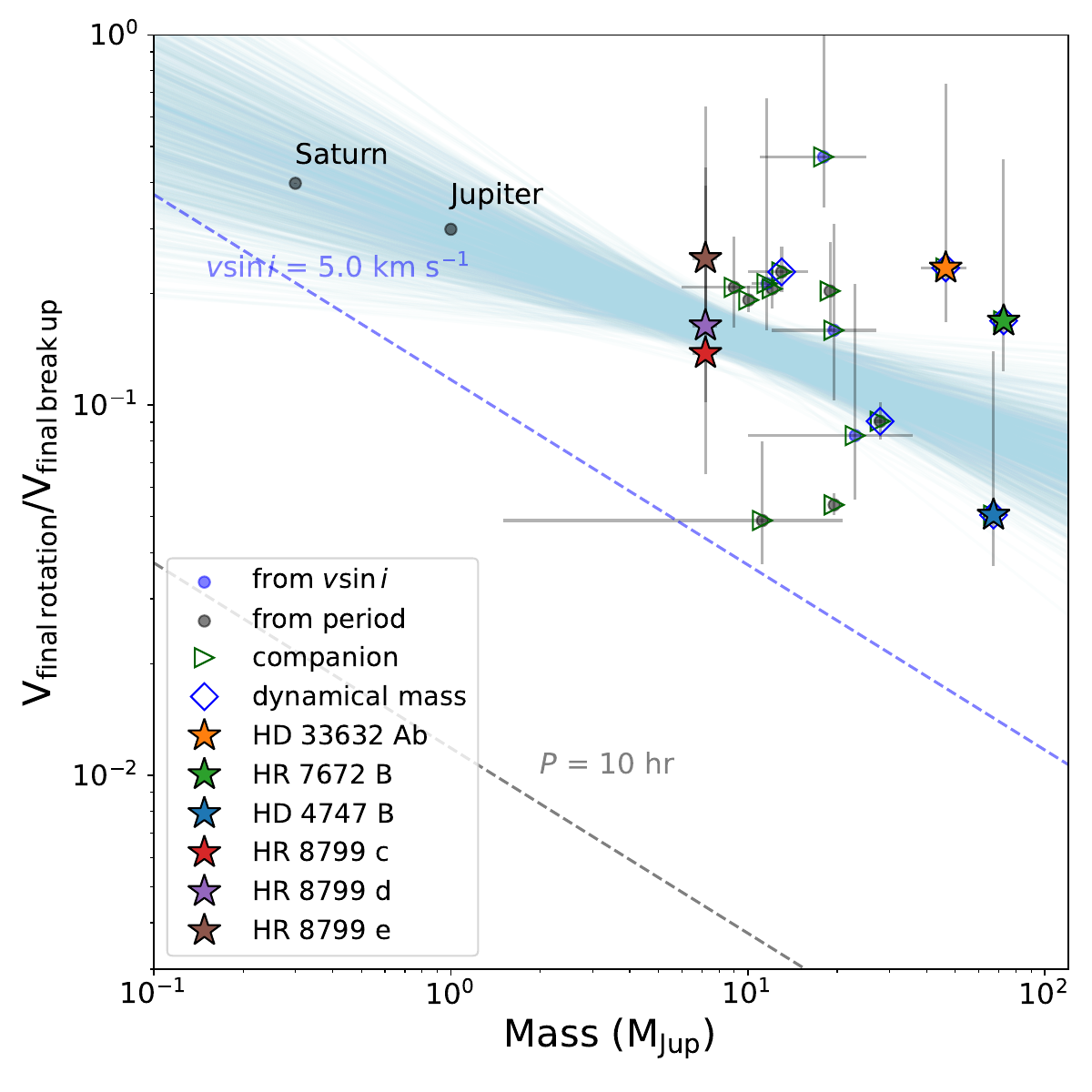}
    \caption{Companion spin versus mass for known low-mass companions ($M<0.1$~M$_{\odot}$) and solar system planets. The spins are parameterized as final rotational velocity over final break-up velocity under constant angular momentum evolution to 5~Gyr. 
    \textit{Left}: Final spin versus mass for all objects in our spin sample, including the free-floating field objects, low-mass companions, and solar system planets.
    The sample is compiled from literature $v\sin{i}$ (blue dots) and rotational period (grey dots) measurements.
    The known companions and dynamical mass measurements are labeled in green triangles and blue diamonds, respectively.
    Spin measurements published with KPIC are labeled in star symbols.
    The detection limits for $v\sin{i}$ ($<5$~{\kms}) and photometry time series ($>10$~hr) are plotted in blue and grey dashed lines, respectively.
    The random draws of 1000 solutions from the best-fit rotational trend from \cite{Wang:2021aa} are depicted in light blue lines.
    \textit{Right}: Final spin versus mass for companions in our spin sample, including the low-mass companions and solar system planets. The rest is the same as the left panel.
    }
    \label{fig:vrot_vbreak_final_companion}
\end{figure*}

Figure~\ref{fig:vrot_vbreak_final_companion} shows the ``final`` rotational velocities and ``final`` breakup velocities against the mass of our sample.
We overplotted the best-fit parameters for the mass versus rotation trend from \cite{Wang:2021aa} on top of our final sample.
In particular, the final rotational velocity and final break-up velocity of HD~33632 Ab are 73$^{+153}_{-20}$~{\kms} and 315$^{+26}_{-27}$~{\kms}, respectively.
The ratio of these two gives the final rotational velocity over break-up velocity of 0.23$^{+0.50}_{-0.07}$, well below the breakup velocity and following the population trend with the parameters in \cite{Bryan:2020ab, Wang:2021aa}.
The larger upper uncertainty highlights the inclination effect, while the more constrained lower uncertainty is partially constrained by our relatively high {\vsini}.
HR~7672B, another relatively fast rotator, follows a similar trend, with its final velocity over break-up velocity of 0.17$^{+0.29}_{-0.05}$.
HD~4747 B, on the other hand, has a much lower final velocity over break-up velocity of 0.050$^{+0.089}_{-0.013}$, due to its lower {\vsini} of 13.2$\pm$1.5~{\kms}.
For HR~8799 cde planets, their masses and {\vsini}s are similar (8.1$\pm$4~{\kms}, 10.1$\pm$2.8~{\kms}, and 15$\pm$2.6~{\kms}, respectively), resulting in similar final velocity over break-up velocity of 0.14$^{+0.25}_{-0.07}$, 0.16$^{+0.27}_{-0.06}$, and 0.25$^{+0.39}_{-0.08}$, respectively.
As pointed out in \cite{Wang:2021aa}, there seems to be no clear trend for rotation versus mass for the free-floating objects (left panel of Figure~\ref{fig:vrot_vbreak_final_companion}), while there is a tentative trend for companions and solar system planets (right panel of Figure~\ref{fig:vrot_vbreak_final_companion}).
However, toward the very low-mass end, only two solar system planets Jupiter and Saturn provide measurements below 5~M$_\mathrm{Jup}$, and the current ground- or space- instruments are unavailable to provide any rotation in this regime.
Future observations with young companions such as AF Lep b (\citealp{Mesa:2023aa, De-Rosa:2023aa, Franson:2023ab, Zhang:2023ab}; $\sim$3--5 M$_\mathrm{Jup}$) as well as the discovery of additional planetary mass objects below the deuterium burning mass limit will help us critically examine this trend.
Even for masses between 5--100~M$_\mathrm{Jup}$, the sample size is still too small, and more rotation measurements for VLM objects in this range are needed with instruments such as KPIC.

\section{Summary} \label{sec:sum}

We present the Keck/KPIC high-resolution $K$-band spectroscopy of the benchmark brown dwarf mass companion HD~33632 Ab, which provided companion radial and projected rotational velocities as well as CO and H$_2$O abundances. 
Our main findings are as follows:

\begin{itemize}
    \item HD~33632 Ab's projected rotational ({\vsini}) and radial (RV) velocities measurements are $53\pm3$~{\kms} and $-8\pm3$~{\kms}, respectively, using a forward-modeling framework that incorporates host star spectra, self-consistent substellar atmospheric models from the BT-Settl \citep{Allard:2012ab} and Sonora Bobcat models \citep{Marley:2021aa}. 
    Our {\vsini} and RV are consistent across different assumptions of clear versus cloudy models.
    \item Using our measured relative RV, updated \textit{Gaia} parallax, and HGCA catalog \citep{Brandt:2021aa}, we measured the updated orbital solutions with the \texttt{orbitize!} package.
    Our measured companion mass $m_2 = 37^{+7}_{-4}$~M$_\mathrm{Jup}$, eccentricity $e = 0.25^{+0.17}_{-0.18}$, period $P = 74^{+34}_{-16}$~yr, and mass ratio $q \, (m_2/m_1) = 0.032^{+0.006}_{-0.004}$ is consistent with the measurements in \cite{Currie:2020aa}.
    \item Carbon monoxide and water vapor are clearly detected in our Keck/KPIC spectra of HD~33632 Ab, both with the cross-correlation approach and the forward retrieval method \added{that utilizes on-axis host star spectra as an empirical star flux model, and substellar molecular templates as the companion models}.
    Using the forward retrieval framework with the \texttt{petitRADTRANS} package under 24 model assumptions, the mass fractions of $\log{\mathrm{CO}}$ and $\log{\mathrm{H_2O}}$ are found to be $-2.3 \pm 0.3$ and $-2.7 \pm 0.2$, respectively (or volume mixing ratios of $-3.34 \pm 0.18$ and $-3.54^{+0.17}_{-0.18}$, respectively).
    \item While methane is possibly present in the L/T transition objects, we did not detect methane in our Keck/KPIC spectra of HD~33632 Ab, using the cross-correlation method against methane molecular templates or forward retrieval models. 
    This could be due to several factors, including the relatively low signal-to-noise ratio of our KPIC spectra, the fast rotation of HD~33632 Ab, disequilibrium chemistry, and possible earlier spectral type than L9.5 previously reported. 
    \item Our inferred metallicity ([C/H] = $0.0 \pm 0.2$~dex) and C/O ratio ($0.58 \pm 0.14$) are consistent with the host star metallicity ([C/H] = $-0.13\pm0.05$; [Fe/H] = $-0.15\pm0.03$) as well as its C/O ratio ($0.39^{+0.12}_{-0.09}$) within 1.5 sigmas, expected for brown dwarf companions formed through gravitational core collapse or disk instability.
    \item HD~33632 Ab is a relatively fast rotator but follows the general trend of brown dwarf rotational demographics. 
    We compared the rotation of HD~33632 Ab with the literature spin measurements, including {\vsini} and photometric periods of 237 low-mass stars, brown dwarfs, and directly imaged exoplanets.
    Assuming uniform distributions of inclinations for a population of field and companion low-mass stars, brown dwarfs, and directly imaged exoplanets, we inferred the ratio of rotation over break-up velocity (`final' $v/v_\mathrm{break}$) at an age of 5~Gyr assuming constant angular momentum evolution.
    We examined if angular momenta for lower mass companions are less lost due to weakened magnetic interactions in the proto-planetary/stellar disks \citep{Batygin:2018aa, Ginzburg:2020aa}. 
    We found a tentative trend with a yet small sample size of 21 sources, similar to the findings in \cite{Bryan:2020ab, Wang:2021aa}.
\end{itemize}

Future observations of relative and absolute astrometry, as well as parallax for HD~33632 Ab (e.g. \textit{Gaia} DR4), will provide a longer and more precise orbital phase coverage to improve the orbital solution.
In the meantime, more advanced speckle nulling methods for KPIC \citep{Xin:2023aa}, fringing modeling, more sophisticated forward/retrieval modeling \citep{Xuan:2024aa}, as well as new hardware upgrades in 2024, will allow us to increase the sensitivity of detection and enable more precise measurements of radial and projected rotational velocities, and abundances.
The current instrument and modeling capabilities, the Keck/KPIC, VLT/CRIRES+ \citep{Follert:2014aa}, VLT/HiRISE \citep{Otten:2021aa}, and Subaru/REACH \citep{Kotani:2020aa} will significantly increase the sample of the spins and abundances for dozens of directly imaged brown dwarf companions and exoplanets, while 
next-generation high-resolution spectrographs such as Keck/HISPEC, TMT/MODHIS \citep{Mawet:2022aa, Konopacky:2023aa}, and ELT/ANDES \citep{Marconi:2022aa} will significantly expand the current rotation and abundances for low-mass companions and directly imaged exoplanets.


\begin{acknowledgments}
The authors thank the anonymous referees for \deleted{his/her/}their useful comments, which improved the original manuscript.
The authors thank Trent Dupuy for providing useful comments, which have improved this manuscript.
The authors thank the Keck observing assistants Heather Hershley, Arina Rostopchina, Tony Connors and support astronomer Carlos Alvarez for their help in obtaining the Keck/KPIC spectra.
J.X. acknowledges support from the NASA Future Investigators in NASA Earth and Space Science and Technology (FINESST) award \#80NSSC23K1434.
This research was supported in part through the computational resources and staff contributions provided for the Quest high-performance computing facility at Northwestern University which is jointly supported by the Office of the Provost, the Office for Research, and Northwestern University Information Technology.
This work used computing resources provided by Northwestern University and the Center for Interdisciplinary Exploration and Research in Astrophysics (CIERA). This research was supported in part through the computational resources and staff contributions provided for the Quest high-performance computing facility at Northwestern University which is jointly supported by the Office of the Provost, the Office for Research, and Northwestern University Information Technology.
Funding for KPIC has been provided by the California Institute of Technology, the Jet Propulsion Laboratory, the Heising-Simons Foundation (grants \#2015-129, \#2017-318, \#2019-1312, and \#2023-4598), the Simons Foundation (through the Caltech Center for Comparative Planetary Evolution), and the NSF under grant AST-1611623.
Data presented herein were obtained at the W. M. Keck Observatory, which is operated as a scientific partnership among the California Institute of Technology, the University of California, and the National Aeronautics and Space Administration. The Observatory was made possible by the generous financial support of the W. M. Keck Foundation. 
The authors recognize and acknowledge the significant cultural role
and reverence that the summit of Maunakea has with the
indigenous Hawaiian community, and that the W. M. Keck
Observatory stands on Crown and Government Lands that the
State of Hawai'i is obligated to protect and preserve for future
generations of indigenous Hawaiians.

\end{acknowledgments}

\vspace{5mm}
\facilities{Keck:II (NIRSPEC), Keck:II (NIRC2)}

\software{\texttt{Astropy} \citep{Astropy-Collaboration:2013aa,Astropy-Collaboration:2018aa}, \texttt{Numpy} \citep{Harris:2020aa}, \texttt{Scipy} \citep{Virtanen:2020aa}, \texttt{Matplotlib} \citep{Hunter:2007aa}, \texttt{DYNESTY} \citep{Speagle:2020aa}, \texttt{orbitize!} \citep{Blunt:2020aa}, \texttt{emcee} \citep{Foreman-Mackey:2013aa}, \texttt{seaborn} \citep{Waskom2021}, \texttt{SMART} \citep{Hsu:2021aa, Hsu:2021ab}
          }

\clearpage

\appendix

\section{Forward Modeling Best-Fit Results}\label{sec:appendix}

We present the remaining best-fit models and posterior probability distributions using the forward-modeling and forward-retrieval modeling methods presented in Sections~\ref{sec:forward_model} and \ref{sec:retrieval}.

\restartappendixnumbering

\begin{figure}[!htb]
    \centering
    \includegraphics[width=\textwidth]{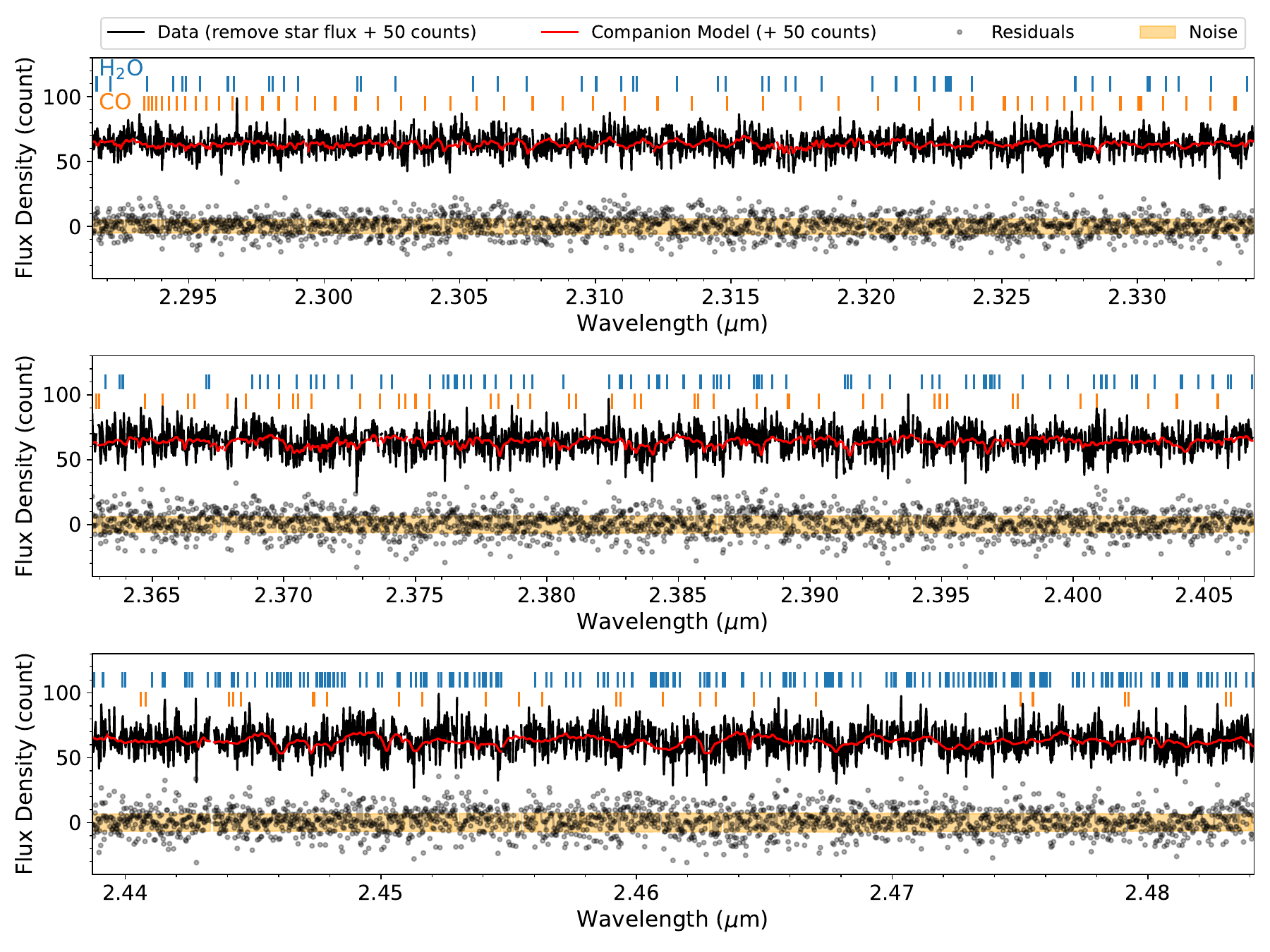}
    \caption{The residual companion spectra after removal of the modeled star flux in Figure~\ref{fig:kpic_spectrum_btsettl}. The companion spectra (black lines) and companion model (red lines) have been shifted by 50 counts for better visualization. The data noise (light yellow shaded regions) and residual (grey dots) are depicted. The CO and H$_2$O features are labeled in vertical orange and light blue lines, respectively.}
    \label{fig:kpic_spectrum_btsettl_no_star_flux}
\end{figure}

\begin{figure}[!htb]
    \centering
    \includegraphics[width=\textwidth]{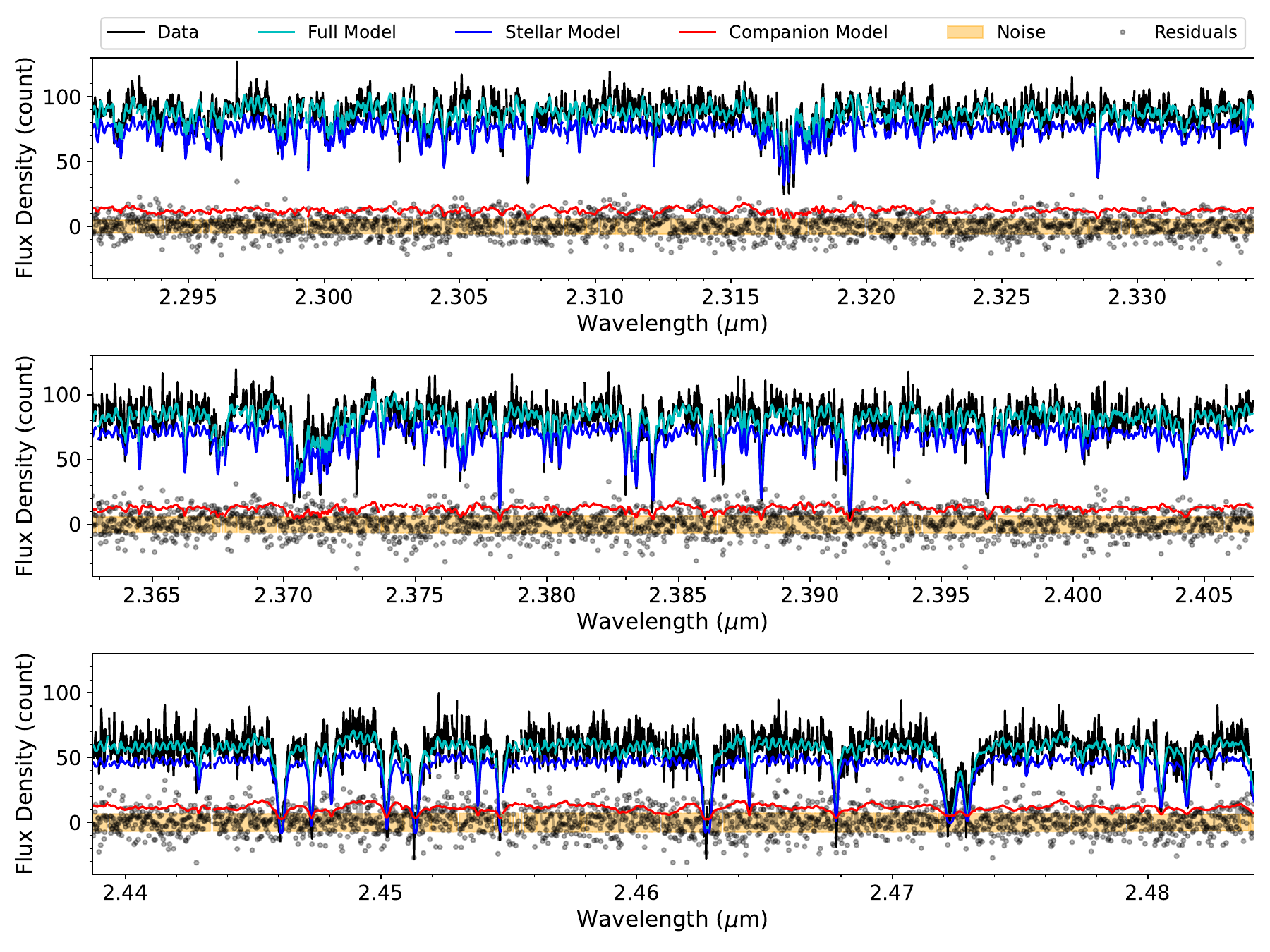}
    \caption{Same as Figure~\ref{fig:kpic_spectrum_btsettl} using the Sonora models.}
    \label{fig:kpic_spectrum_sonora}
\end{figure}

\begin{figure}[!htb]
    \centering
    \includegraphics[width=\textwidth]{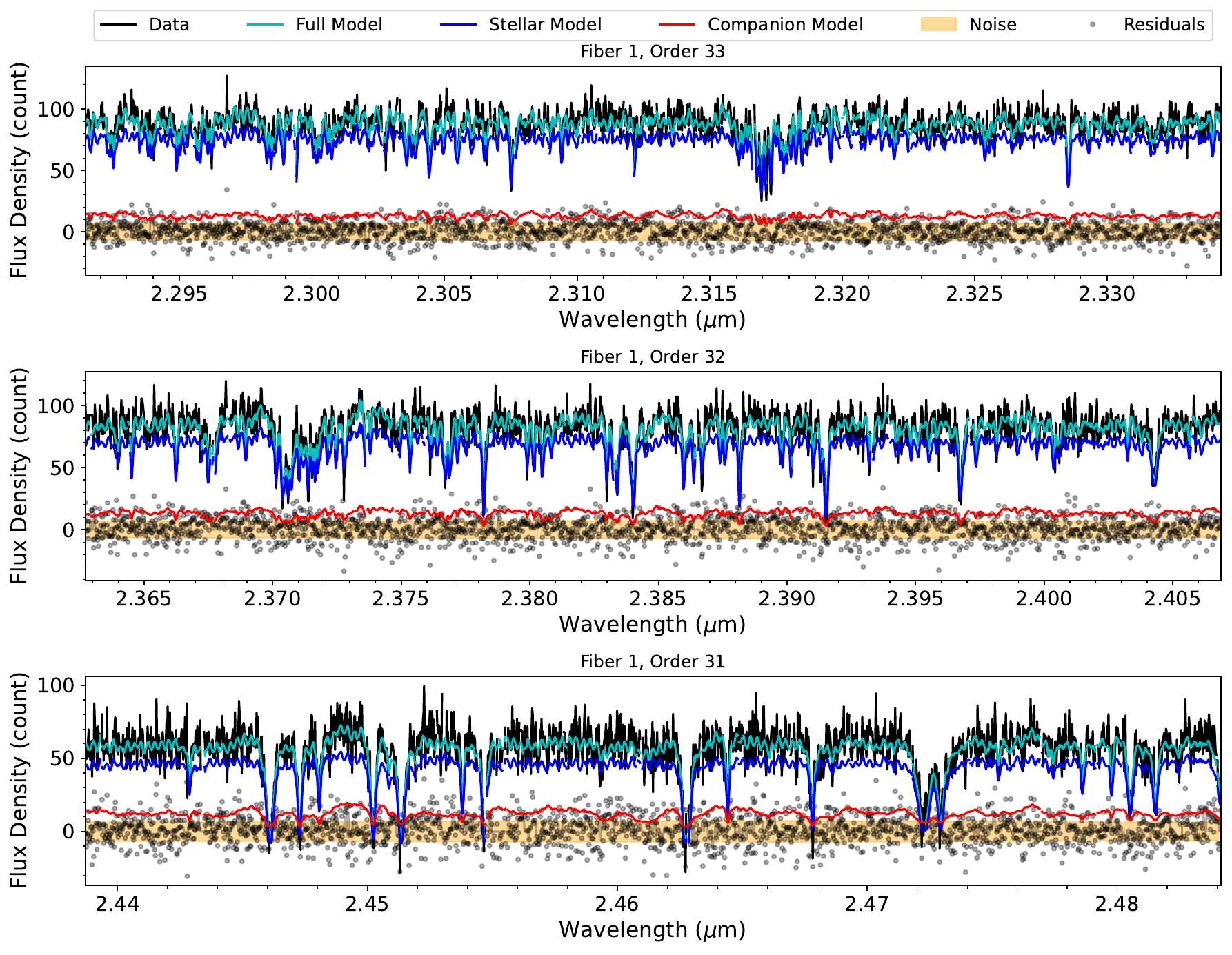}
    \caption{The best-fit baseline forward retrieval model and fiber 1 spectrum of HD~33632 Ab. The labels are identical to Figure~\ref{fig:kpic_spectrum_btsettl}.}
    \label{fig:retrieval_spectrum_baseline}
\end{figure}

\begin{figure*}[ht]
    \centering
    \includegraphics[width=\textwidth]{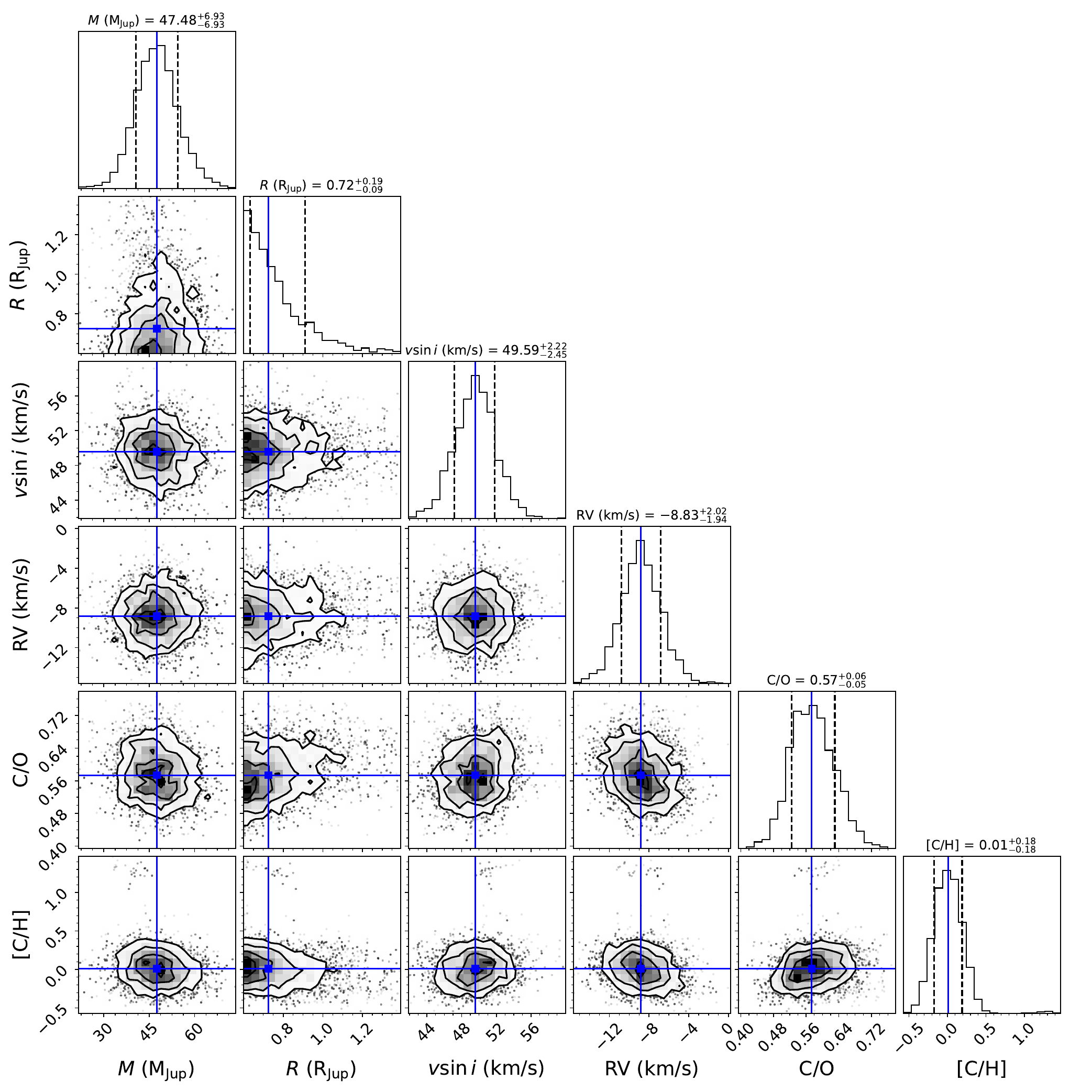}
    \caption{Posterior probability distributions of our baseline forward retrieval model of HD~33632 Ab for the selected substellar parameters.
    The median values are labeled in blue lines, and the 16$^\mathrm{th}$ and 84$^\mathrm{th}$ percentiles are labeled in black dashed lines.
    }
    \label{fig:retrieval_posterior_baseline}
\end{figure*}

\FloatBarrier 
\section{Orbital Fitting}

In Section~\ref{sec:orbit}, we presented the refined orbital solutions using our relative RV, \textit{Gaia} eDR3 astrometry, and the updated Hipparcos-Gaia Catalog of Accelerations \citep{Brandt:2021aa}. Here we present the best-fit orbits and the associated orbital diagnostics and a summary table under various assumptions.

\restartappendixnumbering

\begin{figure*}[!hbp]
    \centering
    \includegraphics[width=1.0\textwidth, trim=0 2cm 0 0]{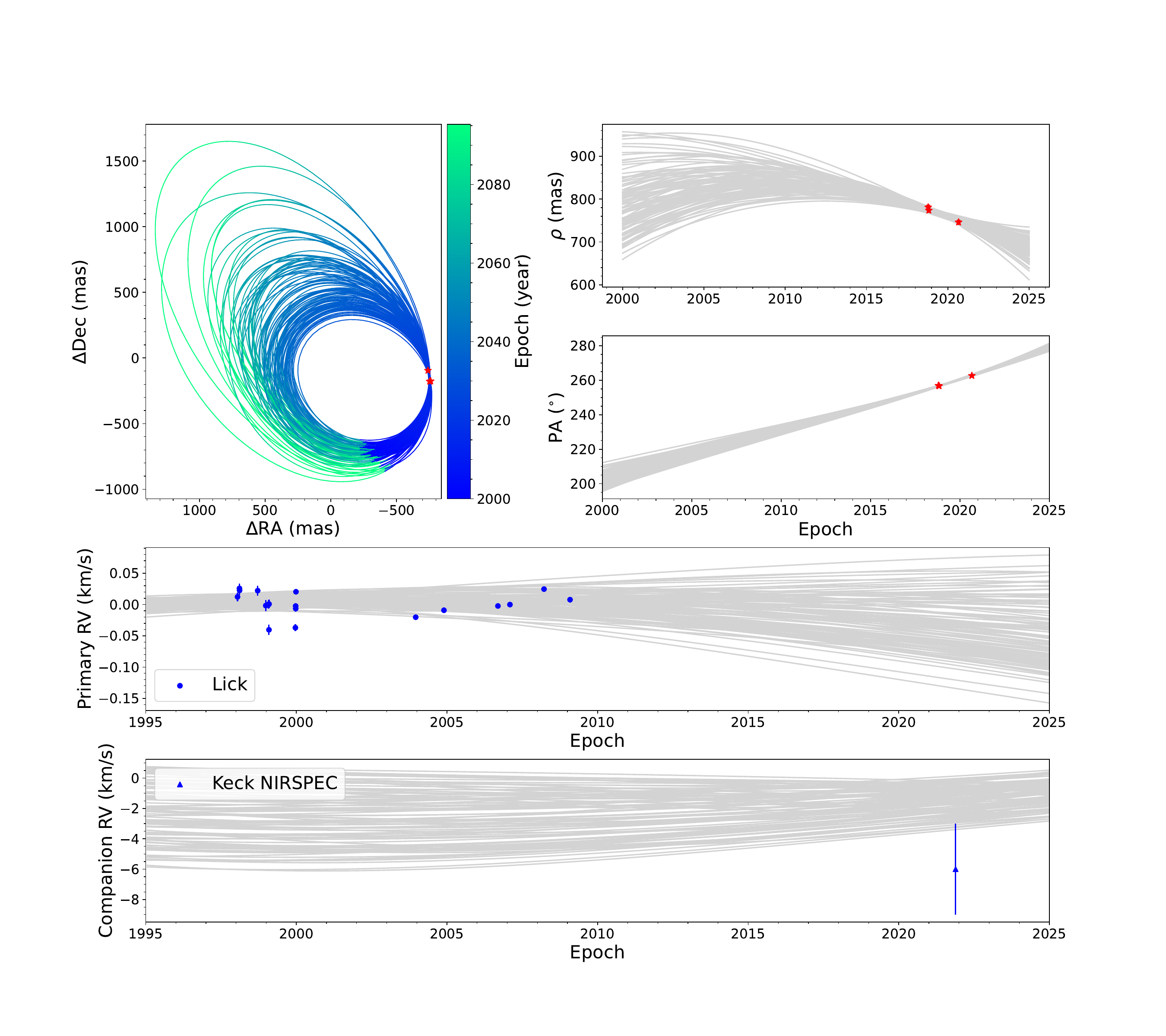}
    \caption{Best-fit 100 random draw orbits for HD~33632 Ab, using host and companion RVs and the Hipparcos-Gaia (EDR3) Catalog of Accelerations (HGCA; \citealp{Brandt:2021aa}) with the \texttt{orbitize!} package. 
    \textit{Upper-left}: best-fit orbits in $\Delta$RA and $\Delta$Dec coordinates for the HD~33632 Ab system (blue/green lines) and the relative astrometry (red stars) from \cite{Currie:2020aa}. 
    \textit{Upper-right; top}: best-fit separation ($\rho$) in mas (grey lines) with relative astrometry (red stars) from \cite{Currie:2020aa}; \textit{Upper-right; bottom}: best-fit position angles (PA) in degrees (grey lines) with the relative astrometry (red stars) from \cite{Currie:2020aa}.
    \textit{Middle}: best-fit RV orbits (grey lines) for the host star HD~33632A using the Lick measurements (blue dots).
    \textit{Bottom}: best-fit RV orbits (grey lines) for the companion HD~33632 Ab using our Keck/KPIC measurements (blue triangle).
    See Section~\ref{sec:orbit} for details.}
    \label{fig:hgca_orbit}
\end{figure*}


\begin{figure*}[!htbp]
\centering
\gridline{\fig{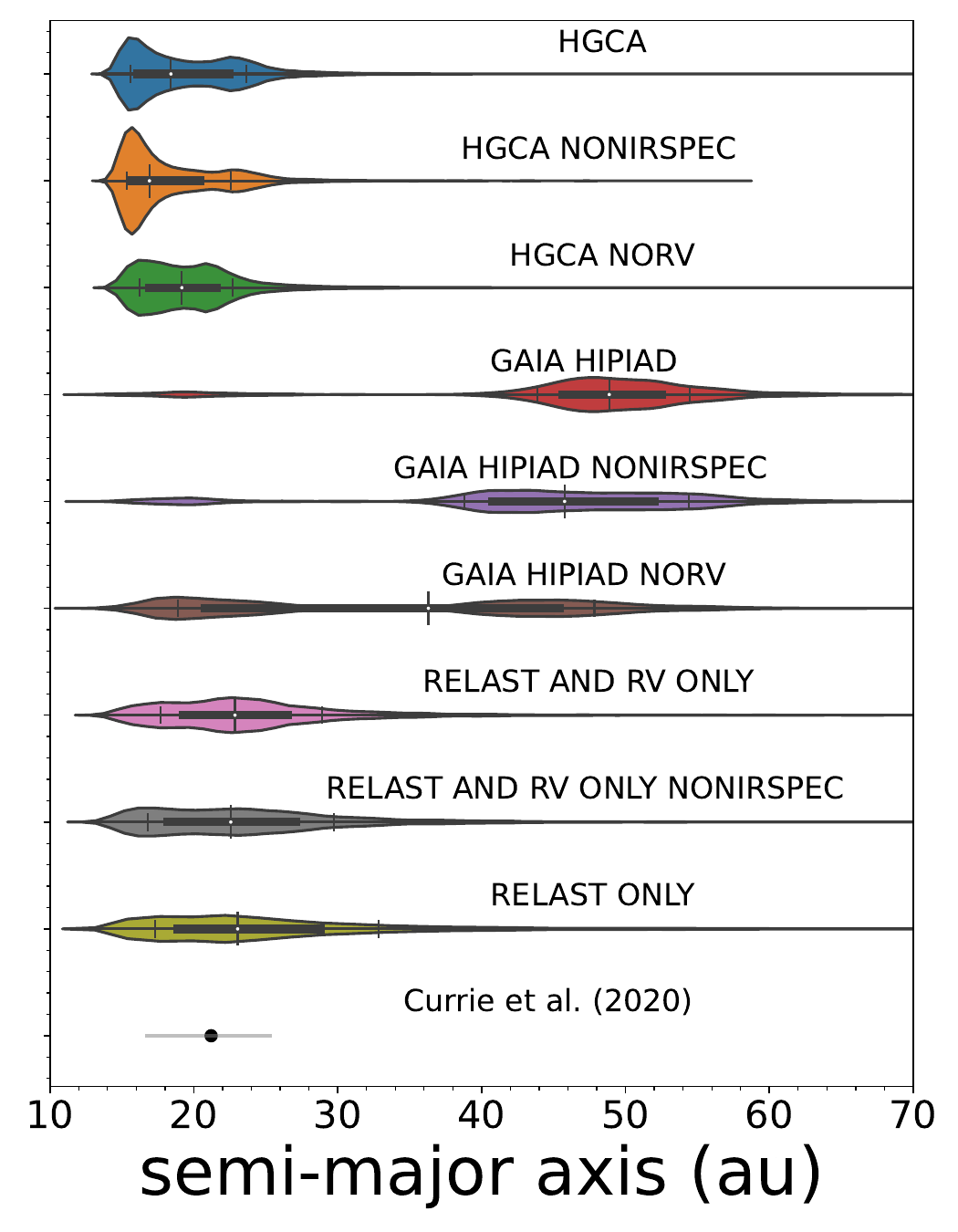}{0.33\textwidth}{}
\fig{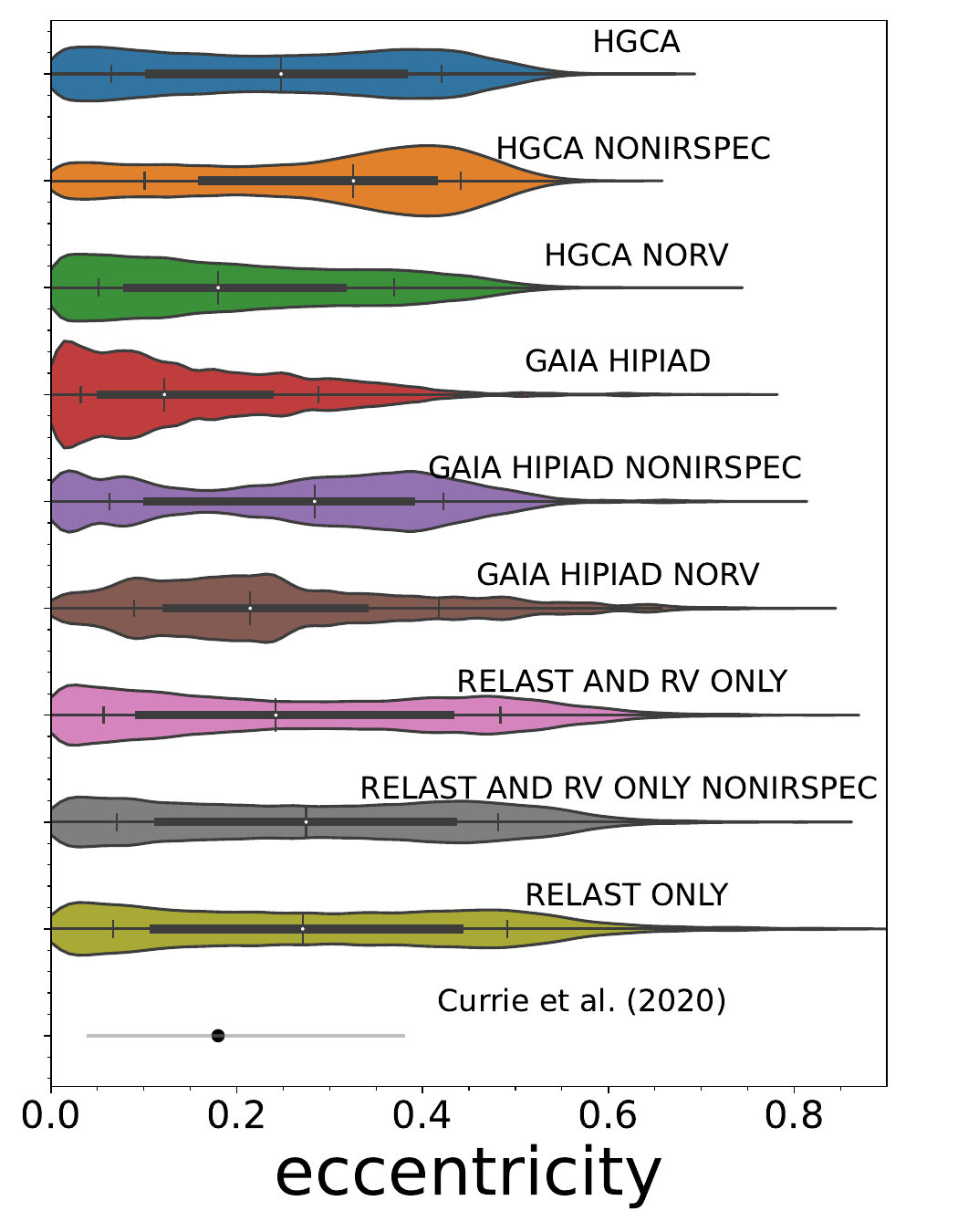}{0.33\textwidth}{}
\fig{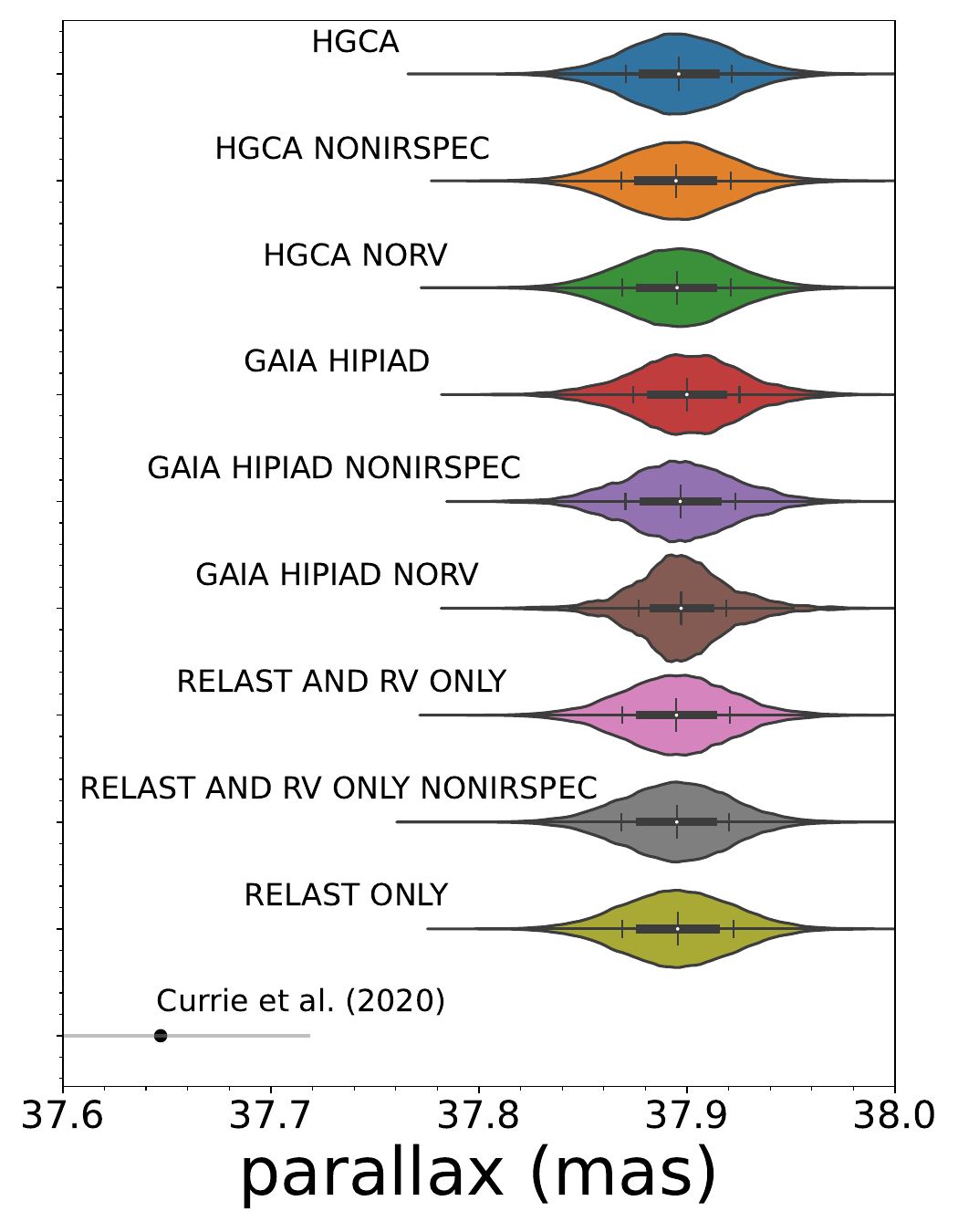}{0.33\textwidth}{}
}
\gridline{
\fig{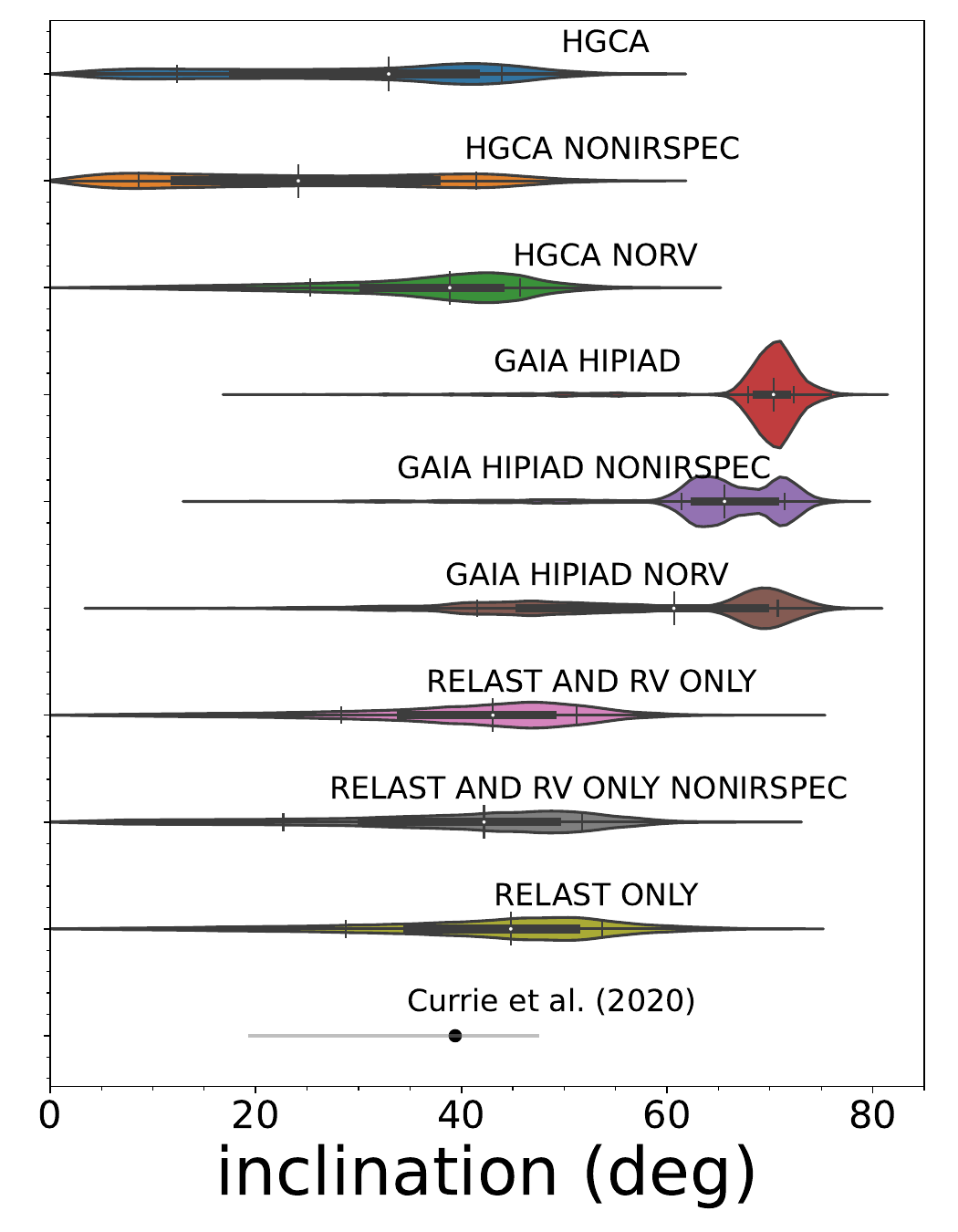}{0.33\textwidth}{}
\fig{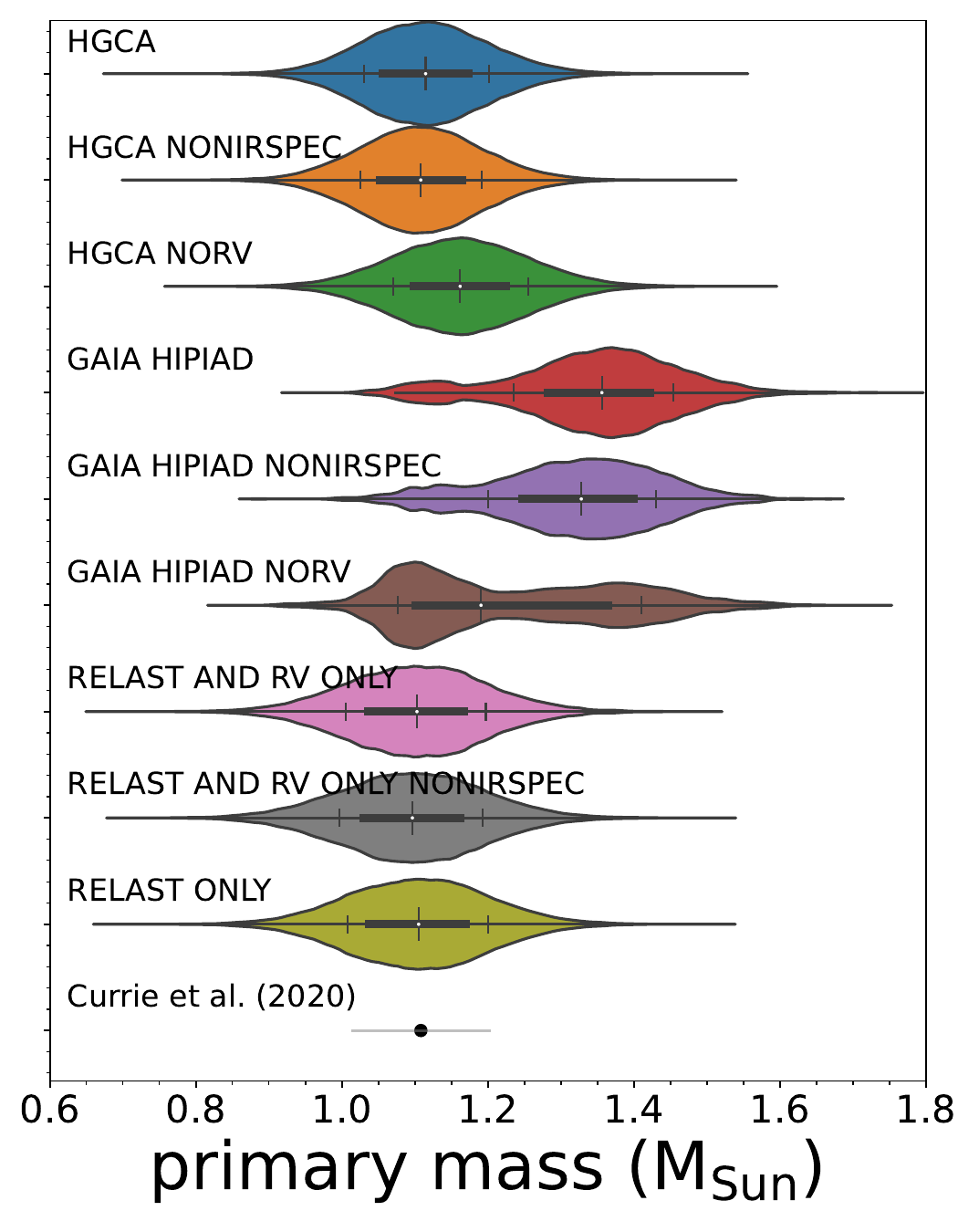}{0.33\textwidth}{}
\fig{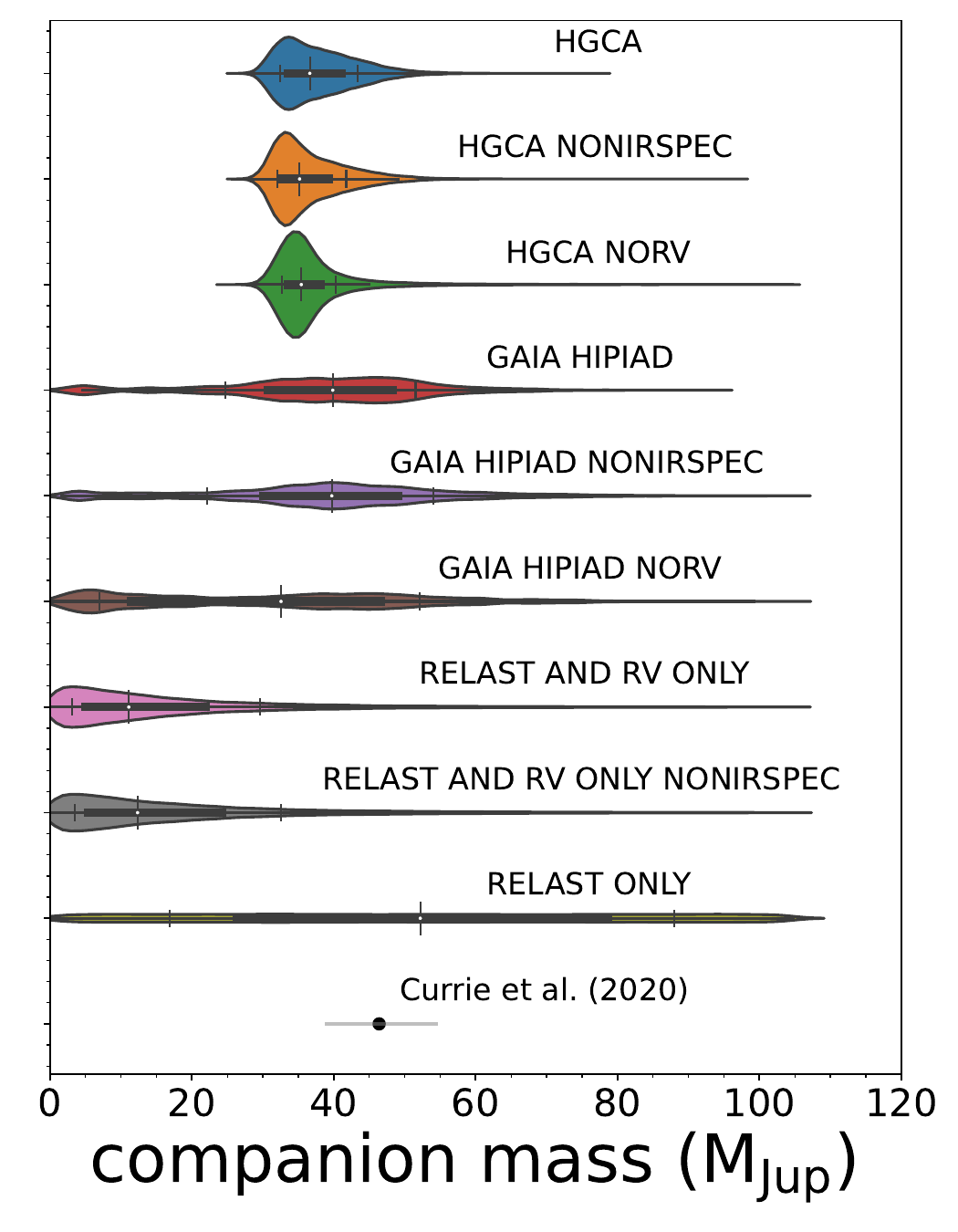}{0.33\textwidth}{}
}
\caption{Violin plots of the best-fit orbital parameters. \textit{Top-left:} semi-major axis (au); \textit{Top-middle:} eccentricity; \textit{Top-right:} parallax (mas); \textit{Bottom-left:} inclination (deg); \textit{Bottom-middle:} primary mass (M$_{\odot}$); \textit{Bottom-right:} secondary mass (M$_\mathrm{J}$).
Each panel, from top to bottom, illustrates (1--3): the joint orbit fit using the EDR3 edition of the Hipparcos–Gaia Catalog of Accelerations (HGCA; \citealp{Brandt:2021aa}); (4--6) the joint orbit fit using \textit{Gaia} and \textit{Hipparcos} IAD data from \cite{van-Leeuwen:2007aa}; (7--9) using relative astrometry measurements in \cite{Currie:2020aa}; and (10) orbital solutions in \cite{Currie:2020aa}.
Under each scenario, we consider three combinations: (1) one with both primary (Lick) and secondary (NIRSPEC) RVs, (2) one with only primary (Lick) RV, and (3) one without RVs.
The major difference illustrates the updated \textit{Gaia} EDR3 version of HGCA and \textit{Gaia} DR3 parallaxes compared to earlier solutions illustrated in \cite{Currie:2020aa}.
\label{fig:orbit_fit_compare}}
\end{figure*}

\begin{longrotatetable}
\begin{deluxetable*}{l|ccc|ccc|ccc|cc}
\tablewidth{700pt}
\tablecaption{HD~33632 Ab Orbital Best-fit Parameters \label{tab:orbit}} 
\tabletypesize{\scriptsize} 
\tablehead{ 
\colhead{Parameter} & 
\multicolumn{3}{c}{HGCA} & 
\multicolumn{3}{c}{\textit{Gaia} and \textit{Hipparcos} IAD} & 
\multicolumn{3}{c}{Relative Astrometry Only} & 
\colhead{Prior\tablenotemark{a}} &
\colhead{Currie+2020} \\
\colhead{} & \colhead{All RVs} & \colhead{No RV$_2$} & \colhead{No RVs} & \colhead{All RVs} & \colhead{No RV$_2$} & \colhead{No RVs} &
\colhead{All RVs} & \colhead{No RV$_2$} & \colhead{No RVs} &
\colhead{} & \colhead{} 
} 
\startdata
\multicolumn{11}{c}{Fitted Parameters}\\
\hline
$a$ (au) & $18^{+5}_{-3}$ & $17^{+6}_{-2}$ & $19^{+4}_{-3}$ & $49^{+6}_{-5}$ & $46^{+9}_{-7}$ & $36^{+12}_{-17}$ & $23^{+6}_{-5}$ & $23^{+7}_{-6}$ & $23^{+10}_{-6}$ & $\mathcal{U}$(0.001, 10000) & 21.2$^{+4.1}_{-4.5}$ \\
$e$ & $0.25^{+0.17}_{-0.18}$ & $0.33^{+0.12}_{-0.22}$ & $0.18^{+0.19}_{-0.13}$ & $0.12^{+0.17}_{-0.09}$ & $0.28^{+0.14}_{-0.22}$ & $0.21^{+0.2}_{-0.12}$ & $0.24^{+0.24}_{-0.19}$ & $0.27^{+0.21}_{-0.2}$ & $0.27^{+0.22}_{-0.2}$ & $\mathcal{U}$(0, 1) & 0.18$^{+0.20}_{-0.14}$ \\
$i$ (deg) & $33^{+11}_{-21}$ & $24^{+17}_{-16}$ & $39^{+7}_{-14}$ & $70^{+2}_{-2}$ & $66^{+6}_{-4}$ & $61^{+10}_{-19}$ & $43^{+8}_{-15}$ & $42^{+10}_{-20}$ & $45^{+9}_{-16}$ & $\mathcal{U}$(Sine) & 39$^{+8}_{-20}$ \\
$\pi$ (mas) & $37.9^{+0.03}_{-0.03}$ & $37.89^{+0.03}_{-0.03}$ & $37.9^{+0.03}_{-0.03}$ & $37.9^{+0.03}_{-0.03}$ & $37.9^{+0.03}_{-0.03}$ & $37.9^{+0.02}_{-0.02}$ & $37.89^{+0.03}_{-0.03}$ & $37.9^{+0.03}_{-0.03}$ & $37.9^{+0.03}_{-0.03}$ & $\mathcal{N}$(37.8953, 0.026) & 37.647$\pm$0.071 \\
$\omega$ (deg) & $65^{+248}_{-47}$ & $47^{+265}_{-30}$ & $134^{+202}_{-113}$ & $105^{+125}_{-26}$ & $203^{+18}_{-106}$ & $93^{+188}_{-39}$ & $233^{+80}_{-194}$ & $219^{+97}_{-194}$ & $225^{+89}_{-191}$ & $\mathcal{U}$(0, $\pi$) & 151$^{+155}_{-131}$ \\
$\Omega$ (deg) & $39^{+21}_{-9}$ & $38^{+38}_{-11}$ & $54^{+17}_{-17}$ & $21^{+148}_{-6}$ & $175^{+4}_{-157}$ & $30^{+64}_{-13}$ & $51^{+64}_{-18}$ & $43^{+68}_{-19}$ & $47^{+63}_{-27}$ & $\mathcal{U}$(0, 2$\pi$) & 38$\pm$7 \\
$\tau$ & $0.37^{+0.29}_{-0.05}$ & $0.35^{+0.15}_{-0.03}$ & $0.4^{+0.3}_{-0.08}$ & $0.53^{+0.08}_{-0.17}$ & $0.25^{+0.29}_{-0.06}$ & $0.4^{+0.13}_{-0.22}$ & $0.29^{+0.41}_{-0.16}$ & $0.3^{+0.24}_{-0.15}$ & $0.3^{+0.38}_{-0.17}$ & $\mathcal{U}$(0, 1) & 0.48$^{+0.2}_{-0.23}$ \\
$m_1$ (M$_{\mathrm{J}}$) & $37^{+7}_{-4}$ & $35^{+7}_{-3}$ & $35^{+5}_{-3}$ & $40^{+12}_{-15}$ & $40^{+14}_{-18}$ & $33^{+20}_{-26}$ & $11^{+19}_{-8}$ & $12^{+20}_{-9}$ & $52^{+36}_{-35}$ & $\mathcal{U}$(0, 0.1) M$_{\odot}$ & 46.4$^{+8.1}_{-7.5}$ \\
$m_0$ (M$_{\odot}$) & $1.11^{+0.09}_{-0.08}$ & $1.11^{+0.08}_{-0.08}$ & $1.16^{+0.09}_{-0.09}$ & $1.36^{+0.1}_{-0.12}$ & $1.33^{+0.1}_{-0.13}$ & $1.19^{+0.22}_{-0.11}$ & $1.1^{+0.09}_{-0.1}$ & $1.1^{+0.1}_{-0.1}$ & $1.1^{+0.09}_{-0.1}$ & $\mathcal{N}$(1.11, 0.09) & 1.11$\pm$0.09 \\
\hline
\multicolumn{11}{c}{Derived Parameters}\\
\hline
$P$ (yr) & $74^{+34}_{-16}$ & $66^{+34}_{-10}$ & $77^{+23}_{-17}$ & $289^{+51}_{-44}$ & $266^{+72}_{-59}$ & $196^{+82}_{-119}$ & $104^{+44}_{-33}$ & $102^{+52}_{-36}$ & $103^{+72}_{-36}$ &  & 91$\pm$27 \\
$m_1$/$m_0$ & $0.032^{+0.006}_{-0.004}$ & $0.031^{+0.006}_{-0.003}$ & $0.029^{+0.005}_{-0.003}$ & $0.028^{+0.008}_{-0.011}$ & $0.028^{+0.011}_{-0.012}$ & $0.024^{+0.014}_{-0.018}$ & $0.01^{+0.017}_{-0.007}$ & $0.011^{+0.018}_{-0.008}$ & $0.045^{+0.031}_{-0.03}$ &  & 0.0402$^{+0.0078}_{-0.0073}$ \\
\enddata
\tablenotetext{a}{$\mathcal{U}$ and $\mathcal{N}$ stand for uniform and normal priors, respectively.}
\tablenotetext{b}{Sampled in the log space}
\tablenotetext{c}{Derived from the periastron time in \cite{Currie:2020aa}}
\end{deluxetable*}
\end{longrotatetable}

\clearpage

\bibliography{mylibrary}{}
\bibliographystyle{aasjournal}

\end{document}